\documentclass[fleqn,usenatbib]{mnras}

\usepackage{newtxtext}
\usepackage[T1]{fontenc}

\DeclareRobustCommand{\VAN}[3]{#2}
\let\VANthebibliography\thebibliography
\def\thebibliography{\DeclareRobustCommand{\VAN}[3]{##3}\VANthebibliography}

\usepackage{graphicx}	% Including figure files
\usepackage{amsmath}	% Advanced maths commands
\usepackage{amssymb}	% Extra maths symbols
\usepackage{subcaption}
\title[Multi-line study of jet-ISM interplay in NGC\,3100]{\centering{The AGN fuelling/feedback cycle in nearby radio galaxies - IV. Molecular gas conditions and jet-ISM interaction in NGC\,3100}}

\author[I. Ruffa et al.]{Ilaria Ruffa,$^{1,2}$\thanks{E-mail: RuffaI@cardiff.ac.uk}
Isabella Prandoni,$^{2}$
Timothy A. Davis,$^{1}$
Robert A. Laing,$^{3}$
Rosita Paladino,$^{2}$
\newauthor
Viviana Casasola,$^{2}$
Paola Parma,$^{2}$
and Martin Bureau$^{4}$
\\
$^{1}$Cardiff Hub for Astrophysics Research \&\ Technology, School of Physics \&\ Astronomy, Cardiff University, Queens Buildings, The Parade, Cardiff, CF24 3AA, UK\\
$^{2}$INAF - Istituto di Radioastronomia, via P.\ Gobetti 101, 40129 Bologna, Italy\\
$^{3}$Square Kilometre Array Organisation, Jodrell Bank Observatory, Lower Withington, Macclesfield, Cheshire SK11 9DL, UK\\
$^{4}$Sub-dept.\ of Astrophysics, Dept.\ of Physics, University of Oxford, Denys Wilkinson Building, Keble Road, Oxford OX1 3RH, UK
}

\date{Accepted XXX. Received YYY; in original form ZZZ}

\pubyear{2020}

\def\cotoh2{CO-to-H$_{2}$}

\defcitealias{Ruffa19a}{Paper~I}
\defcitealias{Ruffa19b}{Paper~II}
\defcitealias{Ruffa20}{Paper~III}

\begin{document}

\label{firstpage}
\pagerange{\pageref{firstpage}--\pageref{lastpage}}
\maketitle

\begin{abstract}
This is the fourth paper of a series investigating the AGN fuelling/feedback processes in a sample of eleven nearby low-excitation radio galaxies (LERGs). In this paper we present follow-up Atacama Large Millimeter/submillimeter Array (ALMA) observations of one source, NGC\,3100, targeting the $^{12}$CO(1-0), $^{12}$CO(3-2), HCO$^{+}$(4-3), SiO(3-2) and HNCO(6-5) molecular transitions. $^{12}$CO(1-0) and $^{12}$CO(3-2) lines are nicely detected and complement our previous $^{12}$CO(2-1) data. By comparing the relative strength of these three CO transitions, we find extreme gas excitation conditions (i.e.\,$T_{\rm ex}\gtrsim50$~K) in regions that are spatially correlated with the radio lobes, supporting the case for a jet-ISM interaction. An accurate study of the CO kinematics demonstrates that, although the bulk of the gas is regularly rotating, two distinct non-rotational kinematic components can be identified in the inner gas regions: one can be associated to inflow/outflow streaming motions induced by a two-armed spiral perturbation; the second one is consistent with a jet-induced outflow with $v_{\rm max}\approx 200$~km~s$^{-1}$ and $\dot{M}\lesssim 0.12$~M$_{\odot}$~yr$^{-1}$. These values indicate that the jet-CO coupling ongoing in NGC\,3100 is only mildly affecting the gas kinematics, as opposed to what expected from existing simulations and other observational studies of (sub-)kpc scale jet-cold gas interactions. HCO$^{+}$(4-3) emission is tentatively detected in a small area adjacent to the base of the northern radio lobe, possibly tracing a region of jet-induced gas compression.
The SiO(3-2) and HNCO(6-5) shock tracers are undetected: this - along with the tentative HCO$^{+}$(4-3) detection - may be consistent with a deficiency of very dense (i.e.\,$n_{\rm crit} > 10^{6}$~cm$^{-3}$) cold gas in the central regions of NGC\,3100.
\end{abstract}


%%%%%%%%%%%%%%%%%%%%%%%%%%%%%%%%%%%%%%%%%%%%%%%%%%

%%%%%%%%%%%%%%%%% BODY OF PAPER %%%%%%%%%%%%%%%%%%

\section{Introduction}\label{sec:intro}
Both theoretical and observational evidence increasingly suggests that the energy released by feedback processes associated with active galactic nuclei (AGN) is a fundamental component in the evolution of their host galaxies \citep[e.g.][]{Combes17,Choi18,Pillepich19}. The complexity of these phenomena, however, makes many of their key aspects still unclear  \citep[e.g.][]{Wylezalek18,Harrison20}.

AGN feedback is currently believed to operate in two (non-exclusive) main modes: radiative (or quasar) mode and kinetic (or jet) mode \citep[e.g.][]{Morganti17}. The former is considered to be dominant in radiatively efficient (quasar- or Seyfert-like) AGN accreting matter at high rates ($\ga 0.01$ \.{M}$_{\rm Edd}$, where \.{M}$_{\rm Edd}$ is the Eddington accretion rate\footnote{$\dot{M}_{\rm Edd} = \dfrac{4\pi \, G \, M_{\rm SMBH} \, m_{\rm p}}{\varepsilon \, c \, \sigma_{\rm T}}$, where $G$ is the gravitational constant, $M_{\rm SMBH}$ is the mass of the central super-massive black hole, $m_{\rm p}$ is the mass of the proton, $\varepsilon$ is the accretion efficiency, $c$ is the speed of light and $\sigma_{\rm T}$ is the cross-section for Thomson scattering.}).
In these objects, the radiation pressure generated by the accretion process can drive powerful outflows (the so-called AGN winds) that are potentially able to expel large quantities of matter from the nuclear regions (\citealp[e.g.][]{King15,Wagner16,Bieri17,Ishibashi18}). Kinetic feedback is instead dominant in AGN accreting matter at low rates \citep[$\ll 0.01$ \.{M}$_{\rm Edd}$; e.g.][]{Cielo18}, whereby the bulk of the energy generated from the accretion process is channelled into collimated outflows of non-thermal plasma (i.e.\,the radio jets). 

The impact that kinetic AGN feedback can have on large (i.e.\,tens of kpc) scales is well established. Expanding radio jets have been often observed to inflate cavities through the hot X-ray emitting atmospheres of galaxies, groups, and clusters, heating the surroundings, balancing radiative losses and preventing the gas from further cooling 
\citep[see, e.g.,][for a review]{McNamara12}. In the past decade, however, jet-mode feedback has been increasingly observed to alter the distribution, kinematics and physics of the surrounding gaseous medium also on (sub-)kpc scales (\citealp[e.g.][see also \citealt{Veilleux20}, for a recent review on this subject]{Combes13,Garcia14,Morganti15,Mahony16,Zovaro19,Murthy19,Santoro20,Venturi20}). Observations find support in 3D hydrodynamical simulations \citep[e.g.][]{Wagner12,Wagner16,Mukherjee16,Mukherjee18a,Mukherjee18b} showing that young and compact radio jets expanding through the surrounding layers of matter can 
produce turbulent cocoons of shocked gas that can be accelerated up to $1000$~km~s$^{-1}$ over a wide range of directions. The extent of jet-induced perturbations has been predicted to vary depending on the radio jet power, relative jet--gas orientation and age of the radio source. Large samples of objects covering wide ranges of jet ages, radio powers, and jet-ISM geometric configurations are then required to gain a better understanding of the incidence and properties of jet-ISM interactions. This is essential information to fully assess the role of jet-induced feedback in galaxy evolution \citep[e.g.][]{Morganti20b}.  

Molecular line observations are powerful tools to investigate the physical conditions of the gas in various environments, as different molecules, isotopologues, and transitions of the same species trace different gas components within the same galaxy. The most used molecular gas tracer is CO: it is the most abundant molecule after H$_{2}$ and its low excitation temperature (i.e.\,$T_{\rm ex}=5.53$~K for the ground $J =1\rightarrow0$ rotational transition; \citealp[e.g.][]{Carilli13}) makes it easily excited even at low temperatures. The critical density (i.e.\,the density at which collisional excitation balances spontaneous radiative de-excitation) for the CO(1-0) transition is $n_{\rm crit} \approx 2200$~cm$^{-3}$. Higher-$J$ transitions require rapidly increasing densities and temperatures to be excited (i.e.\, $n_{\rm crit} \propto J^{3}$ and $T_{\rm ex} \propto J^{2}$; \citealp[see e.g.][]{Bolatto13}), therefore ratios of multiple CO lines can be used to probe the excitation temperature and density of molecular gas (\citealp[e.g.][]{Kam17,Kam18,Saito17}). Ratios of $^{12}$CO and $^{13}$CO isotopologues can also provide estimates of the gas optical depth, as $^{12}$CO emission is generally optically thick, while $^{13}$CO is mostly (but not totally; see e.g.\,\citealt[][]{Cormier18}) optically thin over large areas \citep[e.g.][]{Jimenez17}. The SiO molecule is commonly considered a good tracer of shocks. Indeed, silicon particles are mostly found in the core of dust grains, and can be released only in grain sputtering or evaporation caused by high-velocity shocks. Once Si-bearing material is in the gas phase, it can react with molecular oxygen to form SiO \citep{Schilke97}, which can then be used to trace regions of fast shocks ($v>50$~km~s$^{-1}$; \citealp[e.g.][]{Usero06}). Growing observational evidence suggests that the HNCO molecule is a particularly good tracer of slow ($v<20$~km~s$^{-1}$) shocks: its presence in the outer layers of dust grains means that it is easily sublimated, and it remains detectable only in weakly-shocked regions \citep[e.g.][]{Kelly17}. HCO$^{+}$, along with CS, HCN and HNC, are the most used tracers of the dense ($n_{\rm crit}>10^{4}$~cm$^{-3}$) cold gas component \citep[e.g.][]{Gao04,Costagliola11,Garcia14}. The abundance of these molecules, relative to CO, is commonly used to investigate the density conditions of the cold gas \citep[e.g.][]{Usero15,Topal16,Bigiel16}. A multi-species approach is therefore key to properly characterise the impact of jet-induced feedback on the physics of the surrounding cold gas. 

Despite their relevance, detailed spatially-resolved studies of jet-ISM interactions using multiple molecular transitions have been carried out so far only for a limited number of objects, mostly nearby (Seyfert-like) galaxies hosting (sub-)kpc scale jets \citep[see e.g.][and references therein]{Alatalo13,Combes13,Garcia14,Feruglio15,Oosterloo17}. These few existing studies show that line ratios are drastically modified in the (vicinity of) regions where kinetic AGN feedback is in action, implying great differences between the physical conditions of the quiescent and interacting gas: the latter is usually clumpy, with much higher temperatures and densities. Signs of compression and fragmentation in the molecular gas distribution have been also observed and attributed to the driving mechanism of the interaction.

Our aim is to expand this kind of studies to a specific class of low-power radio galaxies, known as low-excitation radio galaxies (LERG).  LERGs are, by number, the dominant radio galaxy population in the local Universe \citep[e.g.][]{Hardcastle07}, predominantly hosted by red, massive ($M_{*} \geq 10^{11}~M_{\odot}$) early-type galaxies (ETGs; \citealp[e.g.][]{Best12}). These objects mostly have  Fanaroff-Riley type I radio morphologies \citep[FR\,I; ][]{Fanaroff74} and low or moderate radio luminosities ($P_{\rm 1.4GHz}<10^{25}$~W~Hz$^{-1}$). The central super-massive black holes (SMBHs) in LERGs  accrete matter at low rates ($\ll 0.01$ \.{M}$_{\rm Edd}$) and produce almost exclusively kinetic feedback. Few detailed, spatially-resolved studies of LERGs have yet been made, so their triggering mechanisms and associated AGN feedback processes are still poorly understood \citep[e.g.][]{Hardcastle18}: investigating the nature of LERGs is crucial to shed light on the mechanisms determining the observed properties of massive nearby ETGs.

We are carrying out the first, extensive, spatially-resolved, multi-phase study of a small but complete sample of eleven LERGs in the southern sky. In this paper, we present follow-up ALMA observations of one object, NGC\,3100, targeting  $^{12}$CO(1-0) and $^{12}$CO(3-2) transitions, along with HCO$^{+}$(4-3), SiO(3-2) and HNCO(6-5) lines. Coupling these new data with previous $^{12}$CO(2-1) and radio continuum observations, we primarily aim at carrying out an analysis of the molecular gas physics and jet-CO interplay in this object.

The paper is structured as follows. In Section~\ref{sec:sample} we briefly present the overall project and the target of this study.  
In Section~\ref{sec:obs} we describe ALMA observations and the data reduction. The data analysis is presented in Section~\ref{sec:cube_analysis}. We discuss the results in Section~\ref{sec:discuss}, before summarising and concluding in Section~\ref{sec:conclusion}.

Throughout this work we assume a standard $\Lambda$CDM cosmology with H$_{\rm 0}=70$\,km\,s$^{-1}$\,Mpc$^{\rm -1}$, $\Omega_{\rm \Lambda}=0.7$ and $\Omega_{\rm M}=0.3$. This gives a scale of $180$~pc/$''$ at the redshift of NGC\,3100 (see Table~\ref{tab:ngc3100}). All of the velocities in this paper are given in the optical convention and Kinematic Local Standard of Rest (LSRK) frame.

\begin{table}
\centering
\caption{General properties of NGC\,3100.}
\label{tab:ngc3100}
\begin{tabular}{l l c c}
\hline
\multicolumn{1}{c}{ ID } &
\multicolumn{1}{c}{ Parameter} &
\multicolumn{1}{c}{ Value } \\
\hline
(1) & Type & S0\\
(2) & RA (J2000) & 10$^{\rm h}$00$^{\rm m}$40$^{\rm s}$.8\\
(3) & DEC (J2000) & -31$^{\circ}$39$^{'}$52$^{''}$.3\\
(4) &  Redshift &  0.0088 \\
(5) & D$_{\rm L}$ (Mpc) &  38.0 \\
(6) &  Radio Source & PKS 0958$-$314  \\
 (7) & log~$P_{\rm 1.4GHz}$ (W\,Hz$^{-1}$) &  23.0   \\
 (8) & $v_{\rm opt}$ (km~s$^{-1}$) & $2629\pm20$ \\
\hline
\end{tabular}
\parbox[t]{8.5cm}{ \textit{Notes.} $-$ Rows: (1) Morphological type of the galaxy from the HyperLeda database (http://leda.univ-lyon1.fr). (2) and (3) Galaxy sky coordinates from \citet[][]{Petrov05}. (4) Galaxy redshift from \citet[][]{Dopita15}. (5) Luminosity distance derived from the tabulated redshift and assuming the cosmology stated in Section~\ref{sec:intro}. (6) Name of the radio source.  (7) Radio power at 1.4~GHz derived from the flux density given in \citet[][]{Brown11}, including all the radio emission associated with the source. (8) Optical stellar velocity from \citet[][]{Kerr86}.}
\end{table}

\begin{figure*}
\centering
\includegraphics[scale=0.6]{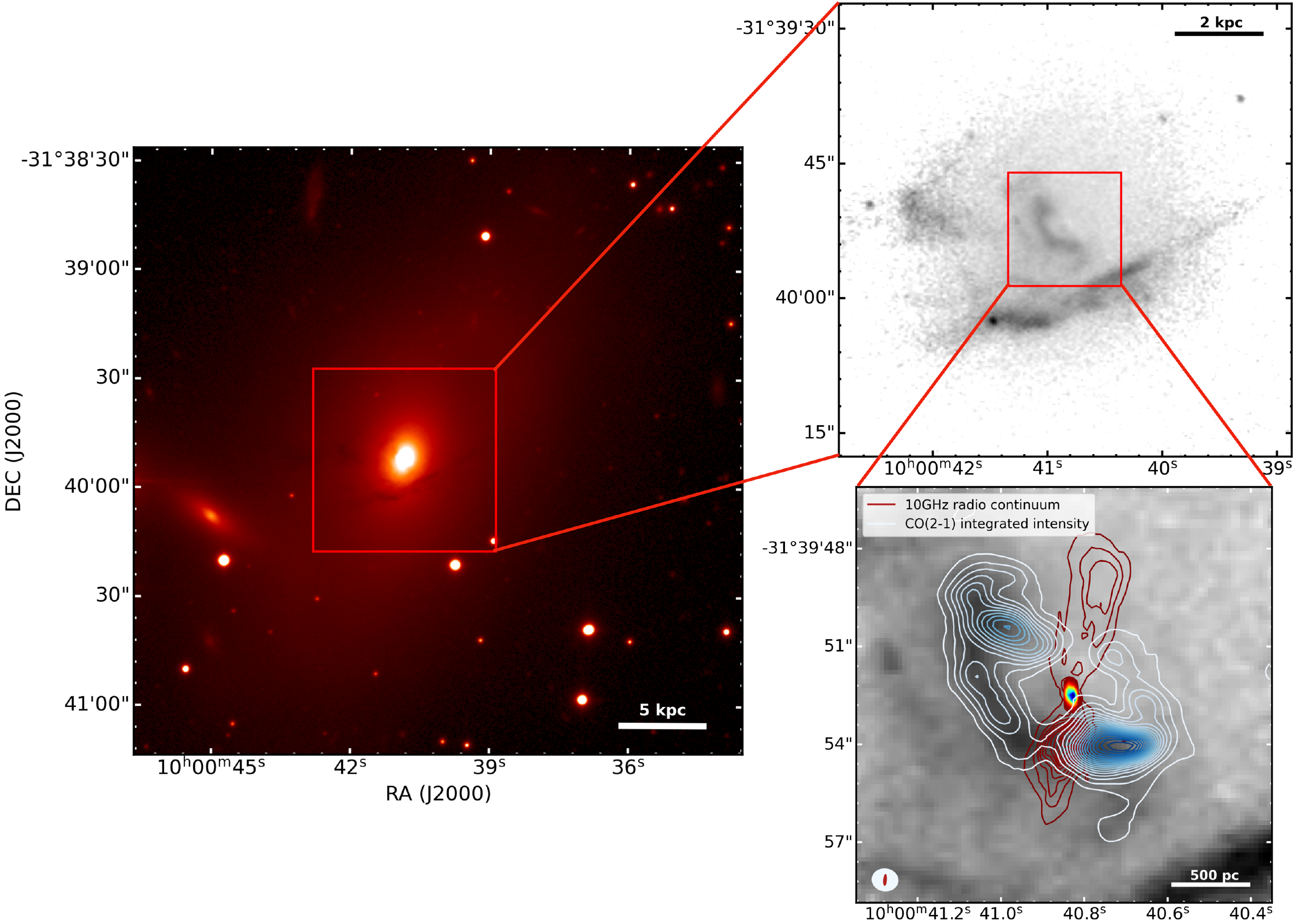}
\caption[]{\small{\textbf{Left panel:} Optical image of NGC\,3100 from the Carnegie-Irvine Galaxy Survey (CGS; \citealt{Ho11}) taken with a blue filter (300-400~nm). The image resolution (i.e.\,seeing FWHM) is $0.77''$. The size of the panel is $167''\times167''$ ($\approx30\times30$~kpc$^{2}$). \textbf{Right panel, top:} $B-I$ colour (dust extinction) map of NGC\,3100 in the central $50''\times50''$ ($\approx9\times9$~kpc$^{2}$; see \citetalias[][for details]{Ruffa19a}). \textbf{Right panel, bottom:} As above, but zoomed in the inner $13''\times13''$ ($\approx2.3\times2.3$~kpc$^{2}$). CO(2-1) integrated intensity contours from our Cycle 3 ALMA observations \citepalias{Ruffa19a,Ruffa19b}, and radio continuum contours from our JVLA data at 10~GHz \citepalias{Ruffa20} are overlaid in light-blue and red contours, respectively. Contours are drawn at 1,3,9....times the 3$\sigma$ rms noise level. } }\label{fig:NGC3100_optical}
\end{figure*}

\section{A southern radio galaxy sample}~\label{sec:sample}
A full description of our project, including details on the sample selection and all the available observations, can be found in \citet[][]{Ruffa19a}; a brief summary is presented here.

Starting from the southern radio galaxy (RG) sample of \citet{Ekers89}, which lists all the sources in the declination range $-17^{\circ}<\delta<-40^{\circ}$ with radio flux $S_{2.7 \rm GHz}\geq 0.25$~Jy and optical apparent magnitude $m_{\rm v}<17.0$, we have selected a complete sub-sample of eleven RGs. Based on the available optical spectroscopy \citep[][]{Tad93,Smith00,Colless03,Coll06,Jones09,Dopita15}, all of the eleven sources can be classified as LERGs. 

Nine sample members were observed during ALMA Cycle 3 in the $^{12}$CO(2-1) line\footnote{Hereafter, we omit the carbon mass number and refer to the $^{12}$CO isotopologue as simply CO.} and 230~GHz continuum \citep[][hereafter \citetalias{Ruffa19a}]{Ruffa19a}. These observations show that rotating (sub-)kpc molecular gas discs are very common in LERGs. The 3D modelling of these discs \citep[][hereafter \citetalias{Ruffa19b}]{Ruffa19b} demonstrates that, although the bulk of the gas appears to be in ordered rotation, low-amplitude perturbations and/or non-circular motions are ubiquitous and strongly suggestive of the presence of radio jet--CO disc interactions in at least two cases. The comparison between the ALMA CO(2-1) data and newly-acquired, high-resolution Karl G. Jansky Very Large Array (JVLA) radio continuum observations at 10~GHz \citep[][hereafter \citetalias{Ruffa20}]{Ruffa20} adds further evidence for the presence of such an interaction in at least one of the two candidates, NGC\,3100.

\begin{table*}
\begin{small}
\setlength{\tabcolsep}{5pt} % Default value: 6pt
\centering
\caption{Main properties of the ALMA observations presented in this paper.}
\label{tab:ALMA observations summary}
\begin{tabular}{l l c c c c c c c c c c}
\hline
\multicolumn{1}{c}{ ALMA } &
\multicolumn{1}{c}{ Observation } &
\multicolumn{1}{c}{ Targeted } & 
\multicolumn{1}{c}{ $\nu$\textsubscript{rest}} &
\multicolumn{1}{c}{ $\nu$\textsubscript{sky}} &
\multicolumn{1}{c}{ Time } & 
\multicolumn{1}{c}{   MRS } & 
\multicolumn{1}{c}{   FoV } & 
\multicolumn{1}{c}{   $\theta$\textsubscript{maj} } & 
\multicolumn{1}{c}{   $\theta$\textsubscript{min}  } & 
\multicolumn{1}{c}{   PA$_{\rm beam}$} & 
\multicolumn{1}{c}{   Scale }\\        
\multicolumn{1}{c}{ Band } &  
\multicolumn{1}{c}{ date } &  
\multicolumn{1}{c}{ line} &   
\multicolumn{1}{c}{    } &      
\multicolumn{1}{c}{   } &          
\multicolumn{1}{c}{  } &          
\multicolumn{1}{c}{  } &
\multicolumn{1}{c}{  } &
\multicolumn{2}{c}{  } &
\multicolumn{1}{c}{    } &
\multicolumn{1}{c}{   } \\
\multicolumn{1}{c}{ } &  
\multicolumn{1}{c}{ } &  
\multicolumn{1}{c}{ } &   
\multicolumn{1}{c}{   (GHz)  } &      
\multicolumn{1}{c}{   (GHz) } &          
\multicolumn{1}{c}{   (min)} &          
\multicolumn{1}{c}{   (kpc) (arcsec)} & 
\multicolumn{1}{c}{   (kpc) (arcsec)} & 
\multicolumn{2}{c}{ (arcsec) } &   
\multicolumn{1}{c}{   (deg) } &
\multicolumn{1}{c}{   (pc)} \\
\multicolumn{1}{c}{   (1) } &   
\multicolumn{1}{c}{   (2) } &
\multicolumn{1}{c}{   (3) } &
\multicolumn{1}{c}{   (4) } & 
\multicolumn{1}{c}{   (5) } &   
\multicolumn{1}{c}{   (6) } &
\multicolumn{1}{c}{   (7) } &               
\multicolumn{1}{c}{   (8) } &
\multicolumn{1}{c}{   (9) } &
\multicolumn{1}{c}{   (10) } &
\multicolumn{1}{c}{   (11) } &
\multicolumn{1}{c}{   (12) } \\

\hline
 Band 3  &  2019-04-23/24 & $^{12}$CO(1-0)  & 115.2712 & 114.2657 & 62 &  3.7 (21)  & 8.9 (50) &  1.4  & 1.0  & -85 & 250 \\ 
\hline
 Band 4 & 2018-12-14/15 & \begin{tabular}[]{@{}c@{}}  HNCO(6-5) \\ SiO(3-2) \end{tabular} & \begin{tabular}[]{@{}c@{}} 131.8857 \\ 130.2686 \end{tabular} & \begin{tabular}[]{@{}c@{}} 130.7352 \\ 129.1322 \end{tabular}  &    188    &  3.4 (19)    & 7.6 (43) &  1.1    &     0.9    &    \begin{tabular}[c]{@{}c@{}} -81 \\ -84 \end{tabular}   &    195    \\
 \hline
 Band 6$^{*}$ &    2016-03-22    & $^{12}$CO(2-1)  & 230.5380  & 228.6299  &    28     &    1.9 (11)    &  4.4 (25) &  0.9    &    0.7    &    -87    &    160   \\
 \hline
 Band 7  & 2019-03-15 &\begin{tabular}[c]{@{}c@{}}    HCO$^{+}$(4-3) \\ $^{12}$CO(3-2) \end{tabular}   & \begin{tabular}[c]{@{}c@{}}    356.7342 \\ 345.7960 \end{tabular}   & \begin{tabular}[c]{@{}c@{}}    353.6224 \\ 342.7835 \end{tabular}  &    32    &    1.2 (7.0)    &   3.0 (17) & 1.1    &     0.6    &    -84    &    195   \\  
\hline
\end{tabular}
%\end{small}
\parbox[t]{1\textwidth}{ \textit{Notes.} $-$ Columns: (1) ALMA frequency band. (2) Date of the observation. (3) Targeted molecular transition. (4) Rest-frame central line frequency. (5) Redshifted (sky) central frequency of the line estimated from the redshift listed in row (4) of Table~\ref{tab:ngc3100}. (6) Total integration time-on-source. (7) Maximum recoverable scale (MRS) in kiloparsec, and corresponding scale in arcseconds in parentheses. (8) Field-of-view (FOV; i.e.\,primary beam FWHM) in kiloparsec, and corresponding scale in arcseconds in parentheses. (9) and (10) Major and minor axis FWHM of the synthesized beam. (11) Position angle of the synthesized beam. (12) Physical scale corresponding to the major axis FWHM of the synthesized beam.\\
$^{*}$ALMA observation presented in \citetalias{Ruffa19a} and reported here for completeness. See Section~\ref{sec:obs} for details.}
\end{small}
\end{table*}

\subsubsection*{A case study: NGC\,3100}
NGC\,3100 is optically classified as a late S0 galaxy, characterised by a patchy dust distribution and a bright nuclear component (\citealp[][]{Sandage79,Lau06}; Figure~\ref{fig:NGC3100_optical}, left panel). It resides in a poor group and forms a pair with the barred spiral galaxy NGC\,3095, located at a projected linear distance of $\approx$120~kpc (\citealp{Devac76}). NGC~3100 is the host galaxy of the radio source PKS~0958-314 \citep[e.g.][\citetalias{Ruffa19a,Ruffa20}]{Ekers89}. A summary of some general target properties is provided in Table~\ref{tab:ngc3100}.

Among the radio galaxies that we imaged in CO(2-1), NGC\,3100 stands out: we detect a well-resolved ring-like central CO structure extending $\approx$1.6 $\times$ 0.5~kpc$^{2}$, observed to be nicely co-spatial with dust (Figure~\ref{fig:NGC3100_optical}, upper- and bottom-right panel). Other structures with patchy morphologies are observed up to $\approx$2~kpc from the central ring (\citetalias{Ruffa19a}). The 3D kinematic modelling of the main CO distribution (\citetalias{Ruffa19b}) demonstrates that, although the bulk of the gas is regularly rotating, there are evident distortions in the velocity field that are best described by the combination of both warps and non-circular motions (with peak velocities around $\pm40$~km~s$^{-1}$, corresponding to $\pm52$~km~s$^{-1}$ deprojected\footnote{Observed velocities are projected along the line of sight. Intrinsic values can be estimated by correcting for the inclination of the gas distribution, i.e.\,$v_{\rm deproj}=v_{\rm obs}/\sin{i}$.}). The 10~GHz JVLA observations provide the highest resolution radio continuum map currently available for NGC\,3100 (Figure~\ref{fig:NGC3100_optical}, bottom-right panel). As extensively discussed in \citetalias[][]{Ruffa20},  these data show no evidence for highly collimated jet structures on scales larger than the unresolved core, so the inner jets must have disrupted and decollimated close to the region where they were launched. This indicates rapid jet deceleration, which is also consistent with the wide opening-angle radio lobes observed on larger scales in our 10~GHz map. Such lobes are rather asymmetric, both in surface brightness and morphology. The northern lobe, in particular, bends at a position that (at least in projection) seems coincident with a distortion in the CO(2-1) disc (see Figure~\ref{fig:NGC3100_optical}, bottom-right panel). The total linear extent of the radio source is $\approx 2$\,kpc and there is no evidence from the available multi-frequency radio data for emission on larger scales. This suggests that the radio source in NGC\,3100 is likely in an early phase of its evolution (\citetalias{Ruffa20}). More in general, all of the observational properties described above strongly suggest the presence of a (sub-)kpc scale jet-ISM interaction in this object.

\section{ALMA observations and data reduction}\label{sec:obs}
The observations presented in this paper were carried out during ALMA Cycle 6 between December 2018 and April 2019 (PI: I.\,Ruffa). Their main properties are summarised in Table~\ref{tab:ALMA observations summary}.
 %The maximum recoverable spatial scale (MRS), together with the major and minor axis full width half maxima (FWHM) and position angle of the synthesized beam for each observation are reported in Table~\ref{tab:ALMA observations summary}.
 
NGC\,3100 was observed in three different ALMA Bands: in Band 3 we targeted the CO(1-0) transition, in Band 4 the HNCO(6-5) and SiO(3-2), and in Band 7 the CO(3-2) and HCO$^{+}$(4-3) lines. We adopted the same spectral configuration for all of the observations: a total of four spectral windows (SPWs); one (or two, depending on the number of lines targeted in the observing block) centred at the line(s) redshifted frequency ($\nu_{\rm sky}$) and divided into 1920 1.129~MHz-wide channels, the other three (or two) used to map the continuum emission and divided into 240 7.813-MHz-wide channels. Between 41 and 48 12-m antennae were used; the maximum baseline length was 783~m for Band 3 and 4 observations, 360~m for Band 7. A standard calibration strategy was adopted: for each session, a single bright quasar was used as both flux and bandpass calibrator, a second one as phase calibrator. Specifically, we used J1037$-$2934 and J1024$-$3234 for Band 3, J1107$-$4449 and J1037$-$2934 for Band 4, and J1058$+$0133 and J1037$-$2934 for Band 7, respectively.

As already mentioned above, these newly-acquired data complement previous Band 6 observations of NGC\,3100 in the CO(2-1) transition, which were obtained in March 2016 as part of ALMA Cycle 3 (PI: I.\,Prandoni). Full details on these observations can be found in \citetalias{Ruffa19a}, a summary is provided in Table~\ref{tab:ALMA observations summary} for completeness.%The CO(2-1) line was observed in one SPW using the high-resolution correlator configuration (1920 1.129~MHz-wide channels); three additional SPWs (divided into 128 31.25-MHz-wide channels) were used to map the 230~GHz continuum.

We manually calibrated all data using the Common Astronomy Software Application \citep[{\sc casa};][]{McMullin07} package, version 5.4.1 (4.7.2 for Band 6 data), reducing each dataset separately with customized \textsc{python} data reduction scripts.

\begin{figure*}
\centering
\begin{subfigure}[t]{.48\textwidth}
\centering
\caption{\textbf{Band 3}}\label{fig:Band3_cont}
\includegraphics[width=\linewidth]{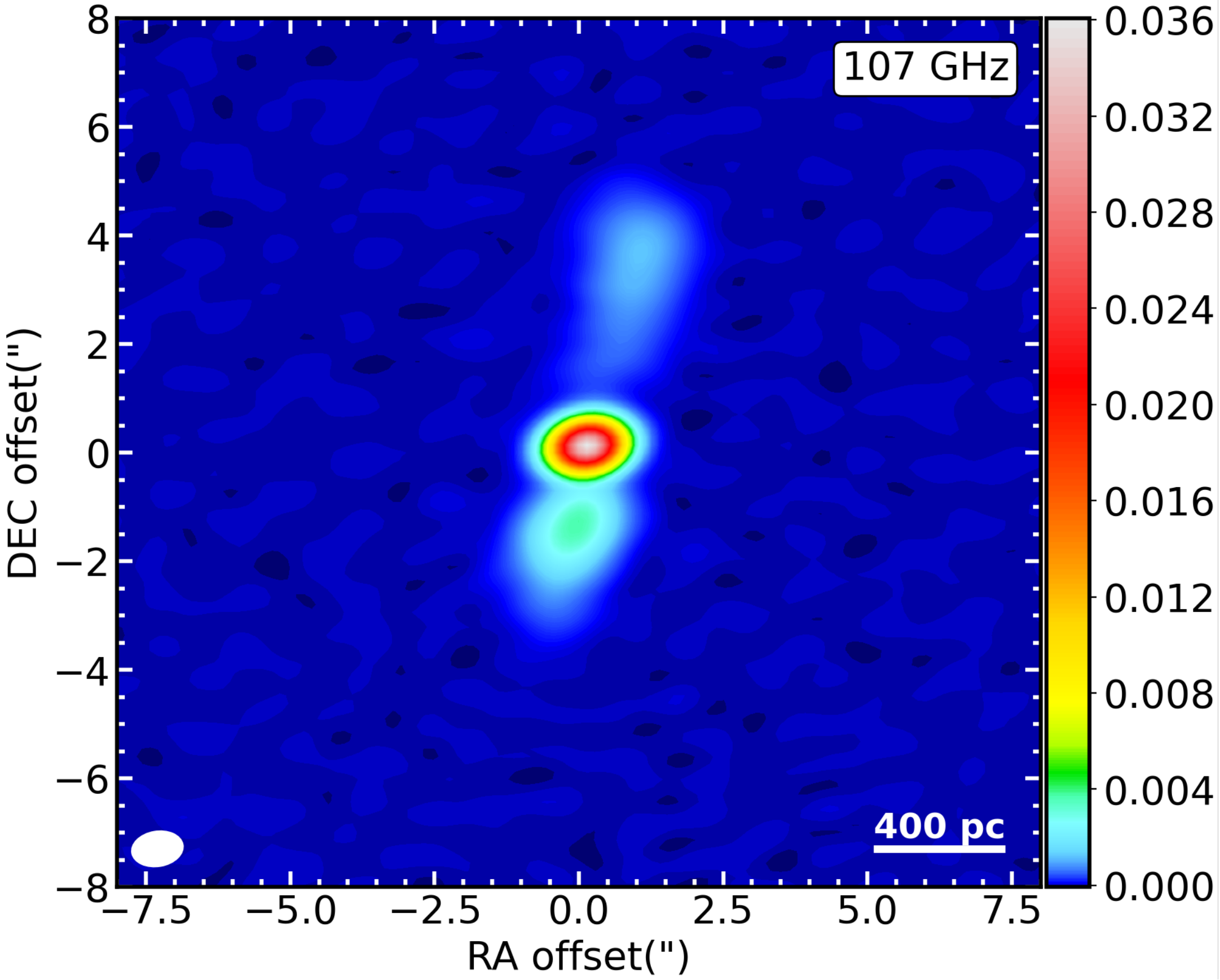}
\end{subfigure}
\hspace{2mm}
\begin{subfigure}[t]{.48\textwidth}
\centering
\caption{\textbf{Band 4}}\label{fig:Band4_cont}
\includegraphics[width=\linewidth]{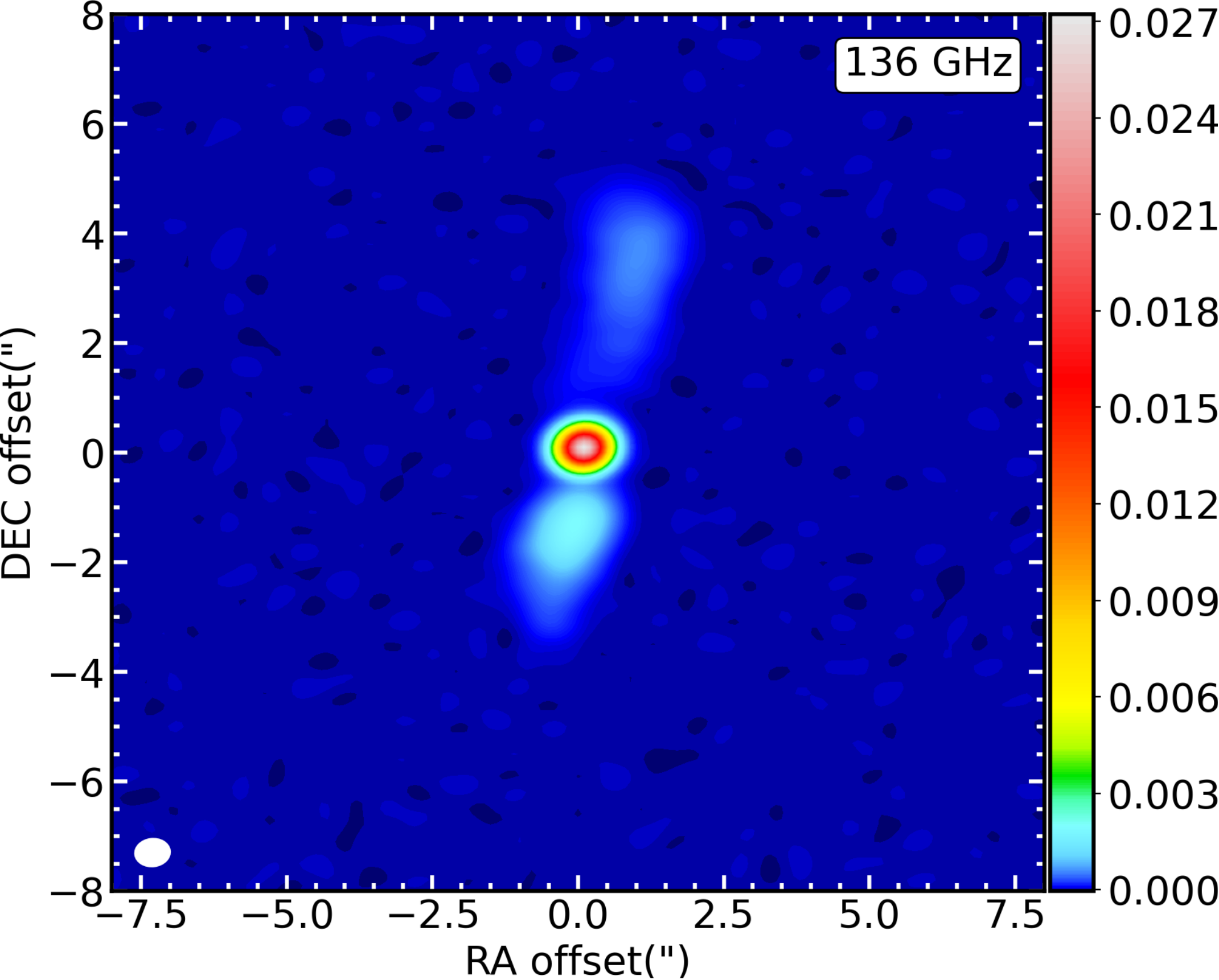}
\end{subfigure}
%\hspace{3mm}

\medskip

\begin{subfigure}[t]{.48\textwidth}
\centering
 \caption{\textbf{Band 6}}\label{fig:Band6_cont}
\includegraphics[width=\linewidth]{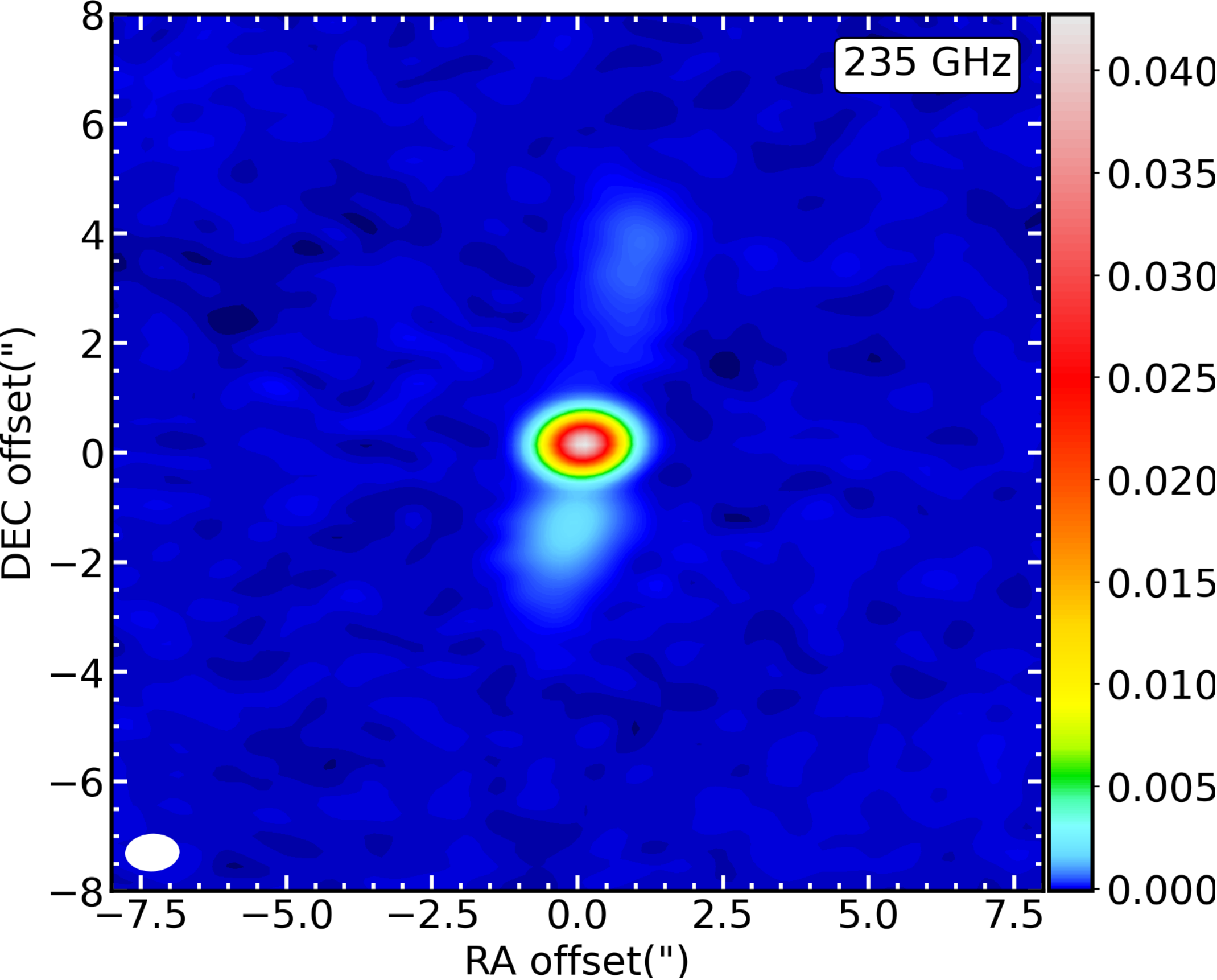}
\end{subfigure}
\hspace{2mm}
\begin{subfigure}[t]{.48\textwidth}
\centering
\caption{\textbf{Band 7}}\label{fig:Band7_cont}
\includegraphics[width=\linewidth]{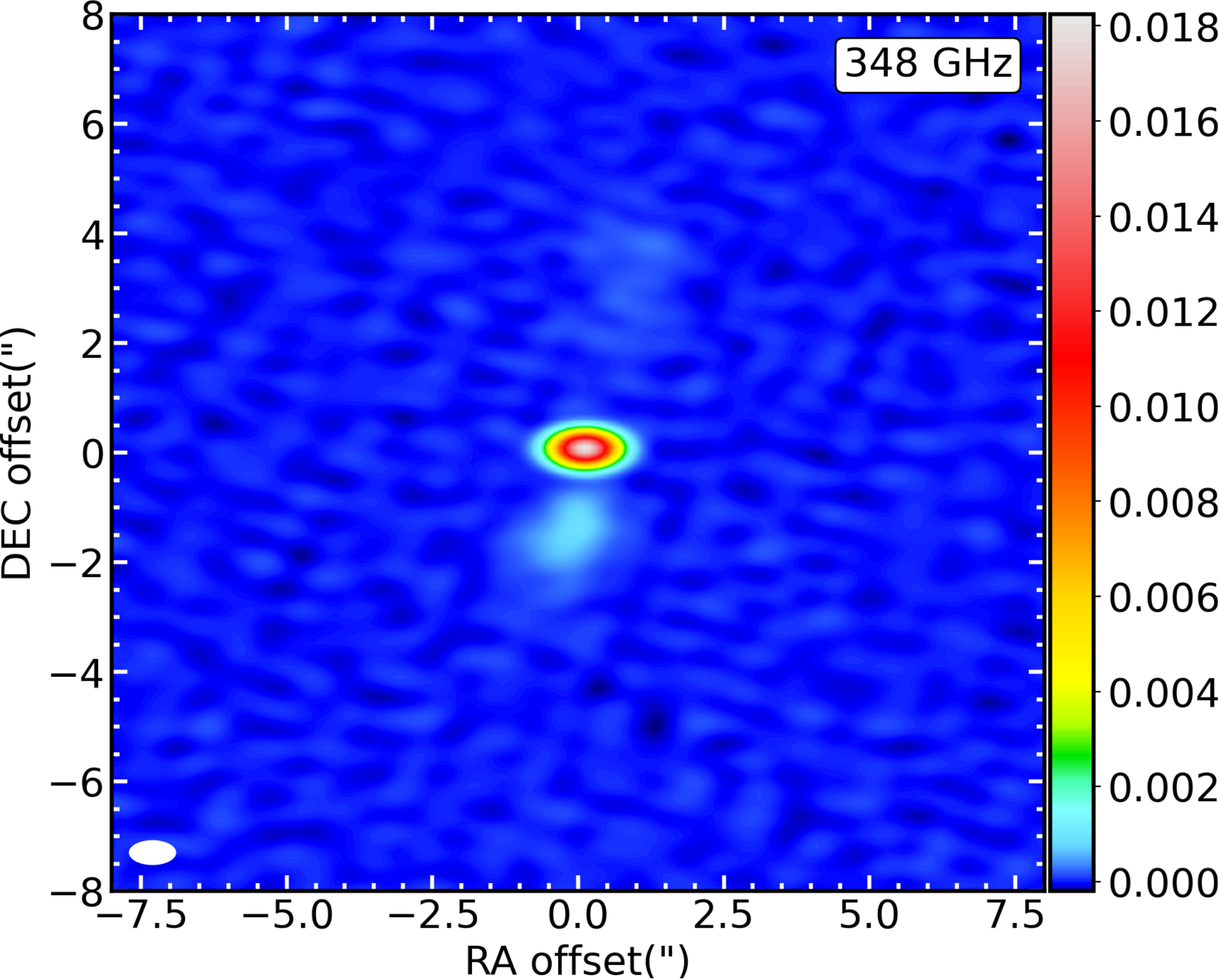}
\end{subfigure}
\caption[]{Naturally-weighted (sub-)mm continuum maps of NGC\,3100 in ALMA Band 3 (panel a), 4 (panel b), 6 (panel c) and 7 (panel d). The Band 6 image is from \citetalias{Ruffa19a}. The reference frequency of each map is indicated in its top-right corner. The bar to the right of each map shows the colour scale in Jy~beam$^{-1}$. Coordinates are given as relative to the image phase centre; East is to the left and North to the top. The synthesised beam is shown in white in the bottom-left corner of each panel; a scale bar is also shown in the bottom-right corner of panel a. The properties of the continuum maps are summarised in Table~\ref{tab:Continuum images}. \label{fig:continuum}}
\end{figure*}

\subsection{Continuum imaging}\label{sec:cont_imaging}
The continuum SPWs and the line-free channels of the line SPW(s) of each observing run were used to produce the continuum maps, using the \texttt{tclean} task in multi-frequency synthesis (MFS) mode \citep{Rau11} with no spectral dependence. All the continuum maps were made using natural weighting in order to maximise the sensitivity, with the goal of imaging extended emission from the radio source. Since continuum emission from the core is detected at high signal-to-noise ratios (S/N) at all frequencies, multiple cycles of phase-only self calibration and one cycle of amplitude and phase self-calibration were performed at all frequencies. This allowed us to obtain root-mean square (rms) noise levels ranging from 9 to 50~$\mu$Jy~beam$^{-1}$ for synthesized beams of 0.6 -- 0.9 arcsec full-width at half-maximum (FWHM). %Two-dimensional Gaussian fits were performed within the regions covered by the continuum emission to estimate the spatial extent of the observed components (when spatially resolved). 
Table~\ref{tab:Continuum images} summarises the main properties of the millimetre continuum maps, which are shown in Figure~\ref{fig:continuum}.  For completeness, we include the 230~GHz continuum map and its properties (originally presented in \citetalias{Ruffa19a}) in Figure~\ref{fig:continuum} and Table~\ref{tab:Continuum images}, respectively. All the continuum maps are primary-beam corrected.

\begin{table}
\centering
\caption{Properties of the continuum images.} 
\label{tab:Continuum images}
\begin{tabular}{l c c c c c c c c c c c}
\hline
\multicolumn{1}{c}{ ALMA} & 
\multicolumn{1}{c}{ Continuum } &
\multicolumn{1}{c}{ rms }&
\multicolumn{1}{c}{ $S_{\rm \nu}$ } \\
%\multicolumn{2}{c}{ Size } &
%\multicolumn{1}{c}{ PA }\\
\multicolumn{1}{c}{ Band } &
\multicolumn{1}{c}{ component } &
\multicolumn{1}{c}{  } &
\multicolumn{1}{c}{ } \\
%\multicolumn{2}{c}{ FWHM } &
%\multicolumn{1}{c}{  } \\
\multicolumn{1}{c}{  } & 
\multicolumn{1}{c}{  } &
\multicolumn{1}{c}{  ($\mu$Jy~beam$^{-1}$) } &
\multicolumn{1}{c}{ (mJy) } \\
%\multicolumn{1}{c}{ (arcsec$^{2}$) } &
%\multicolumn{1}{c}{  (pc$^{2}$) } &
%\multicolumn{1}{c}{ (deg) } \\
\multicolumn{1}{c}{  (1) } &
\multicolumn{1}{c}{  (2) } &
\multicolumn{1}{c}{  (3) } &
\multicolumn{1}{c}{  (4) } \\
%\multicolumn{1}{c}{  (5) } &
%\multicolumn{1}{c}{  (6) } &
%\multicolumn{1}{c}{  (7) } \\
\hline
Band 3 & \begin{tabular}[]{@{}c@{}}  Core \\ N jet \\ S jet \end{tabular} & 15 &  \begin{tabular}[]{@{}c@{}}  37.0$\pm$3.7 \\5.9$\pm$0.6 \\ 10$\pm$1.0 \end{tabular} \\ % &  \begin{tabular}[]{@{}c@{}} (0.16$\times$0.06) \\(2.9$\times$0.9) \\ (1.8$\times$0.7) \end{tabular} &  \begin{tabular}[]{@{}c@{}} (28$\times$11) \\(520$\times$160)\\ (320$\times$125) \end{tabular} &   \begin{tabular}[]{@{}c@{}} 173$\pm$82 \\168$\pm$4 \\ 162$\pm$4\end{tabular} \\
\hline
Band 4 & \begin{tabular}[]{@{}c@{}}  Core \\ N jet \\ S jet \end{tabular}  & 9  & \begin{tabular}[]{@{}c@{}}  28$\pm$2.8 \\ 4.5$\pm$0.4 \\ 7.7$\pm$0.8\end{tabular}  \\ % \begin{tabular}[]{@{}c@{}}   (0.08$\times$0.05) \\ (2.4$\times$0.9) \\ (1.5$\times$0.7) \end{tabular}   &  \begin{tabular}[]{@{}c@{}}  (15$\times$11) \\ (430$\times$160) \\ (270$\times$125) \end{tabular} &  \begin{tabular}[]{@{}c@{}}  173$\pm$85 \\ 170$\pm$2 \\ 160$\pm$3 \end{tabular} \\
\hline
Band 6$^{*}$ & \begin{tabular}[]{@{}c@{}}  Core \\ N jet \\ S jet \end{tabular} & 20  &  \begin{tabular}[]{@{}c@{}}  43.0$\pm$4.3 \\ 2.2$\pm$0.2 \\ 4.3$\pm$0.4 \end{tabular} \\ % \begin{tabular}[]{@{}c@{}}  (0.11$\times$0.05) \\ (2.4$\times$1.0)  \\ (1.6$\times$0.7) \end{tabular}  & \begin{tabular}[]{@{}c@{}}  (20$\times$10) \\ (440$\times$180) \\ (290$\times$130) \end{tabular}  & \begin{tabular}[]{@{}c@{}} 170$\pm$89 \\ 167$\pm$2  \\ 164$\pm$4 \end{tabular}  \\
\hline
Band 7 &\begin{tabular}[]{@{}c@{}}  Core \\ N jet$^{\dagger}$ \\ S jet \end{tabular} & 50  & \begin{tabular}[]{@{}c@{}} 19$\pm$1.9 \\ $-$ \\ 3.4$\pm$0.3 \end{tabular}   \\ %  \begin{tabular}[]{@{}c@{}}  (0.11$\times$0.06) \\ $-$ \\ (1.3$\times$0.6) \end{tabular}   &  \begin{tabular}[]{@{}c@{}}  (20$\times$11) \\ $-$ \\ (230$\times$110) \end{tabular} &  \begin{tabular}[]{@{}c@{}}  91$\pm$18 \\ $-$ \\ 155$\pm$6 \end{tabular} \\
\hline
\end{tabular}
\parbox[]{1\columnwidth}{ \textit{Notes.} $-$ Columns: (1) ALMA frequency band. (2) Component identified in the mm continuum maps. (3) Rms noise level measured from emission-free regions of the cleaned continuum map. (4) Integrated continuum flux density of the components listed in column (2) and corresponding uncertainties. These latter are estimated as $\sqrt{{\rm rms}^{2} + (0.1 \times S_{\rm \nu})^{2}}$, where the second term dominates in all cases. \\ % (5) Size (FWHM) deconvolved from the synthesized beam. The sizes were estimated by performing 2D Gaussian fits to the continuum components listed in Column (2). (6) Spatial extent of each component corresponding to the angular sizes in column (5). (7) Position angle of the corresponding component, defined North through East.\\
$^{*}$Continuum properties of the Band 6 continuum map from \citetalias{Ruffa19a}. See Section~\ref{sec:obs} for details.\\
$^{\dagger}$Undetected component.}
\end{table}

\subsection{Line imaging}\label{sec:line_imaging}
After applying the continuum self-calibration, the CO line emission was isolated in the $uv$-plane using the {\sc casa} task {\tt uvcontsub}. This forms a continuum model from linear fits in frequency to line-free channels, and then subtracts this model from the visibilities. 

We created the data cubes of the CO transitions using the {\tt tclean} task with natural weighting and channel widths of 10~km~s$^{-1}$. The channel velocities were computed in the source frame with zero-points corresponding to the redshifted frequency of the corresponding line (i.e.\,$\nu$\textsubscript{sky}; see Table~\ref{tab:ALMA observations summary}). The continuum-subtracted dirty cubes were cleaned in regions of line emission (identified interactively) to a threshold equal to 1.5 times the rms noise level (determined from line-free channels), and then primary-beam corrected. CO(1-0) and CO(3-2) emission is clearly detected, with peak S/N of 23 and 83, respectively, and a rms noise (determined from line-free channels) of 0.6~mJy~beam$^{-1}$ in both cases. 

SiO(3-2), HNCO(6-5) and HCO$^{+}$(4-3) emission (and consequently line-free channels) could not be identified in the SPWs centred on the corresponding redshifted line frequencies; the continuum was thus modelled using the continuum SPWs only. Continuum-subtracted data cubes of these transitions were created with a conservative channel width of 105~km~s$^{-1}$ and natural weighting. The resulting dirty cubes were then cleaned down to 1.5 times the expected rms noise, yielding cleaned channel maps with $1\sigma$ rms noises ranging from 0.05 to 0.2~mJy~beam$^{-1}$. The SiO(3-2) and HNCO(6-5) lines are undetected, while HCO$^{+}$(4-3) emission is tentatively detected at $\approx5\sigma$.

Table~\ref{tab:line images} summarises the properties of the imaged data cubes (including CO(2-1) from \citetalias[]{Ruffa19a}); for non-detections, $3\sigma$ upper limits are tabulated.

\begin{table}
\centering
\caption{Properties of the imaged data cubes.}
\label{tab:line images}
\begin{tabular}{ l c c c c c c}
\hline
\multicolumn{1}{c}{ Transition } &
\multicolumn{1}{c}{ rms } &
\multicolumn{1}{c}{ Peak flux } &
\multicolumn{1}{c}{ S/N  } &
\multicolumn{1}{c}{ $\Delta v_{\rm chan}$  } \\
\multicolumn{1}{c}{  } &
\multicolumn{1}{c}{ (mJy~beam$^{-1}$) } &
\multicolumn{1}{c}{ (mJy~beam$^{-1}$) } &
\multicolumn{1}{c}{  } &
\multicolumn{1}{c}{ (km~s$^{-1}$) } \\
\multicolumn{1}{c}{ (1) } &
\multicolumn{1}{c}{ (2)} &
\multicolumn{1}{c}{ (3) } &
\multicolumn{1}{c}{ (4) } &
\multicolumn{1}{c}{ (5) } \\
\hline
 $^{12}$CO(1-0)  &  0.6  &  13.8  &   23 &   10  \\
 HNCO(6-5)    &  0.05  &  $<0.15$   &   $-$  &  105  \\
  SiO(3-2)  &  0.05  &  $<0.15$  &   $-$  &   105  \\
  $^{12}$CO(2-1)$^{*}$  &  0.6  &  28.3 &  45 &  10  \\
  HCO$^{+}$(4-3)   &  0.2  &  1.0 &   5  &  105 \\
 $^{12}$CO(3-2)  &  0.6   &  50.0  &   83  &   10  \\
\hline
\end{tabular}
\parbox[t]{1\columnwidth}{ \textit{Notes.} $-$ Columns: (1) Line transition. (2) Rms noise measured from line-free channels with the widths listed in column (5). (3) Peak flux of the line emission. (4) Peak signal-to-noise ratio of the detection. (5) Channel width of the data cube (km\,s$^{-1}$ in the source frame).\\
$^{*}$Properties of the CO(2-1) data cube presented in \citetalias{Ruffa19a}. See Section~\ref{sec:obs} for details.}
\end{table}

\section{Image cube analysis and results}\label{sec:cube_analysis}
\subsection{Moment maps}\label{sec:mom_maps}
\begin{figure*}
\centering
\begin{subfigure}[t]{1.0\textwidth}
\centering
\caption{\textbf{CO(1-0) moments}}\label{fig:CO10_moms}
\vspace{-0.1cm}
\includegraphics[width=\linewidth]{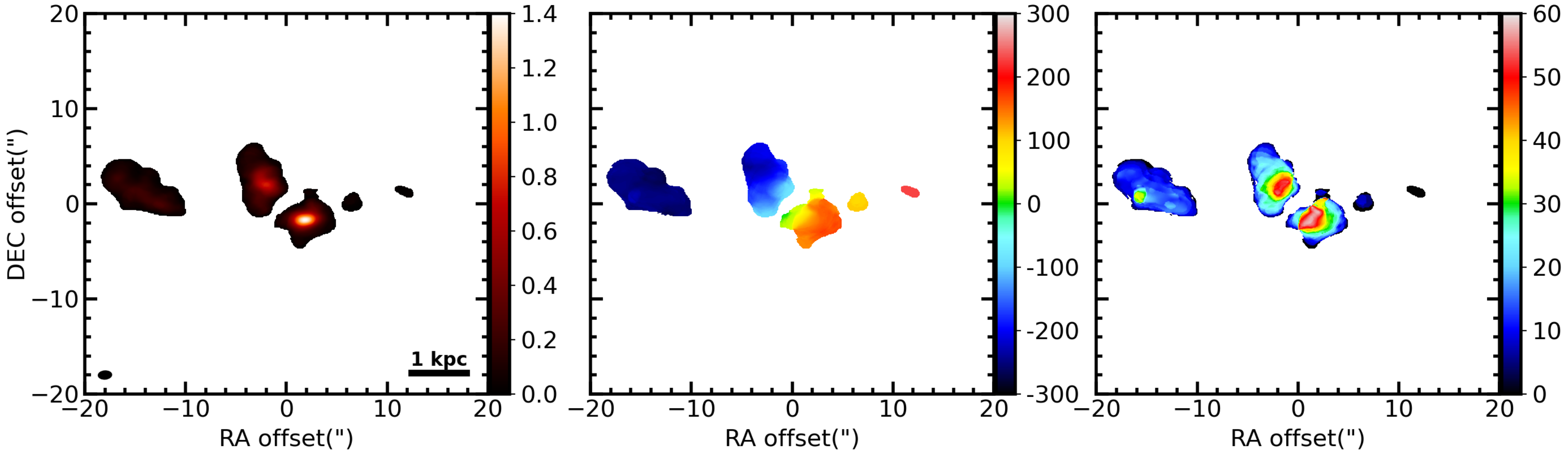}
\end{subfigure}

\medskip

\begin{subfigure}[t]{1.0\textwidth}
\centering
 \caption{\textbf{CO(2-1) moments}}\label{fig:CO21_moms}
 \vspace{-0.1cm}
\includegraphics[width=\linewidth]{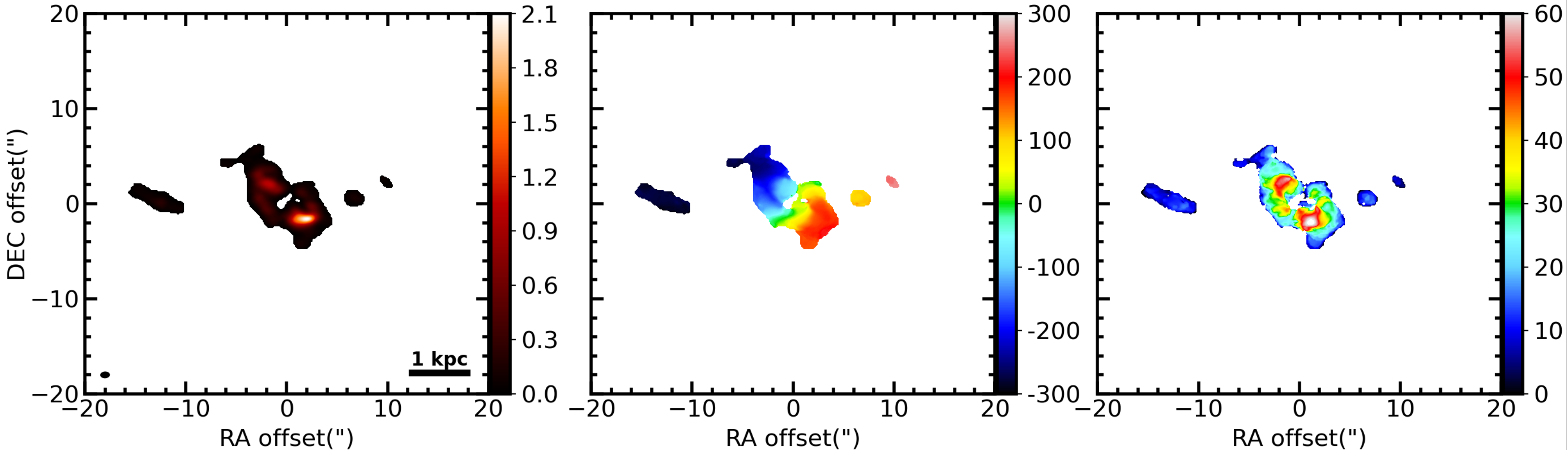}
\end{subfigure}

\medskip

\begin{subfigure}[t]{1.0\textwidth}
\centering
\caption{\textbf{CO(3-2) moments}}\label{fig:CO32_moms}
\vspace{-0.1cm}
\includegraphics[width=\linewidth]{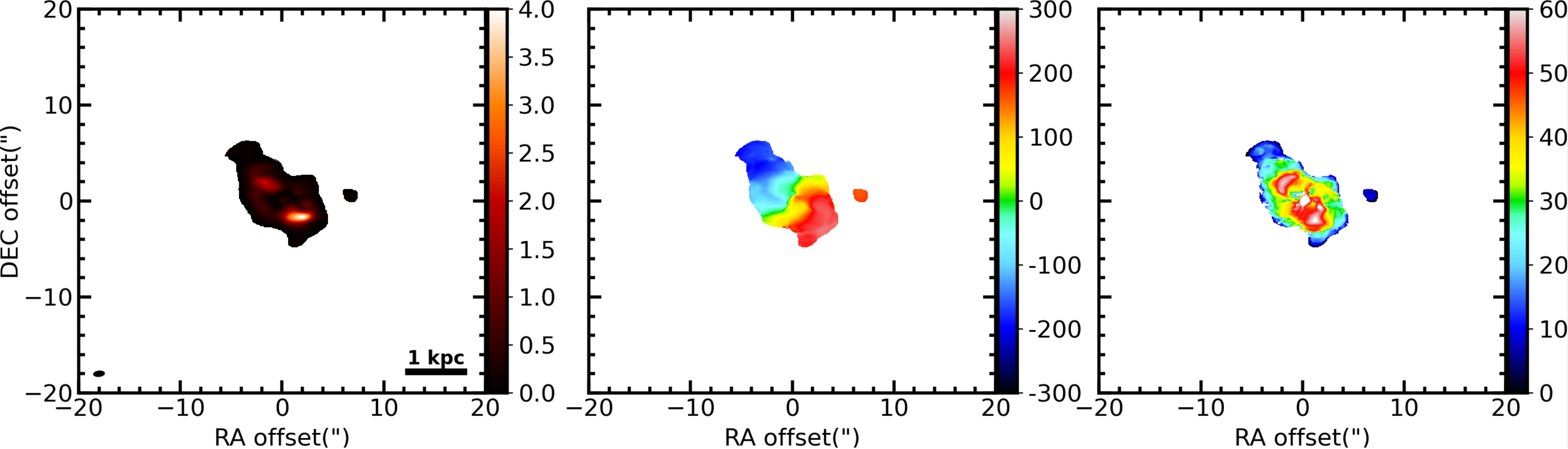}
\end{subfigure}
\caption[]{NGC\,3100 moment 0 (integrated intensity; left panels), moment 1 (intensity-weighted mean line-of-sight velocity; middle panels) and moment 2 (intensity weighted line-of-sight velocity dispersion; right panels) maps of the CO(1-0), CO(2-1) and CO(3-2) transitions (top, middle and bottom row, respectively). The maps were created with the masked moment technique described in Section~\ref{sec:mom_maps} using data cubes with a channel width of 10~km~s$^{-1}$ (see Table~\ref{tab:line images}). The synthesised beam and a scale bar are shown in the bottom-left and bottom-right corner, respectively, of each moment 0 map. The bar to the right of each map shows the colour scales (in Jy~beam$^{-1}$~km~s$^{-1}$ and km~s$^{-1}$ for moment 0 and moment 1/2 maps, respectively). Coordinates are with respect to the image phase centre; East is to the left and North to the top. Velocities are measured in the source frame and the zero-point corresponds to the intensity-weighted centroid of the CO emission. \label{fig:CO_moments}}
\end{figure*}

Integrated intensity (moment 0), mean line-of-sight velocity (moment 1), and velocity dispersion (moment 2) maps of the CO lines were created from the continuum-subtracted data cubes using the masked moment technique as described by \citet[][see also \citealt{Bosma81a,Bosma81b,Kruit82,Rupen99}]{Dame11}. In this technique, a copy of the cleaned data cube is first Gaussian-smoothed spatially (with a FWHM equal to that of the synthesised beam) and then Hanning-smoothed in velocity. A three-dimensional mask is then defined by selecting all the pixels above a fixed flux-density threshold. This threshold is chosen so as to recover as much flux as possible while minimising the noise (higher thresholds usually need to be set for noisier maps; see also \citetalias[][]{Ruffa19a}). Given the high significance of our CO detections (see Table~\ref{tab:line images}), a threshold of 1.2$\sigma$ (where $\sigma$ is the rms noise level measured in the un-smoothed data cube) has been used in all the cases. The moment maps were then produced from the un-smoothed cubes using the masked regions only \citep[e.g.][]{Davis17}. The resulting moment maps of the CO transitions (including those of the CO(2-1) from \citetalias{Ruffa19a}) are shown in Figure~\ref{fig:CO_moments}. 

It is clear from the left panels of Figure~\ref{fig:CO_moments} that for all the CO transitions the bulk of the emission arises from two bright peaks centred approximately at RA offsets of $1^{\prime \prime}$ ($\approx190$~pc) to the north-east (NE) and south-west (SW) of the core. These features are surrounded by fainter emission extending up to $11^{\prime \prime}$ ($\approx2$~kpc) along the major axis of the CO(1-0) distribution, a bit less ($\approx1.6$~kpc) in the higher-$J$ transitions. %Such features, often called ’twin peaks’ in the literature, have been found in other galaxies (e.g., NGC 3351, NGC 6951) and might be caused by the confluence of the inward gas transported by the bar and the gas in roughly circular orbits inside the Inner Linblad Resonance (ILR) of the bar (Kenney 1994; van der Laan et al. 2011). The two peaks seen in CO(2-1) are spatially coincident with the two arms seen in the V-H color map (see Fig. 6a, and also Erwin & Sparke 2003). Then, if there is a nuclear bar, the peaks can be interpreted as molecular gas in the leading edges of this secondary bar.  
In general, substantial differences can be observed between the three lines in term of the relative brightness of the inner and outer regions: higher-$J$ transitions are brighter towards the core, whereas lower-$J$ transitions are brighter towards the outskirts. As a result, also the gas morphology changes: a gap is observed at the centre of the CO(1-0) distribution, whereas the CO(2-1) and CO(3-2) lines shows a ring-like and full disc morphology, respectively.
%The maximum linear extent of the rotating structure is $\sim$2~kpc along the major axis, when observed in the CO(1-0) line, a bit smaller ($\sim 1.6$~kpc) when observed at higher-J transitions (see Table~\ref{tab:line_parameters}). %The shape becomes ring-like in the CO(2-1) line, while a full disc is observed in the CO(3-2) integrated intensity map, both extending up to 1.6~kpc along their major axis (see Table~\ref{tab:line_parameters}). 
Additional detached CO structures with lower surface brightness are visible at $\approx$1~kpc west and $\approx$2~kpc east from the outer edges of the central CO distribution (Figure~\ref{fig:CO_moments}). The latter, in particular, is clearly more prominent in the 1-0 than in the 2-1 transition, while it is not observed in CO(3-2). This seems to confirm the general trend for lower-$J$ lines to be brighter farther away from the centre, although we note that (at least some) of the CO(2-1) emission of the eastern patch may be lost due to significant primary beam attenuation in that region at 235~GHz (FoV$=25^{\prime \prime}$; see Table~\ref{tab:ALMA observations summary}). Primary beam attenuation/limitation may also explain the non-detection of any emission from the eastern patch in Band 7 observations (FoV$=17^{\prime \prime}$; see Table~\ref{tab:ALMA observations summary}). The relative spatial distribution of the CO lines suggests higher gas excitation in the central regions. This is further analysed in Section~\ref{sec:ratios_results} and discussed in Section~\ref{sec:interaction_discuss}.

The velocity pattern of the three CO transitions (Figure~\ref{fig:CO_moments}, middle panels) show that the gas is rotating, but with evident kinematic distortions (i.e.\,s-shaped iso-velocity contours). Such disturbances usually trace the presence of unrelaxed substructures in the gas disc, possibly caused by either warps or non-circular motions, or a combination of both. Accurate 3D kinematic modelling can help differentiating between the two. This is illustrated in Section~\ref{sec:kin_mod}.

The line-of-sight gas velocity dispersion ($\sigma_{\rm gas}$; Figure~\ref{fig:CO_moments}, right panels) varies significantly across the gas sky distributions. In all the transitions, the gas is observed as being dynamically cold at the outskirts of the main CO structure and at larger scales, with $\sigma_{\rm gas}\approx10-20$~km~s$^{-1}$. These values are in agreement with the typical ranges measured in nearby cold gas clouds \citep[e.g.][]{Voort18}. The gas velocity dispersion then progressively increases towards the central gas regions, but differently from lower- to higher-$J$ transitions. In both the 1-0 and 2-1 lines, the highest values ($\geq50$~km~s$^{-1}$) are almost exclusively concentrated in areas spatially coincident with the location of the two line peaks. In the CO(3-2), we instead observe large line broadening ($\sigma_{\rm gas}\geq40$~km~s$^{-1}$) throughout the central regions of the main gas structure. Such large values would imply that an ongoing perturbation (such as deviations from purely circular motions) is affecting the dynamical state of the gas, inducing turbulence within the gas clouds. Caution is needed, however, as observed line-of-sight velocity dispersions can be significantly overestimated due to observational effects. Among these, beam smearing (i.e.\,contamination from partially resolved velocity gradients within the host galaxy) usually dominates in regions of large velocity gradients, such as the inner regions of galaxies. Full kinematic modelling is the only way to derive reliable estimates of the intrinsic gas velocity dispersion (see Section~\ref{sec:kin_mod}).  %As detailed in Section~\ref{sec:kin_mod}, we model the gas velocity dispersion assuming it is constant throughout the disc. From our best-fitting models, the average dispersion increases from 29~km~s$^{-1}$ in CO(1-0) to 41~km~s$^{-1}$ in CO(3-2) (see Table~\ref{tab:line_parameters}), reproducing the observed trend. 

%The relative spatial distribution of the CO and the radio continuum shown in Figure~\ref{fig:NGC3100_optical} (bottom right panel) suggests a link between the different excitation conditions of the gas in the central regions and the presence of the radio jets. Indeed, the area where we observe a gap in the ground transition, then filled in the CO(3-2) line (Figure~\ref{fig:CO_moments}), seems closely spatially connected (at least in projection) with the radio continuum. A similar behaviour has been observed in the few existing analogous studies (albeit focused on different - Seyfert-like - type of objects; \citealp[see e.g.][]{Garcia14,Oosterloo17}), and interpreted as a strong indication for the jet interacting with the surrounding medium and affecting its excitation conditions. This is discussed more in detail in Section~\ref{sec:gas_excitation}.

\begin{figure}
\centering
\includegraphics[scale=0.38]{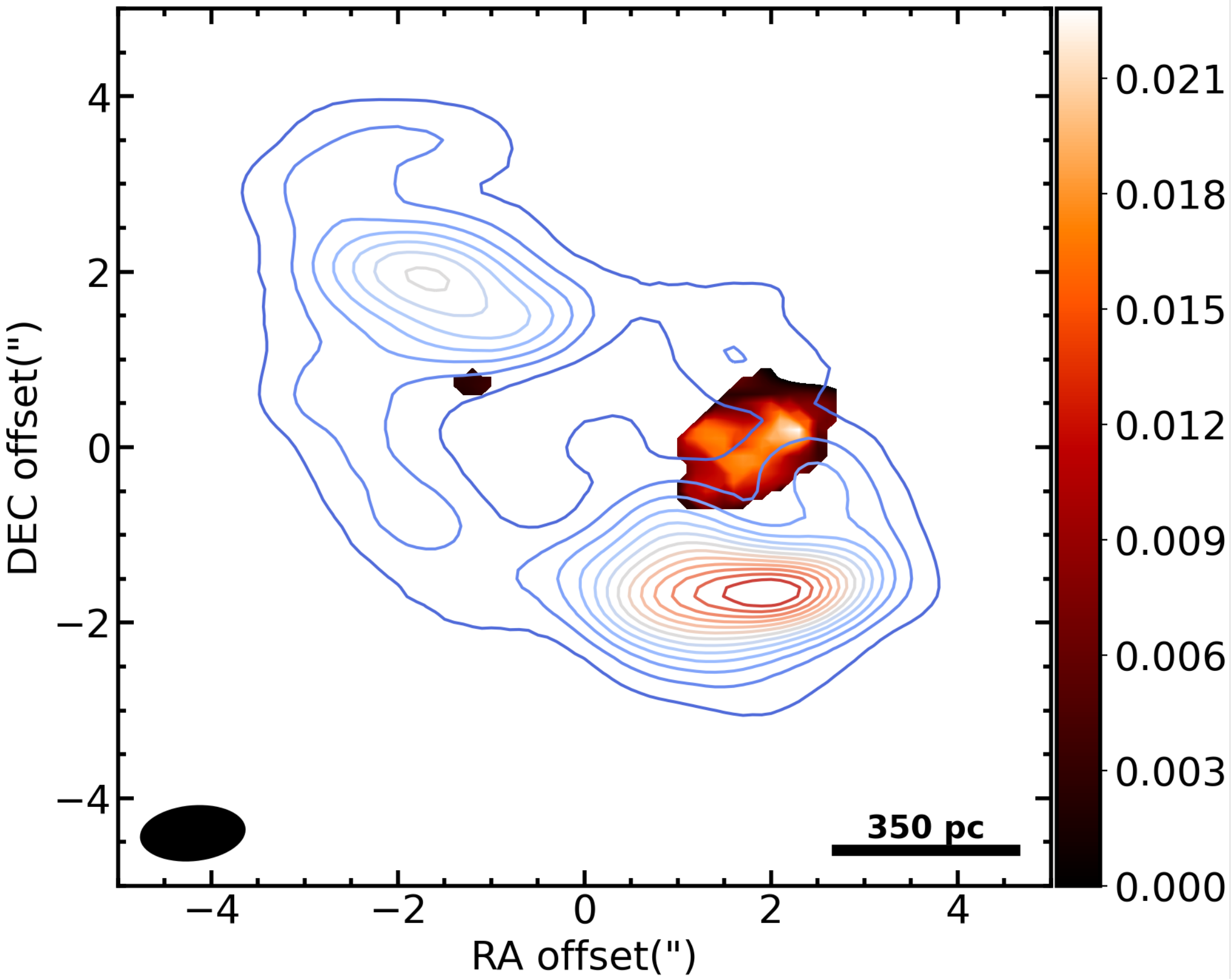}
\caption[]{NGC\,3100 moment 0 (integrated intensity) map of the HCO$^{+}$(4-3) transition, with CO(3-2) contours overlaid in a cool-to-warm colour scale. The map was created with the masked moment technique described in Section~\ref{sec:mom_maps} using a data cube with a channel width of 105~km~s$^{-1}$ (see Table~\ref{tab:line images}). The synthesised beam is shown in the bottom-left corner, and a scale bar in the bottom-right corner. The colour bar to the right shows the flux density scale in Jy~beam$^{-1}$~km~s$^{-1}$. Coordinates are relative to the image phase centre; East is to the left and North to the top.}\label{fig:HCO_mom0}
\end{figure}

A masked moment 0 map of the detected HCO$^{+}$(4-3) emission is shown in Figure~\ref{fig:HCO_mom0}, with CO(3-2) contours overlaid. The significance of the detection is not sufficient to produce reliable higher moment maps (line emission can be identified only in four channels of the data cube, with a peak S/N $=5$; see Table~\ref{tab:line images} and Section~\ref{sec:line_profiles}). The faint HCO$^{+}$(4-3) detection is restricted to a single unresolved blob-like structure at 2$''$ ($\approx350$~pc) west of the core, not coincident with the position of any of the CO peaks. Interestingly, HCO$^{+}$(4-3) emission is detected only at blueshifted velocities (see Section~\ref{sec:line_profiles}), while it is located in a small region of the receding (i.e.\,redshifted) CO side. This suggests that the two molecular species have opposite velocity gradients (i.e.\,they are kinematically-decoupled) in that region. This is discussed in Section~\ref{sec:HCO_discuss}.  %We hypothesize that they could have observed a warp in the galaxy disk in denser molecular gas tracers.

%{\bf The extent of the detected molecular gas distributions was estimated by performing 2D Gaussian fits to the moment 0 maps within the regions covered by the detected emission. The estimated sizes, given as major and minor axis FWHM deconvolved from the synthesized beam, are listed in the last column of Table~\ref{tab:line_parameters}.}

%{\bf FROM Dominguez-Fernandez+2020: In the central region, the CO(2-1) emission shows a two-arm mini-spiral morphology starting from the edges of the disk-like structure. The nuclear spiral morphology has also been re- ported by Riffel & Storchi-Bergmann (2011). The concave sides of the arms show strong reddening, suggesting that the spiral trails. Finally, the more prominent dust lanes are in the SW indi- cating that this is the near side of the galaxy.}

\subsection{Line widths and profiles}\label{sec:line_profiles}
\begin{figure}
\centering
\vspace{-0.5cm}
\begin{subfigure}[t]{0.96\columnwidth}
\centering
\caption{\textbf{CO spectra}}\label{fig:COspectra}
\includegraphics[width=\linewidth]{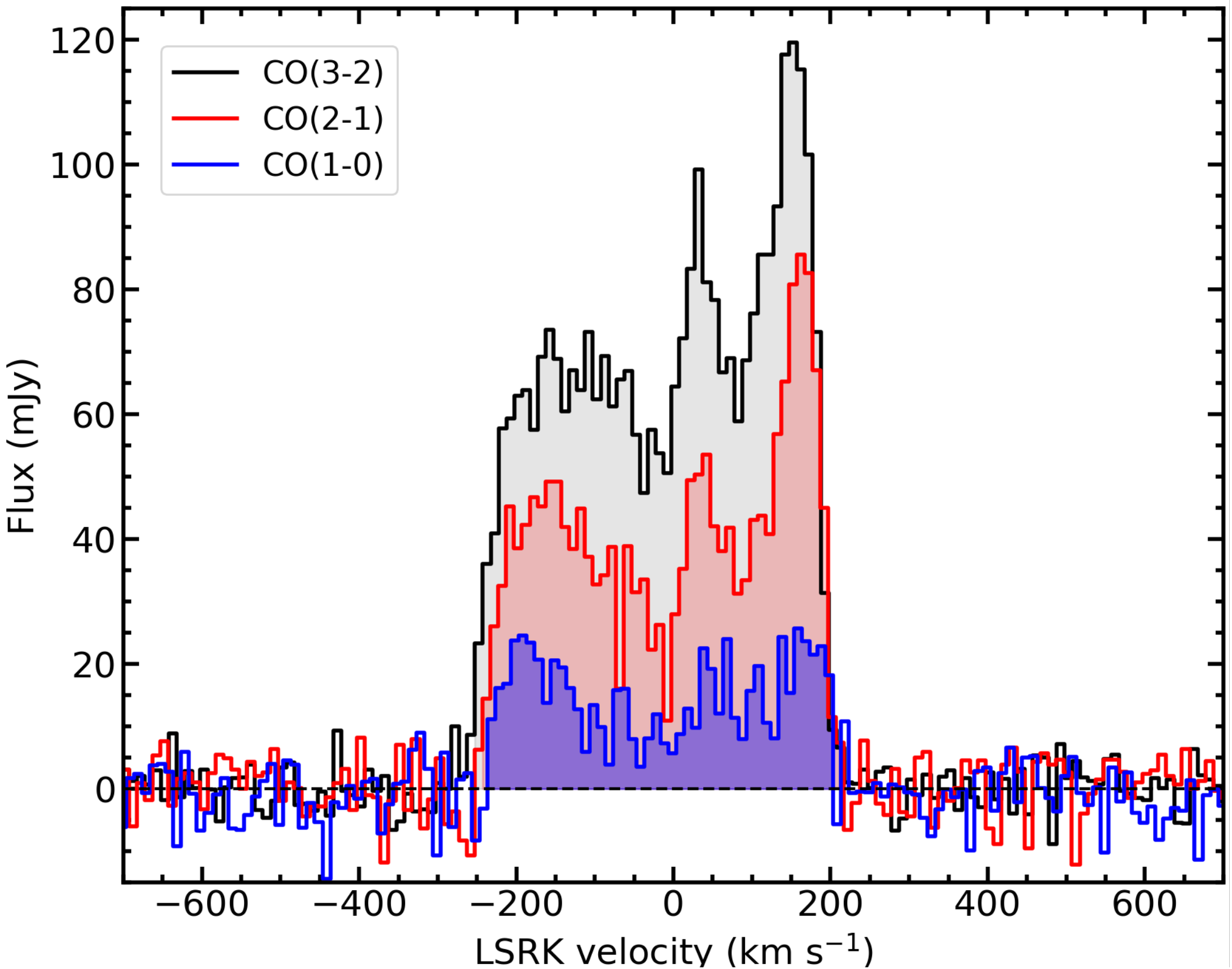}
\end{subfigure}
\begin{subfigure}[t]{1\columnwidth}
\vspace{0.3cm}
\centering
\caption{\textbf{HCO$^{+}$(4-3) spectrum}}\label{fig:HCO_spectum}
\includegraphics[width=\linewidth]{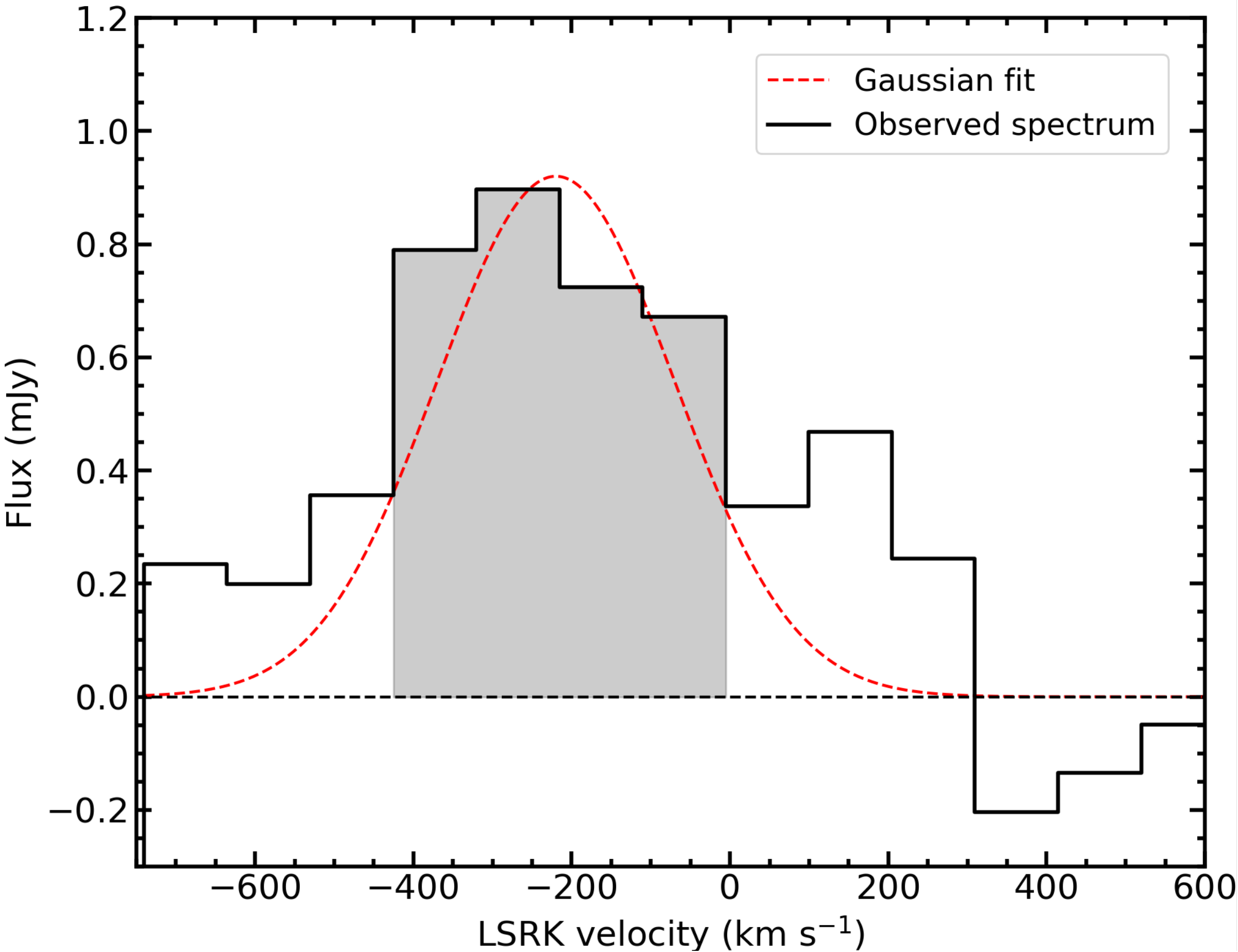}
\end{subfigure}
%\vspace{0.2cm}
\caption[]{{\bf Top panel}: CO(3-2), CO(2-1) and CO(1-0) spectral profiles (colours as labelled in the top-left corner) extracted within boxes of $10''\times11''$, including all the emission of the main CO structure visible in Figure~\ref{fig:CO_moments}. {\bf Bottom panel}: HCO$^{+}$(4-3) spectrum (black solid line) extracted within a box of $1.5''\times1.5''$, including all the detected HCO$^{+}$ emission. The best-fitting Gaussian profile is overlaid as a red dashed line. In both panels, the black dashed horizontal line indicates the zero flux level. The shaded regions highlight the spectral channels used to estimate the line FWZI (see Section~\ref{sec:line_profiles} for details).}\label{fig:CO_spectra}
\end{figure}

The spectral profiles of the central CO(1-0) and CO(3-2) structures were extracted from the cleaned data cubes within boxes covering the same $10''\times11''$ central area of each cube. These spectra are shown in Figure~\ref{fig:COspectra}, together with the CO(2-1) spectrum from \citetalias{Ruffa19a} (extracted within the same region). The spectral profile of the detected HCO$^{+}$(4-3) emission, extracted from the cleaned data cube within a $1.5''\times 1.5''$ box centred on the emission peak, is shown in Figure~\ref{fig:HCO_spectum}. Line widths were measured as both full-width at zero-intensity (FWZI) and FWHM. The former was defined as the full velocity range covered by channels (identified interactively in the channel maps) with line intensities $\geq3\sigma$. These channels are highlighted by the shaded regions in Figure~\ref{fig:CO_spectra}. The integrated flux densities listed are calculated over this range. For all the CO transitions, these values include the flux densities of the larger-scale eastern/western CO structures calculated over the FWZI range of their respective spectral profiles (not shown here).

Different methods were adopted to estimate the line FWHM, depending on the shape of the observed line profile. For the CO lines, showing the classic double-horned shape of a rotating disc, each FWHM was defined directly from the integrated spectrum as the velocity difference between the two channels with intensities exceeding half of the line peak most distant from the line centre. For the HCO$^{+}$(4-3) line, that is centrally peaked, we estimated the FWHM by performing a Gaussian fit on the integrated spectrum (see Figure~\ref{fig:HCO_spectum}).

For the undetected SiO(3-2) and HNCO(6-5) shock tracers, 3$\sigma$ upper limits on the velocity-integrated flux densities were calculated from the relation \citep[e.g.][]{Koay15}:
\begin{eqnarray}\label{eq:upper_limits}
  S \Delta {\rm v} < 3\sigma \Delta {\rm v}_{\rm FWHM} \sqrt{\dfrac{\Delta {\rm v}}{\Delta {\rm v}_{\rm FWHM}}},
\end{eqnarray}   
where $\Delta$v is the channel width of the data cube in which the rms noise ($\sigma$) is measured (105~km~s$^{-1}$ in these cases; Table~\ref{tab:line images}), and $\Delta$v\textsubscript{FWHM} is the expected line FWHM. We assumed the same FWHM of the CO(2-1) line (see Table~\ref{tab:line_parameters}). The factor $\sqrt{\Delta {\rm v} / \Delta {\rm v}_{\rm FWHM}}$ accounts for the expected decrease in noise level with increasing bandwidth \citep{Wrobel99}. We note that Equation~\ref{eq:upper_limits} is only valid if all of the molecular gas is concentrated within the synthesized beam ($\approx190 \times 160$~pc$^2$ for Band 4 observations; Table~\ref{tab:ALMA observations summary}). If the molecular gas is actually distributed on larger scales, the quoted upper limits would be underestimated.

The measured FWZI, FWHM and integrated flux densities of the detected lines, along with the upper limits of non-detections, are reported in Table~\ref{tab:line_parameters}.

\begin{table}
\centering
\caption{Main integrated parameters of the observed molecular transitions.}
\label{tab:line_parameters}
\begin{tabular}{l c c  c  c }
\hline
\multicolumn{1}{c}{Transition} &
\multicolumn{1}{c}{ Line FWHM }&
\multicolumn{1}{c}{ Line FWZI  }&
\multicolumn{1}{c}{ $S \Delta{\rm v}$}&
%\multicolumn{1}{c}{ M\textsubscript{mol} } &
\multicolumn{1}{c}{ R\textsubscript{J1}  } \\
%\multicolumn{1}{c}{ PA\textsubscript{CO} } &
%\multicolumn{1}{c}{ Size FWHM } \\
\multicolumn{1}{c}{  } &
\multicolumn{1}{c}{ (km~s$^{-1}$) } &
\multicolumn{1}{c}{ (km~s$^{-1}$) } &
\multicolumn{1}{c}{ (Jy km~s$^{-1}$) } &
%\multicolumn{1}{c}{ (M$_{\rm \odot}$) } &
\multicolumn{1}{c}{  }  \\
%\multicolumn{1}{c}{ (deg) } &
%\multicolumn{1}{c}{ (kpc$^{2}$) } \\
\multicolumn{1}{c}{ (1) } &
\multicolumn{1}{c}{ (2) } &
\multicolumn{1}{c}{ (3) } &
\multicolumn{1}{c}{ (4) } &
\multicolumn{1}{c}{ (5) } \\
%\multicolumn{1}{c}{ (6) } \\
%\multicolumn{1}{c}{ (7) } \\
%\multicolumn{1}{c}{ (9) } \\
\hline
 $^{12}$CO(1-0)  &  350  &  440  &  9.3$\pm$1.1  &  $-$ \\ %& (2.0$\pm$0.6)$\times$(0.6$\pm$0.2) \\
 HNCO(6-5)  &  340  &  $-$  &  $<28$ \\ %& $-$ & $-$ \\
 SiO(3-2)  &  340  &  $-$  &  $<28$ \\ %& $-$ & $-$ \\
$^{12}$CO(2-1) &  340 &  440  &  20.0$\pm$2.1$^{*}$ & 2.14$\pm0.4$ \\ %&  (1.6$\pm$0.3)$\times$(0.5$\pm$0.1)  \\
HCO$^{+}$(4-3)  &  150  &  420  &  0.32$\pm0.2$ & $-$ \\ %& <0.2 \\
$^{12}$CO(3-2) &  350  &  460  &  31.3$\pm$3.2  & 3.35$\pm0.6$ \\ % (1.6$\pm$0.2)$\times$(0.4$\pm$0.1)  \\
\hline
\end{tabular}
\parbox[t]{1\columnwidth}{ \textit{Notes.} $-$ Columns: (1) Molecular transition. (2) Line FWHM. For the CO detections, this is defined directly from the spectral profiles as the velocity difference between the two channels with intensities $\geq$50\% of the line peak most distant from the line centre. For the HCO$^{+}$(4-3) line, the FWHM is estimated from a Gaussian fit to the integrated spectrum (red dashed line in Figure~\ref{fig:HCO_spectum}). (3) Full velocity range covered by channels (identified interactively in the channel maps) with intensities $\geq$3$\sigma$ (shaded regions in Figure~\ref{fig:CO_spectra}). (4) Flux density integrated over all the channels in the range defined by the FWZI (for CO transitions, the emission of the detached larger-scale eastern/western CO structures visible in Figure~\ref{fig:CO_moments} is included). For non-detections, velocity-integrated upper limits (in units of Jy~beam$^{-1}$~km~s$^{-1}$) are estimated assuming that the emission is unresolved and the line FWHM is the same as that of the CO(2-1) (see Section~\ref{sec:line_profiles}). (5) Average ratio of the higher-$J$ CO lines over the 1-0 transition (i.e.\,ratio of the integrated flux density at each transition).\\ % (7) Size (FWHM, deconvolved from the beam) of the detected line emission. For CO transitions the reported sizes refer only to the extension of the central ring/disc-like gas structure visible in Figure~\ref{fig:CO_moments}.\\
$^{*}$Due to a more accurate numerical integration, the integrated CO(2-1) flux density reported here is slightly larger than that tabulated in \citetalias{Ruffa19a}. }
\end{table}

\subsection{Molecular gas mass}\label{sec:mol_masses}
An estimate of the molecular gas mass of NGC\,3100 was already provided in \citetalias{Ruffa19a}, based on our previous CO(2-1) observations. The successful detection of the CO(1-0) transition, however, allows us to obtain a more accurate estimate by removing the uncertainties introduced by the assumed CO(2-1) to CO(1-0) average flux density ratio (i.e.\,the ratio of the integrated flux density at each transition).% $\langle $R$\textsubscript{21} \rangle$. 

We adopted the following relation to estimate the total molecular gas mass, including contributions from heavy elements \citep[$M_{\rm mol}$;][]{Bolatto13}:

\begin{multline}\label{eq:gas mass}
M_{\rm mol}=1.05\times10^{4}~\left(\dfrac{X_{\rm CO}}{2\times10^{20}~\dfrac{{\rm cm}^{-2}}{{\rm K~km~s}^{-1}}}\right) \\ \times \left(\dfrac{1}{1+z}\right)~\left(\dfrac{S_{\rm CO}\Delta\nu}{{\rm Jy~km~s}^{-1}}\right)~\left(\dfrac{D_{{\rm L}}}{{\rm Mpc}}\right)^{2},
\end{multline}
where $S$\textsubscript{CO}$\Delta$v is the CO(1-0) integrated flux density, $z$ the galaxy redshift, $D$\textsubscript{L} the luminosity distance, and $X$\textsubscript{CO} the \cotoh2\ conversion factor. This latter depends on the molecular gas conditions (e.g.\ excitation, dynamics, geometry and stability of single molecular clouds) and other ISM properties (e.g.\,metallicity), and is therefore likely to vary systematically between different galaxy types \citep[see][]{Bolatto13}. As introduced in Section~\ref{sec:mom_maps} (see also Section~\ref{sec:ratios_results} and \ref{sec:interaction_discuss}), the molecular gas in NGC\,3100 displays different excitation conditions in the inner areas of the gas distribution. However, this effect is confined in the regions where CO(1-0) is mostly undetected and can be thus neglected for the purpose of estimating the molecular gas mass from this transition. More in general, little is known about $X$\textsubscript{CO} in nearby ETGs, as most of the studies carried out so far are focused on late-type disc galaxies. In the very few massive nearby ETGs in which accurate measurements have been made, the average conversion factor has been found to be comparable to that estimated in the Milky Way \citep[see e.g.][]{Utomo15}, which is thus the value commonly used for local radio galaxies \citep[e.g.][]{Tremblay16}. Therefore, we adopt an average Milky Way \cotoh2\ conversion factor of $X$\textsubscript{CO}$=2 \times 10^{20}$\,cm$^{-2}$(K\,km\,s$^{-1}$)$^{-1}$.

The estimated molecular gas mass is $(1.4\pm0.2\pm0.3)\times10^{8}$~M\textsubscript{$\odot$}. This is fully consistent with that estimated from CO(2-1) data ($M_{\rm mol}=(1.2\pm0.1\pm0.3)~\times~10^{8}$~M\textsubscript{$\odot$}; \citetalias{Ruffa19a}), despite the average flux density ratio assumed for that calculation (i.e.\,$R$\textsubscript{21}$=2.32$; \citetalias{Ruffa19a}) was slightly larger than that measured ($R$\textsubscript{21}$=2.14\pm0.4$; Table~\ref{tab:line_parameters}). This overestimation, however,  was counterbalanced by a slight underestimation of the CO(2-1) integrated flux density. A new  estimate of the latter, obtained from a more accurate numerical integration, is reported in Table~\ref{tab:line_parameters}. We note that the first error quoted on the molecular gas mass is random and largely dominated by the uncertainty on the integrated flux density, while the second is systematic and accounts for a 10\% uncertainty in the assumed distance. An estimate of the systematic uncertainty introduced by our assumption on $X$\textsubscript{CO} is not included in the quoted error budget. This can be quantified by assuming a conservative scatter of $\pm0.4$~dex (about a factor of 2.5; \citealp[see][]{Bolatto13}) around the adopted Milky Way value, leading the molecular gas mass to vary within the range $0.5-3.5 \times 10^{8}$~M$_{\odot}$.

%Another important uncertainty concerns the CO(2-1) to CO(1-0) flux ratio, $R$\textsubscript{21}, which depends on optical depth and excitation conditions of the molecular gas \citep[e.g.][]{Braine92}. $R$\textsubscript{21} can vary significantly between objects.Measurements of  $R$\textsubscript{21} have been made for gas-rich disc galaxies \citep[e.g.][]{Sandstrom13} and for radio-quiet ETGs \citep[e.g.][]{Young11}, but little is known about local radio-loud ETGs. The presence of a radio-loud AGN can significantly affect the conditions of the molecular gas in the surrounding regions \citep[e.g.][]{Oosterloo17}, but it is not clear whether such phenomena are common.
\begin{figure*}
\begin{subfigure}[t]{.46\textwidth}
\centering
\caption{\textbf{\large{R$_{\rm 21}$}}}\label{fig:CO21_CO10}
\includegraphics[width=\linewidth]{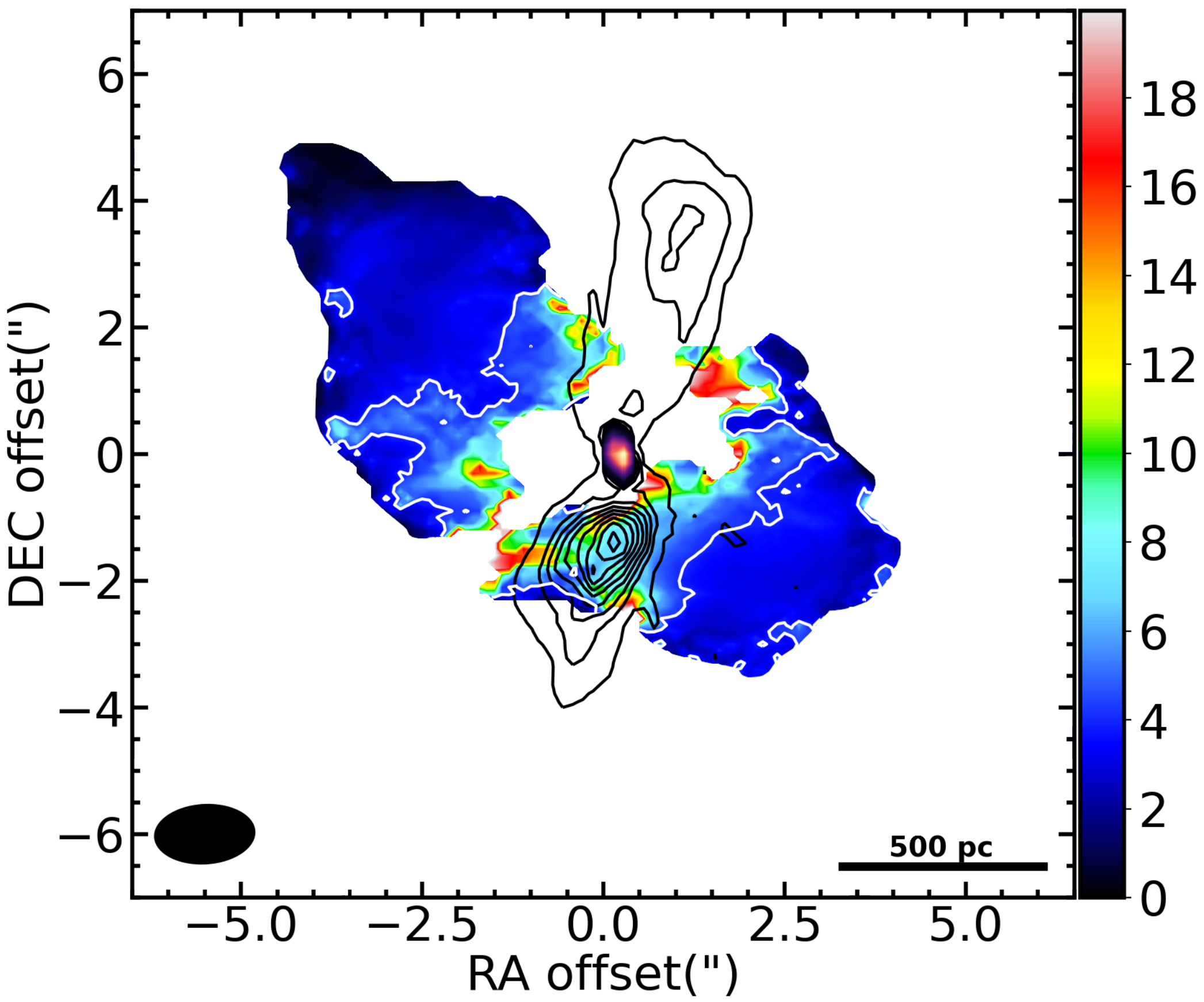}
\end{subfigure}
\hspace{2mm}
\begin{subfigure}[t]{.445\textwidth}
\centering
 \caption{\textbf{\large{R$_{\rm 31}$}}}\label{fig:CO32_CO10}
\includegraphics[width=\linewidth]{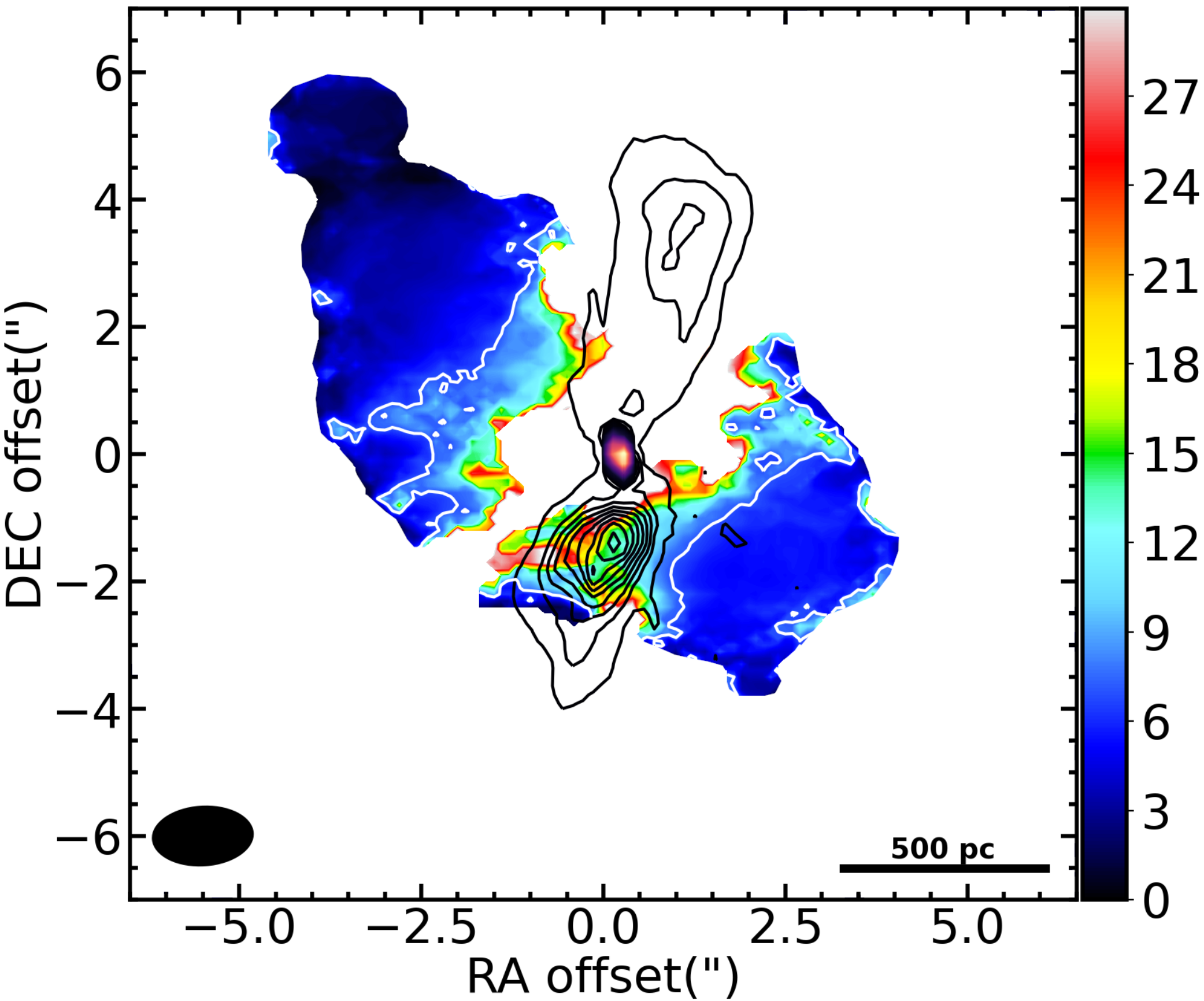}
\end{subfigure}
\caption[]{$R_{\rm 21} \equiv S_{\rm CO(2-1)}/S_{\rm CO(1-0)}$ (left panel) and $R_{\rm 31} \equiv S_{\rm CO(3-2)}/S_{\rm CO(1-0)}$ (right panel) maps with 10~GHz continuum contours overlaid. The white contour in each map roughly marks the transition between the outer low-excitation component with $R_{\rm 21}<4$ and  $R_{\rm 31}<9$, and the inner high-excitation component with $R_{\rm 21}\gtrsim4$ and  $R_{\rm 31}\gtrsim9$. The latter correspond to brightness temperature ratios $>1$ and thus likely to an optically thin gas regime (see the text for details). The bar to the right of each map show the colour scale. In both panels, the beam size and scale bar are shown in black in the bottom-left and bottom-right corners, respectively. Coordinates are given as relative to the image phase centre; East is to the left and North to the top.\label{fig:CO_ratios}}
\end{figure*}
\begin{figure}
\centering
\includegraphics[width=\linewidth]{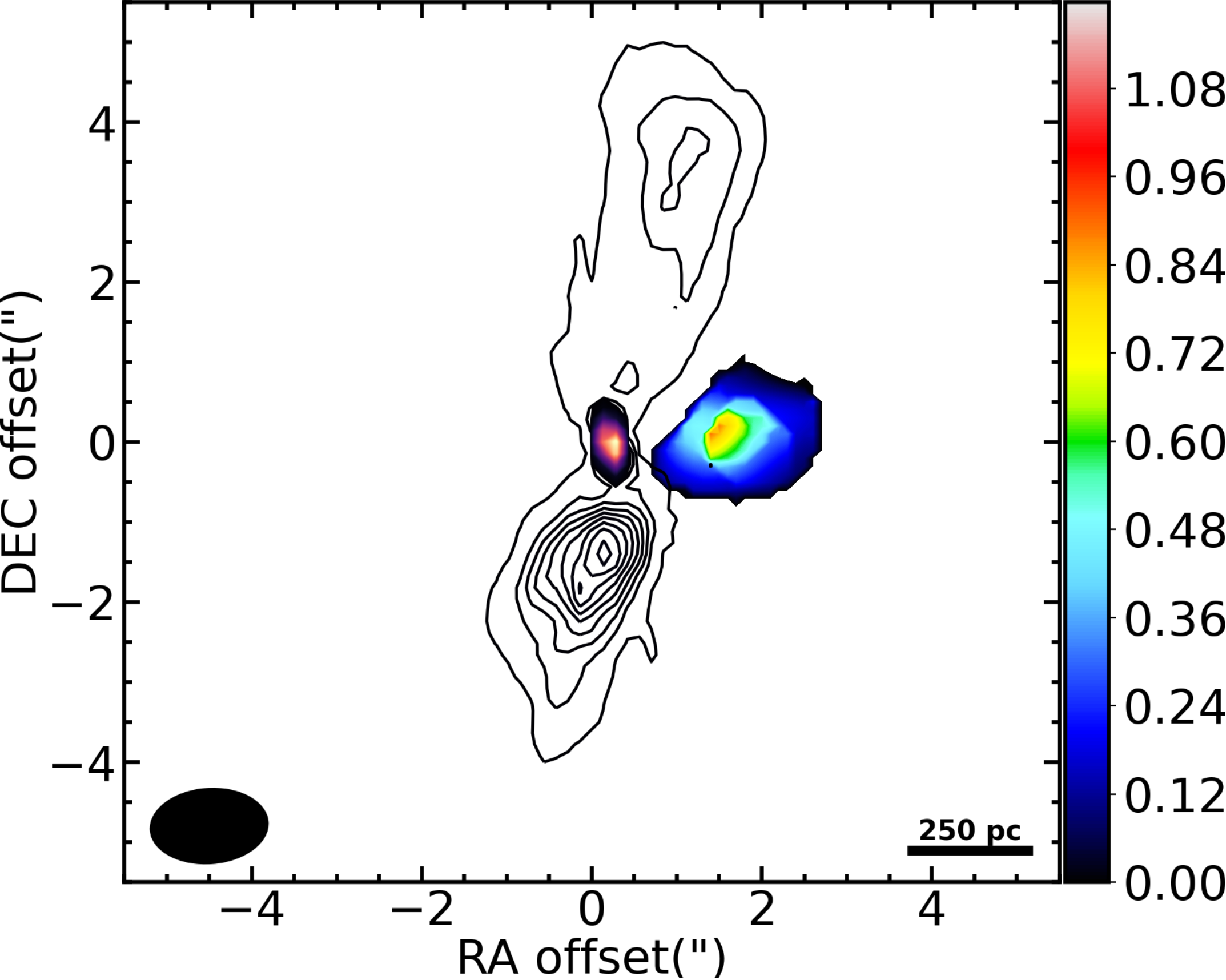}
\caption[]{R$_{\rm HCO} \equiv$ S$_{\rm HCO^{+}(4-3)}$/$S_{\rm CO(2-1)}$ map with 10~GHz continuum contours overlaid. The bar to the right show the colour scale. The beam size and scale bar are shown in black in the bottom-left and bottom-right corners, respectively. Coordinates are given as relative to the image phase centre; East is to the left and North to the top. \label{fig:HCO_ratios}}
\end{figure}
\subsection{Spatially-resolved line ratios}\label{sec:ratios_results}
One of the main aim of the present work is to carry out a spatially-resolved analysis of the strength of the observed higher-$J$ CO transitions relative to the lowest order transition. To do this, the data cubes used to make ratio maps need to be identical. We thus re-imaged the three CO transitions making new data cubes with the same number of channels, channel width (i.e.\,10~km~s$^{-1}$), pixel and image size. We also corrected for the different $uv$ coverages by smoothing to a common resolution (i.e.\,selecting a common $uv$ range and convolving to a common synthesized beam). This resulted in data cubes with synthesized beams of $1^{\prime \prime}.4 \times 0^{\prime \prime}.9$ (corresponding to $\approx250 \times 160$~pc$^{2}$). Masked integrated intensity maps of these smoothed data cubes were produced, and then used to make ratio maps using the {\sc CASA} task {\tt immath}. The resulting $R_{\rm 21} \equiv S_{\rm CO(2-1)}/S_{\rm CO(1-0)}$ and $R_{\rm 31} \equiv S_{\rm CO(3-2)}/S_{\rm CO(1-0)}$ maps\footnote{We note that in this work we use integrated flux density ratios (expressed in Jy~km~s$^{-1}$). Indeed, although naturally thought to be dimensionless, these are different from integrated brightness temperature ratios (expressed in K~km~s$^{-1}$). This is because,  based on the Rayleigh-Jeans formula, the conversion between K and J is not linear:
\begin{eqnarray*}\label{eq:Jy_K_conversion}
  \left(\dfrac{T}{{\rm K}}\right)= 1.22 \times 10^{3} \left(\dfrac{S}{{\rm Jy}}\right)~\left(\dfrac{\nu}{{\rm GHz}}\right)^{-2}~\left(\dfrac{\theta_{\rm maj} \times \theta_{\rm min}}{{\rm arcsec \times arcsec}}\right)^{-1},
\end{eqnarray*}
where $T$ is the brightness temperature, $S$ is the flux density, $\nu$ is the observing frequency, and $\theta_{\rm maj}$ and $\theta_{\rm min}$ are the major and minor axis FWHM of the synthesized beam, respectively.} are shown in Figure~\ref{fig:CO_ratios}, with 10~GHz continuum contours overlaid in black. As already clear from the relative spatial distribution of the three CO lines (see Section~\ref{sec:mom_maps}), we can identify at least two distinct molecular gas components from Figure~\ref{fig:CO_ratios}: an outer low-excitation component with flux density ratios mostly between $\approx1$ and $\approx4$, and an inner high-excitation component with $R_{\rm j1}\geq8$ and $R_{\rm 31}$ reaching values $\geq12$ (Figure~\ref{fig:CO32_CO10}). The former are compatible with the CO gas being on average in sub-thermal excitation condition \citep[e.g.][]{Cormier18}; in some cases, similar ratios have been also associated to gas in an optically thick regime \citep[e.g.][]{Dasyra16}. The large inner ratios instead imply that the gas is likely thermalised (i.e.\,at densities above 3000~cm$^{-3}$) and optically thin, since they are significantly above the maximum values of $R_{\rm 21}=4$ and $R_{\rm 31}=9$ for optically thick emission \citep[$\approx1$ for ratios in brightness temperature; see e.g.][]{Dasyra16,Oosterloo17}. Assuming local thermodynamic equilibrium (LTE) and optically thin conditions, CO ratios $\gtrsim12$ imply gas excitation temperatures $T_{\rm ex} \gtrsim50$~K \citep[e.g.][]{Oosterloo17}. Such extreme conditions are very similar to those observed in the regions of jet-ISM interactions in the few (mostly Seyfert-like) objects in which a multi-line study like the one presented in this paper has been carried out, such as NGC\,1068 \citep[][]{Garcia14} and IC\,5063 \citep{Dasyra16,Oosterloo17}. By making a spatial comparison between the regions where the highest line ratios are observed and the location of the radio lobes, it is clear that also in NGC\,3100 the expanding radio plasma is likely responsible for such different gas conditions in the inner areas. More in general, it is clear from Figure~\ref{fig:CO_ratios} that the gas becomes progressively less excited at increasing distance from the compact radio structure. 

The HCO$^{+}$(4-3) line requires much larger gas densities to be excited ($n_{\rm crit}>1.8\times 10^{6}$ cm$^{-3}$; \citealp[][]{Greve09}), compared to CO transitions ($n_{\rm crit} > 2.1 \times 10^{3}$ cm$^{-3}$; \citealt[][]{Carilli13}). As such, comparing their relative strength can provide clues on the gas density conditions in the region where they are co-spatial. It is therefore interesting to analyse also the HCO$^{+}$/CO flux density ratio, even though we detected the HCO$^{+}$(4-3) emission at much lower level and in a substantially smaller region than the CO transitions. To avoid unnecessary repetitions and - at the same time - allow a coherent comparison with HCO$^{+}$/CO ratios reported in other studies of jet-ISM interactions \citep[e.g.][]{Oosterloo17}, we made a sole representative R$_{\rm HCO} \equiv S_{\rm HCO^{+}(4-3)}$/S$_{\rm CO(2-1)}$ map. We used the same method described above, but binned the CO(2-1) data cube to a channel width of 105~km~s$^{-1}$. The resulting R$_{\rm HCO}$ map is shown in Figure~\ref{fig:HCO_ratios} with 10~GHz continuum contours overlaid in black. The R$_{\rm HCO}$ values are found to vary from $\approx0.1$ to $\approx 0.7$, much higher than those typically observed in unperturbed gas regions \citep[$<0.045$; e.g.][]{Oosterloo17}. Enhanced dense gas emission can be induced by various type of perturbations (such as merger-induced shocks; \citealp[see e.g.][]{Konig18}). However, Figure~\ref{fig:HCO_ratios} clearly shows that the faint HCO$^{+}$(4-3) emission arises from a region adjacent to the base of the southern radio lobe, making a jet-induced gas compression the most likely scenario. 

In general, all the results presented above provide important constraints on the radio jet-cold gas interplay in NGC\,3100. This is discussed in detail Section~\ref{sec:interaction_discuss}. %Thus, the conversion from brightness temperature to flux density ratios is:
%\begin{eqnarray*}\label{eq:ratios_conversion}
  %\dfrac{S_{i}}{S_{j}} = \dfrac{T_{i}}{T_{j}} %\left(\dfrac{\nu_{i}}{\nu_{j}}\right)^{2}, %\dfrac{(\theta_{\rm maj} \times %\theta_{\rm min})_{i}}{(\theta_{\rm maj} \times \theta_{\rm min})_{j}}
%\end{eqnarray*}
%where the letters i and j identify the two transitions involved, and the source sizes cancel out as, by construction, they are the same for both transitions.}.
%{\bf FROM YOUNG et al. 2021: Line ratios between data taken in different ob- serving setups and different tunings, such as 12 CO/13 CO (see Table 1), have an additional uncertainty due to this absolute flux calibration. However, line ratios between data taken simultaneously using the same setup (e.g. 13 CO/C18 O or HCN/HCO+ ) do not have this additional uncertainty.}

\subsection{CO kinematics}\label{sec:kin_mod}
A detailed study of the molecular gas kinematics in NGC\,3100 was presented in \citetalias{Ruffa19b}, based on our previous CO(2-1) observations. Among other interesting results, this analysis demonstrated the presence of radial motions (estimated to be $\leq10$\% of the underlying rotational velocity) in the plane of the CO disc. We tentatively identified these as inflow streaming motions likely induced by the two-armed spiral perturbation that we found in the inner CO regions (see \citetalias{Ruffa19b} for details). The hole at the centre of the CO(2-1) distribution, however, made these results somewhat uncertain. It is thus interesting to repeat the same analysis on the CO(3-2) line, which is the sole among the observed CO transitions tracing the gas up to the very centre of the source. The procedure adopted to carry out the kinematic modelling is the same as that followed for the CO(2-1) transition. This was extensively described in \citetalias{Ruffa19b}, hence only an outline is provided in the following.
\begin{figure*}
\centering
\begin{subfigure}[t]{.45\textwidth}
\centering
\caption{\textbf{Model moment one}}\label{fig:CO32_modmom1}
\includegraphics[width=\linewidth]{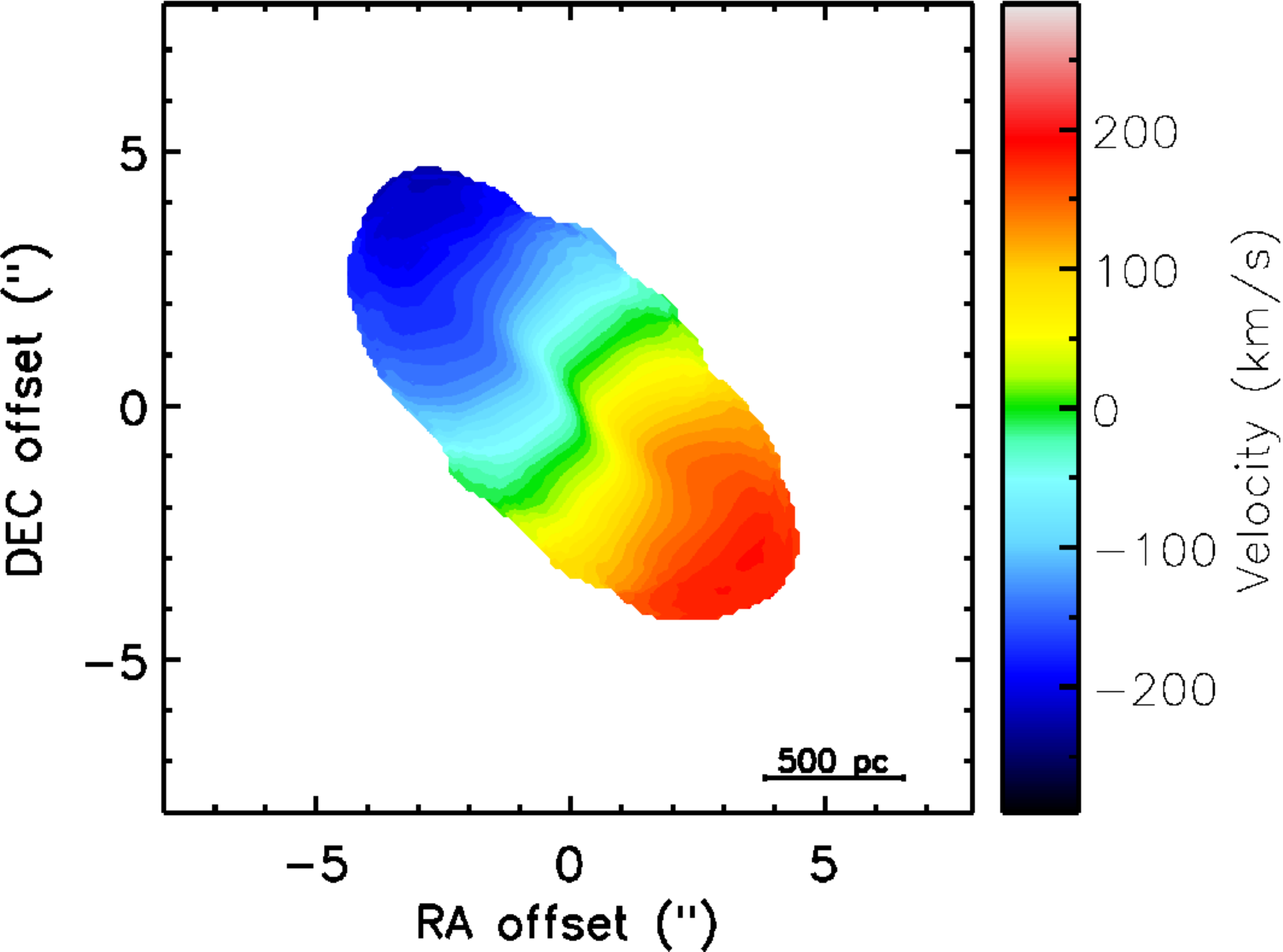}
\end{subfigure}
\hspace{3mm}
\begin{subfigure}[t]{.42\textwidth}
\centering
\caption{\textbf{Data - Model moment one}}\label{fig:CO32_resmom1}
\includegraphics[width=\linewidth]{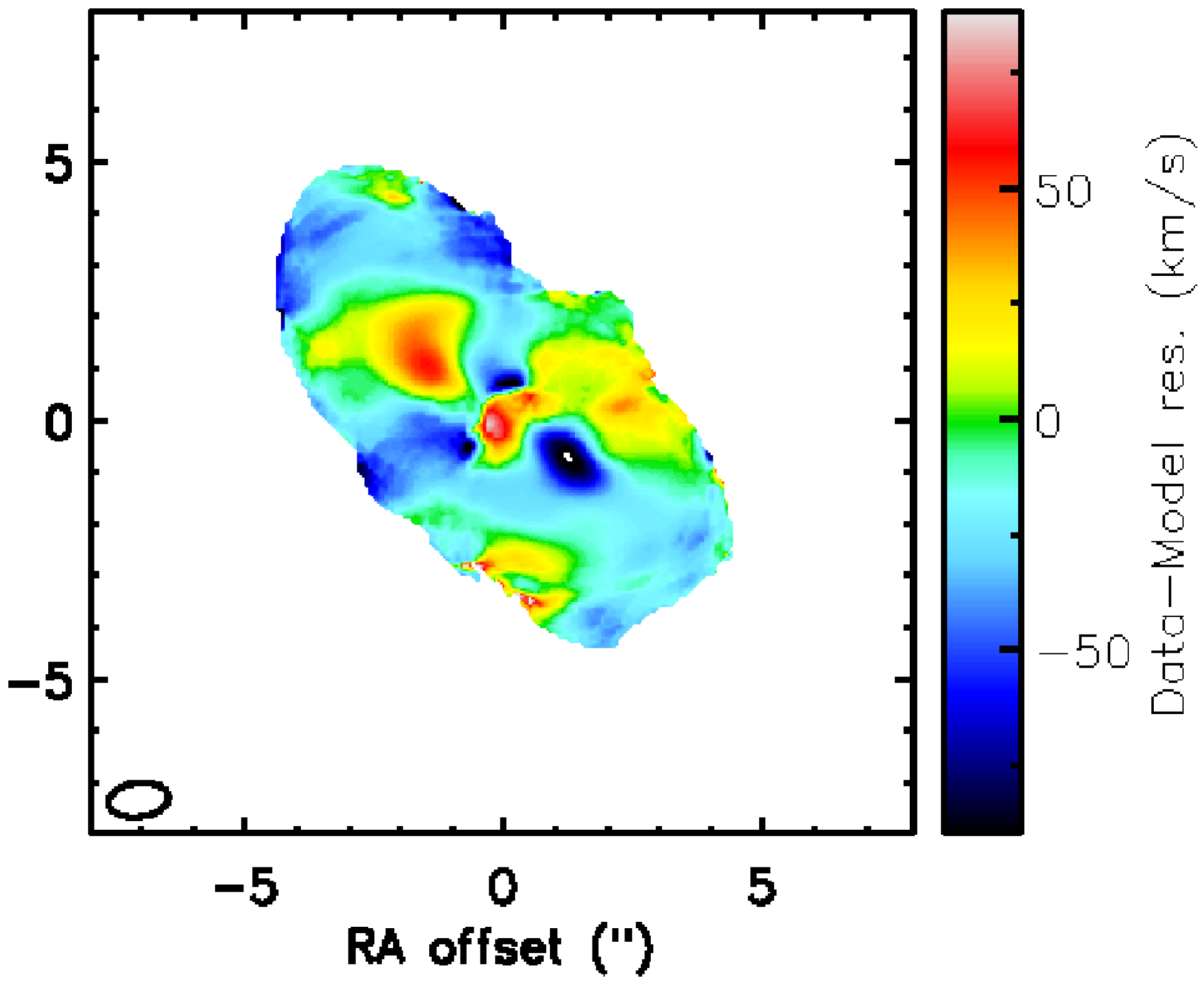}
\end{subfigure}
\medskip
\vspace{3mm}
\begin{subfigure}[t]{.45\textwidth}
\centering
\caption{\textbf{Major axis PVD}}\label{fig:CO32_majPVD}
\includegraphics[width=\linewidth]{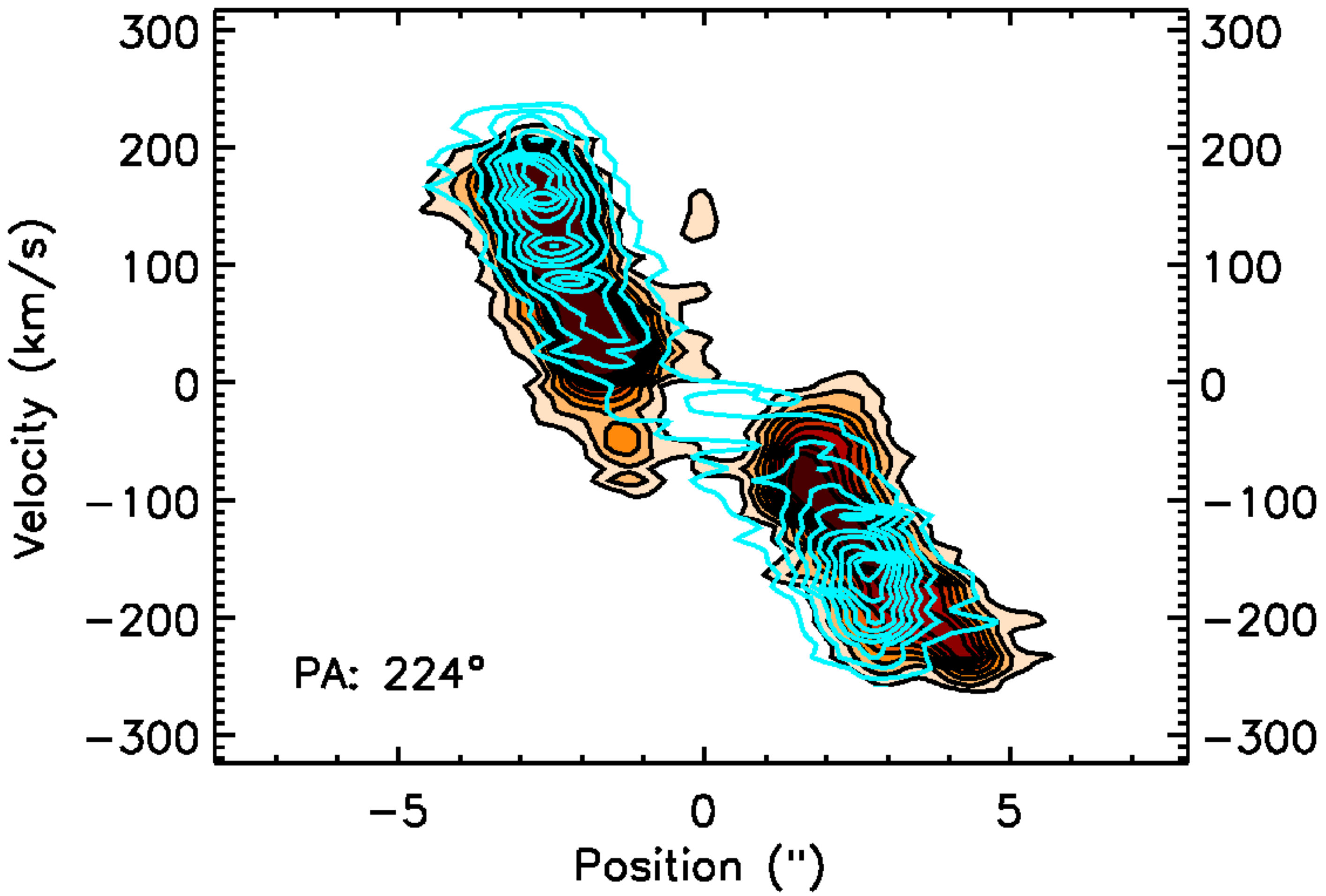}
\end{subfigure}
\hspace{3mm}
\begin{subfigure}[t]{.445\textwidth}
\centering
\caption{\textbf{Minor axis PVD}}\label{fig:CO32_minPVD}
\includegraphics[width=\linewidth]{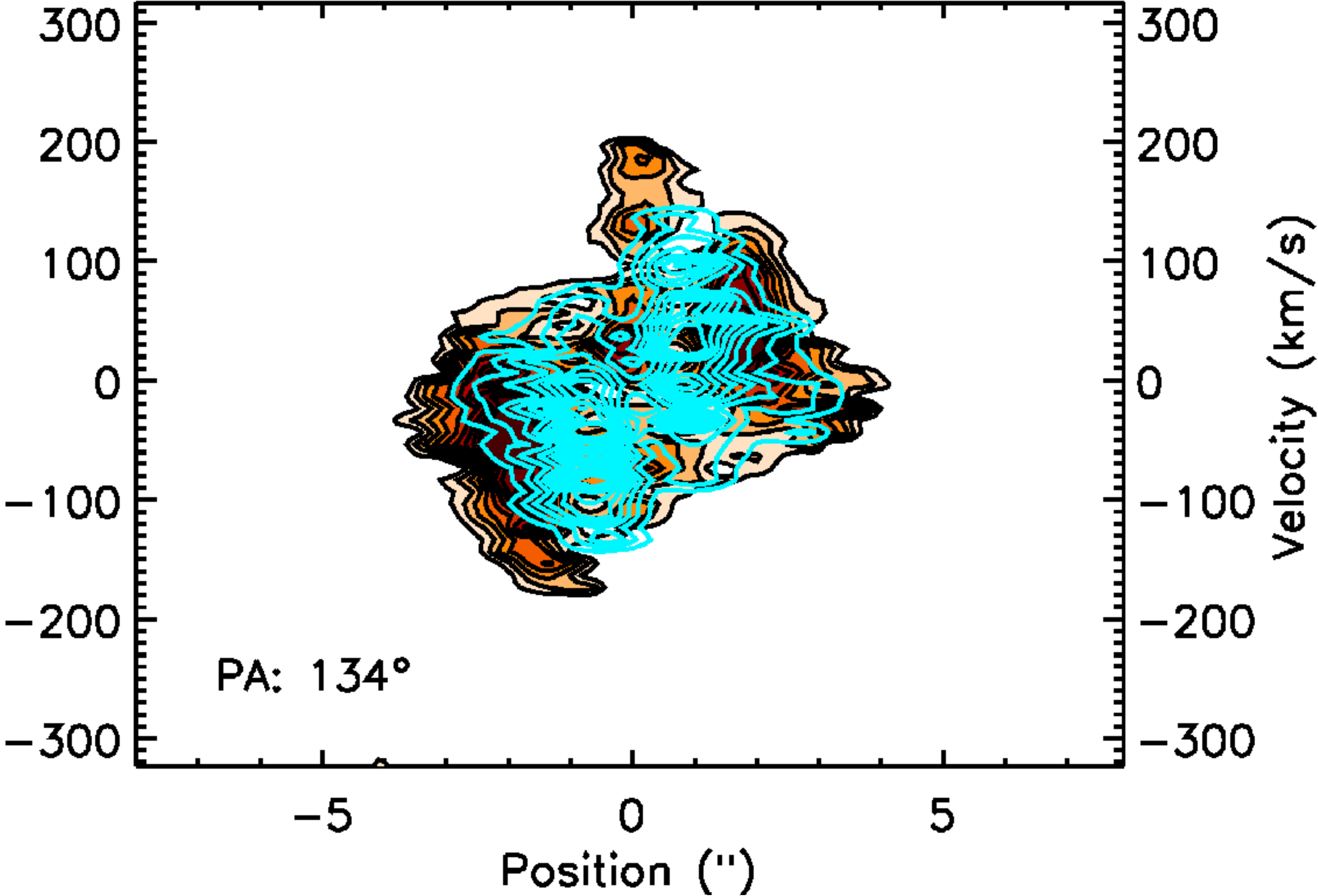}
\end{subfigure}
\caption[]{CO(3-2) model and residual (data - model) mean velocity maps (panels {\bf a} and {\bf b}, respectively). The bars to the right show the colour scale in km~s$^{-1}$. East is to the left and North to the top. Velocities are measured in the source frame and the zero-point corresponds to the intensity-weighted centroid of the CO emission. The maps are created with the masked moment technique described in Section~\ref{sec:mom_maps}, using a data cube with a channel width of 10~km~s$^{-1}$. The CO(3-2) major and minor axis PVDs are shown in panel {\bf c} and {\bf d}, respectively. The former is extracted within a rectangular area whose long axis is orientated according to the kinematic position angle indicated in the bottom-left corner of the panel (i.e.\,along the CO major axis); the same rectangular area is rotated by $90^{\circ}$ to extract the PVD along the kinematic minor axis. The contours of the best-fitting model are overlaid in cyan. The contours are drawn at 3, 9,... times the 1$\sigma$ rms noise level (listed in Table~\ref{tab:line images}). \label{fig:CO32mod}}
\end{figure*}

We analysed the CO kinematics using the {\sc Integrated Development Language} ({\sc IDL}) version of the \textsc{Kinematic Molecular Simulation} tool \citep[{\sc KinMS}\footnote{https://github.com/TimothyADavis/KinMS};][]{Davis13}. This routine allows to input arbitrary functions describing the gas distribution and kinematics, and constructs the corresponding mock data cube by calculating the line-of-sight projection of the circular velocity for a defined number (typically $10^{5}-10^{6}$) of point-like sources representing the gas distribution. As for the CO(2-1) transition, we modelled the CO(3-2) distribution with a nuclear two armed spiral structure (very well reproducing the observed CO peaks at the south-west/north-east side of the core; Figure~\ref{fig:CO_moments}, left panels) embedded in a disc. The former was parametrised in {\sc KinMS} as a uniform brightness distribution along loci whose positions from the phase centre are described by a logarithmic spiral function; the latter as an axisymmetric exponential disc (a form that has been shown to be appropriate in many ETGs; \citealp[e.g.][]{Davis13}). The s-shaped velocity iso-contours visible in the rotation pattern (Figure~\ref{fig:CO_moments}, middle column) may be caused by either warps or non-circular motions, as both phenomena produce the same kinematic features. In NGC\,3100, however, we found that they were best reproduced by adding both a position angle (PA) and an inclination ($i$) warp (i.e.\,allowing a linear radial variation of both parameters between a lower and an upper value). The gas was then assumed to be in purely circular motion and its rotation curve was modelled using a simple arctangent function \citep[see e.g.][]{Swinbank12,Voort18}. This approach still allows to identify possible non-circular motions, if significant residuals (i.e.\,larger than the data cube channel width) are found once such simple rotating model is subtracted from the observed velocity field. We also included in our model a velocity dispersion parameter ($\sigma$\textsubscript{gas}). Since our aim is simply to obtain an average value for the intrinsic gas velocity dispersion across the gas disc, $\sigma$\textsubscript{gas} was assumed to be spatially constant. The so-simulated data cube was then fitted to the observed one using the Markov Chain Monte Carlo (MCMC) code \textsc{KinMS\_MCMC}\footnote{https://github.com/TimothyADavis/KinMS\_MCMC}, which explores the full space defined by the input model and outputs the full Bayesian posterior probability distribution, identifying the best-fitting set of parameters. %To ensure our fitting process converges, each model parameter was let to vary within reasonable ranges (e.g.\,$\sigma$\textsubscript{gas} is constrained to be less than the maximum line-of-sight velocity widths in the right panels of Figure~\ref{fig:CO_moments}; the PA is allowed to vary by $\pm20^{\circ}$ around the values estimated by eye from the middle panels of Figure~\ref{fig:CO_moments}, etc).
These latter are fully in agreement with that obtained from the CO(2-1) kinematic modelling, and thus not further discussed here (see Table~\ref{tab:kin_mod_param} and \citetalias{Ruffa19b} for details). We only note that the best-fitting $\sigma$\textsubscript{gas} was found to vary from $\approx34\pm4$ to $\approx41\pm3$~km~s$^{-1}$ from the 2-1 to the 3-2 transition\footnote{Due to a more accurate processing of the uncertainties in the $\chi^{2}$ statistics, the error associated to the best-fitting CO(2-1) $\sigma$\textsubscript{gas} reported here is larger than that tabulated in \citetalias{Ruffa19b}.}. Although still consistent within their respective errors, these two values imply that the average gas velocity dispersion is intrinsically large, thus confirming the scenario inferred from the visual inspection of the line-of-sight mean velocity dispersion maps (i.e.\,the dynamical state of the cold gas can be considered turbulent on average; see Section~\ref{sec:mom_maps}).

\begin{table}
\centering
\caption{Best-fitting parameters from the kinematic modelling of the CO(3-2) transition.}
\label{tab:kin_mod_param}
\begin{tabular}{l l c c}
\hline
\multicolumn{1}{c}{ ID } &
\multicolumn{1}{c}{ Parameter} &
\multicolumn{1}{c}{ Unit} &
\multicolumn{1}{c}{ Value } \\
\hline
(1) &  Kinematic position angle, PA & (deg)  & 190$-$250  \\
(2) & Disc inclination, $i$ & (deg)  &  20$-$80 \\
(3) & Disc scale length, R$_{\rm disc}$ & (arcsec) & 3.1$\pm0.3$ \\
(4) & Max circular velocity, v$_{\rm flat}$ & (km s$^{-1}$) & 345$\pm$5\\
(5) & Rotation curve turn over radius, r$_{\rm turn}$ & (arcsec) & 2.6$\pm0.2$\\
(6) & Velocity dispersion, $\sigma_{\rm gas}$ & (km~s$^{-1}$) &  41$\pm3$\\
\hline
\end{tabular}
\parbox[t]{8.5cm}{ \textit{Notes.} $-$ Rows: (1) Range of the CO disc kinematic position angle (i.e.\,the PA measured counterclockwise from North to the approaching side of the velocity field). (2) Range of the CO disc inclination. (3) Scale length of the CO disc surface brightness profile. (4) Asymptotic (or maximum) circular velocity in the flat part of the CO rotation curve (corresponding to the major axis PVD shown in Figure~\ref{fig:CO32_majPVD}). (5) Effective radius at which the CO rotation curve turns over. (6) Gas velocity dispersion.}
\end{table}

The best-fitting CO(3-2) mean line-of-sight velocity map shown in Figure~\ref{fig:CO32_modmom1} illustrates that our simple rotating model provides a good overall representation to the observed kinematics (Fig.\,\ref{fig:CO32_moms}, middle panel). The distortions in the rotation pattern are also reasonably well reproduced by the combination of warps in both position angle and inclination included in the model, in agreement with what found from the CO(2-1) modelling. Our main aim here, however, is to look for kinematic signatures that can confirm and/or complement our previous finding of non-circular motions in the plane of the CO disc. Particularly interesting in this regard are the mean velocity residuals visible in Figure~\ref{fig:CO32_resmom1}, with amplitudes $\leq 70$~km~s$^{-1}$ in absolute value (or $\leq 91$~km~s$^{-1}$ deprojected). These are slightly larger than those found in the CO(2-1) (see Section~\ref{sec:sample} and \citetalias{Ruffa19b}) and - differently from them - can be clearly divided into two main groups, each lying along a different kinematic axis. One group consists of two regions along the kinematic major axis with peaks centred at RA offsets of $\approx 2^{\prime \prime}.0$ and $1^{\prime \prime}.5$ to the NE and SW of the core, respectively, clearly spatially correlated with the positions of the two CO(3-2) peaks visible in Figure~\ref{fig:CO32_moms} (left panel).
The two regions exhibit mostly redshifted/blueshifted velocities, respectively: knowing from the B-I colour map of NGC\,3100 (Figure~\ref{fig:NGC3100_optical}, top-right panel) that the eastern is the near side of the disc (see also \citetalias{Ruffa19b}), the observed residual velocities would imply gas inflow. 

A second group of significant residuals can be identified in Figure~\ref{fig:CO32_resmom1} along the kinematic minor axis. These are particularly interesting as the gas rotational component is mostly negligible along the minor axis and thus significant velocity components are likely to be ascribed to radial motions. The presence of such motions can also be clearly inferred from the minor axis PVD in Figure~\ref{fig:CO32_minPVD}, showing a gradient with velocities up to $\approx \pm 200$~km~$^{-1}$ not reproduced by our model (overlaid in cyan). The minor axis PVD of gas in pure circular motions is indeed expected to show nearly flat emission around the systemic velocity, with typical widths of a few tens of km~s$^{-1}$ (due to a combination of intrinsic gas velocity dispersion and beam smearing effects; \citealp[e.g.][]{Garcia14}). The residual velocities (both in the residual map and the minor axis PVD) are preferentially blueshifted to the SE and redshifted to the NW, implying gas outflow when compared with the near side of the CO disc.

\begin{figure*}
\centering
\includegraphics[width=\linewidth]{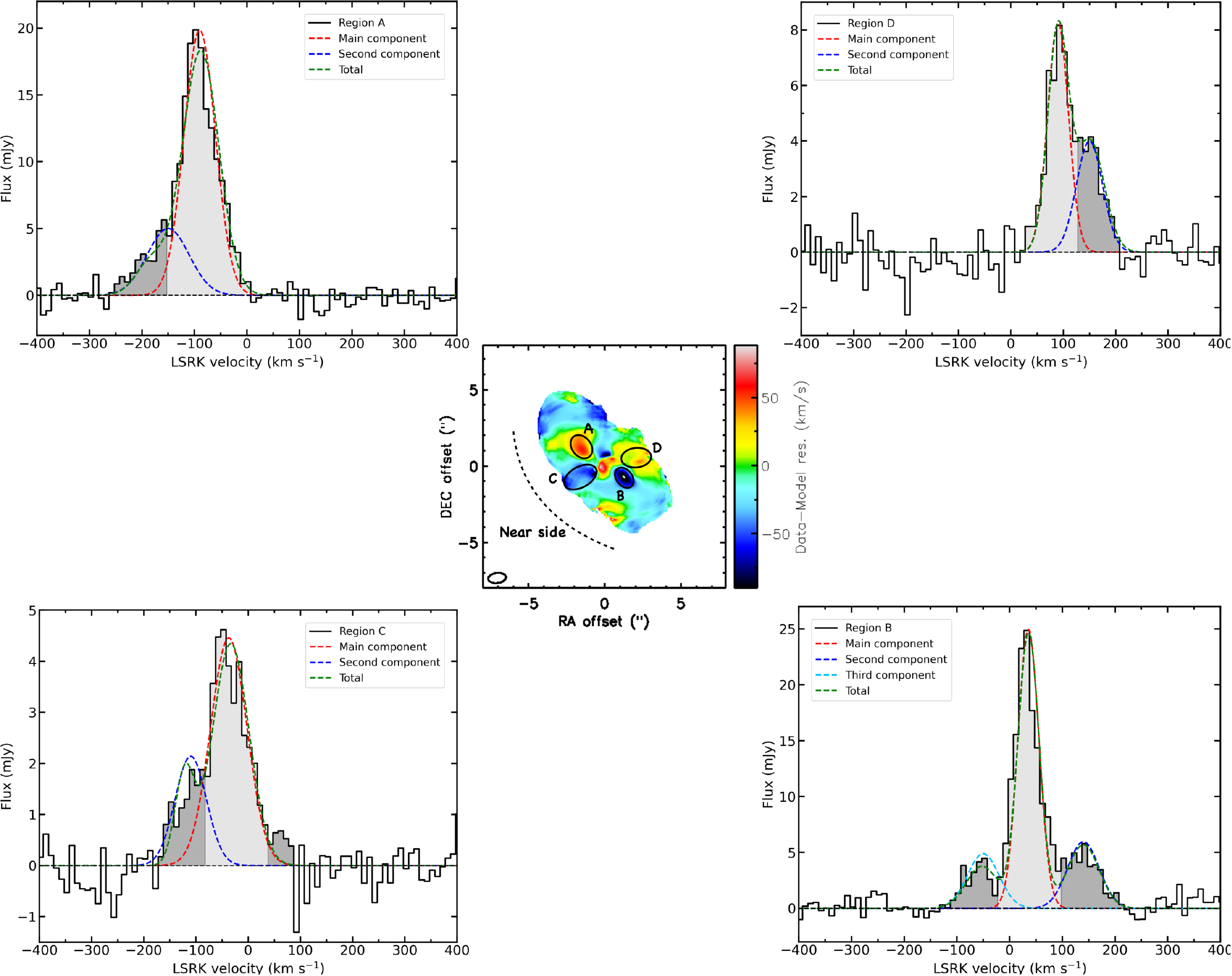}
\caption[]{CO(3-2) spectral profiles extracted from the cleaned data cube within boxes enclosing the regions of the most significant data-model velocity residuals (central panel). The two regions to the NE and SW of the core and along the kinematic major axis are labelled as A and B, respectively; those along the kinematic minor axis are labelled as C and D (see the text for details). The curved black dashed line in the central panel marks the near side of the CO disc. In each spectral profile, the best-fitting Gaussians identifying the main, secondary and total kinematic components are overlaid as red, blue and green dashed lines, respectively. In the spectrum of Region B (bottom-right corner) the cyan dashed line illustrates the best-fitting Gaussian profile of the third additional kinematic component. The light and dark grey shaded regions highlight the spectral channels resulting as part of the main and secondary kinematic components, respectively. The black dashed horizontal line indicates the zero flux level. \label{fig:Res_spectra}}
\end{figure*}

Caution is needed, however, as interpreting the gas motions from the sole residual map may lead to draw inaccurate conclusions. This is because - by definition - the mean line-of-sight velocities are flux density-weighted and thus biased towards the brighter components. As a further check, we then extracted spectral profiles from the cleaned CO(3-2) data cube within the regions enclosing the most significant velocity residuals described above. If the gas motion in the area of the spectral extraction is dominated by regular rotation, we should observe a single spectral component with a centroid close to the systemic velocity. Additional, higher-velocity spectral components are likely related to local gas flows and indicate departures from pure circular rotation \citep[see e.g.][]{Dominguez20}. The obtained spectral profiles and corresponding regions of extraction are illustrated in Figure~\ref{fig:Res_spectra}. 

In all the cases, we can clearly distinguish multiple kinematic components. The spectral profile in the top-left corner of Figure~\ref{fig:Res_spectra} is extracted from the area to the NE of the core with large redshifted residuals (labelled as region A). This shows both a main narrow emission component tracing the local gas circular motion, and a secondary component observed as a blushifted excess with velocities up to $\approx -250$~km~s$^{-1}$. We note that this latter has exactly opposite velocities compared to those expected from just the inspection of the mean line-of-sight velocity residual map, stressing the importance of analysing the velocity residuals together with the local spectral profiles. The panel in the bottom-right corner of Figure~\ref{fig:Res_spectra} shows the spectrum of the region to the SW of the core (labelled as region B) with large blueshifted residuals. In this case, both a blueshifted and a redshifted excess (with velocities up to $\approx -120$ and $\approx +90$~km~s$^{-1}$, respectively) are observed in addition to the main narrow component. By comparing the velocities of the broader kinematic components in region A and B spectra with the near side of the disc, we conclude that they are likely tracing both inflow/outflow motions. As mentioned above, these are clearly spatially correlated with the positions of the two CO peaks, and thus - in turn - with the location of the inner spiral perturbation. Spiral shocks and the associated turbulence can induce streaming motions, causing the gas to lose or gain angular momentum and move inwards or outwards, respectively \citep[e.g.][]{deVen10}. We then more robustly confirm and complement the scenario only tentatively inferred from the previous CO(2-1) kinematic modelling (\citetalias[]{Ruffa19b}): the non-circular motions traced by the velocity residuals at the NE and SW of the core and along the CO kinematic major axis are likely in the form of inflow/outflow streaming motions induced by the inner spiral perturbation. 

The spectral profiles of the areas along the CO kinematic minor axis with the most significant velocity residuals (labelled as regions C and D) are shown in the bottom-left and top-right corners of Figure~\ref{fig:Res_spectra}, respectively\footnote{We note that we also attempted to extract the spectrum of the region at the centre of the gas distribution showing the brightest redshifted residuals, but the poor S/N ratio in that area does not allow to clearly distinguish any spectral feature and carry out a reliable analysis.}. They show blueshifted (C) and redshifted (D) excesses with velocities up to -170 and +200~km~s$^{-1}$, respectively, in addition to the main kinematic component\footnote{A faint redshifted excess is also visible in the spectrum of Region C (bottom-left corner of Figure~\ref{fig:Res_spectra}), and highlighted as well in a dark grey colour. However, the S/N ratio of the corresponding channels is too low to allow a proper Gaussian fitting and thus robustly identifying it as an additional kinematic component.}. Given the orientation of the gas disc with respect to our line-of-sight, the observed velocity excesses are compatible with an outflow. Interestingly, the spatial comparison between the outflow region and the location of the radio lobes (see bottom-right panel of Figure~\ref{fig:NGC3100_optical}) strongly suggests that the former is jet-driven. A rough estimate of the outflowing gas mass can be obtained by calculating the flux densities of the blueshifted/redshifted kinematic components integrated over the velocity ranges highlighted by the dark-grey shaded channels in regions C and D spectra, and using Equation~\ref{eq:gas mass} to convert the integrated fluxes into molecular gas masses. We note that, by comparing Figures~\ref{fig:CO_ratios} and \ref{fig:Res_spectra}, it is evident that the outflowing gas is also the gas showing the largest line ratios and thus in a optically thin regime (see Section~\ref{sec:ratios_results}). In such condition, the \cotoh2\ conversion factor, $X$\textsubscript{CO}, depends on the gas excitation temperature \citep[see][]{Bolatto13}. Assuming $T_{\rm ex}=50$~K for an average $R_{\rm 31}\approx12$ in the high-excitation regions, we get $X_{\rm CO}= 2.5\times10^{19}$~cm$^{2}$~(K~km~s$^{-1}$)$^{-1}$ and $M_{\rm out} = 5.6 \times 10^{4}$~M$_{\odot}$. This mass is only a small fraction of the total CO mass (see Section~\ref{sec:mol_masses}), thus the effect of the jet-induced outflow in NGC\,3100 appears very localised and its impact on the host galaxy as a whole is likely to be negligible. Making the simple assumption of spherical geometry for the outflowing region, an upper limit on the mass outflow rate can also be obtained from the relation $\dot{M} \lesssim 3v_{\rm max}M_{\rm out}/R_{\rm max}$ \citep[e.g.][]{Maiolino12}, where $M_{\rm out}$ is the mass of the outflowing gas estimated above, $R_{\rm max}$ and $v_{\rm max}$ are the maximum extent of the outflow region and maximum outflow velocity, respectively. Assuming $R_{\rm max}=300$~pc (i.e.\,the maximum extent of regions C and D in Figure~\ref{fig:Res_spectra}) and $v_{\rm max}=200$~km~s$^{-1}$, we obtain $\dot{M} \lesssim 0.12$~M$_{\odot}$~yr$^{-1}$. This is at least one order of magnitude lower than that estimated in other known cases of CO outflows induced by young radio jets \citep[e.g.][]{Garcia14,Mahony16,Oosterloo17}, definitely indicating that - at least in terms of the induced kinematic perturbations - the jet-ISM coupling in NGC\,3100 has to be much less extreme. We discuss these results in Section~\ref{sec:interaction_discuss}.

\begin{figure}
\centering
\includegraphics[scale=0.38]{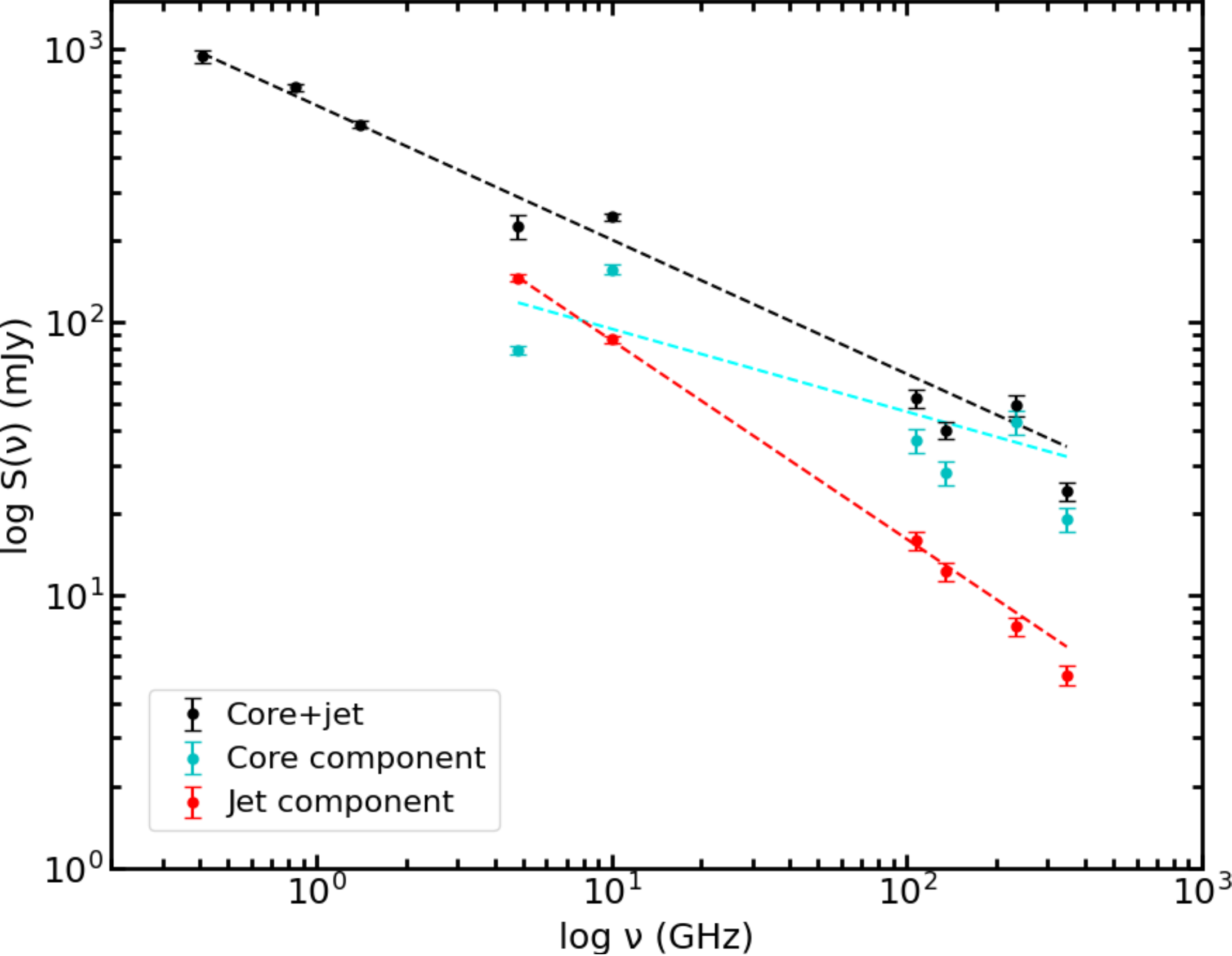}
\caption[]{Radio to sub-mm spectrum of NGC\,3100. Black circles are for flux density measurements encompassing emission from both the nucleus and the jets; these include data from our ALMA and JVLA observations (\citetalias{Ruffa19a,Ruffa20}, and this work) and also 1.4~GHz, 843 and 408~MHz measurements taken from NED. Cyan and red circles are used for flux densities including emission from a single component only (core and jet, respectively); in this case, only spatially-resolved and matched-resolution continuum data are considered (i.e.\,the 5 and 10~GHz VLA and all ALMA continuum data from \citetalias{Ruffa19a}, \citetalias{Ruffa20} and this work). A best-fitting power law is overlaid in the corresponding colour.}\label{fig:SED}
\end{figure}

\section{Discussion}\label{sec:discuss}
\subsection{Continuum emission}\label{sec:cont_discuss}
Our multi-frequency (107, 136, 235 and 348~GHz) mm continuum images show the same compact core-double lobe structure seen in the radio continuum at 5 and 10 GHz (maximum linear size $\approx$2~kpc; see \citetalias{Ruffa19a} and \citetalias{Ruffa20}), although emission from the extended radio structures becomes progressively fainter at increasing frequencies. The great morphological similarity between the continuum emission over this wide range of frequencies suggests that the continuum of NGC\,3100 is dominated by non-thermal synchrotron emission from the core and extended radio structures up to at least 348~GHz. Thermal emission from the Rayleigh–Jeans tail of the dust component is expected to contribute to the continuum at millimetre wavelengths. We find no evidence of such contribution up to the highest frequencies imaged with ALMA (i.e.\,235~GHz and 348~GHz; see Figures~\ref{fig:Band6_cont} and \ref{fig:Band7_cont}), although we cannot exclude some contribution from this mechanism to the unresolved core emission.

Figure~\ref{fig:SED} shows the radio--(sub-)mm spectrum of NGC\,3100, where different colours are used for the data points of different components: core, jet and total (core$+$jet). This latter include data from our ALMA and JVLA observations (\citetalias{Ruffa19a,Ruffa20}, and this work) and also measurements from unresolved 1.4~GHz, 843 and 408~MHz data taken from the NASA/IPAC extragalactic database (NED; https://ned.ipac.caltech.edu). Such combination of high- and low-resolution data in the total radio to sub-mm spectrum is possible in this case because the radio source in NGC\,3100 is compact and there is no evidence from the available multi-frequency radio data for emission on larger scales (see Section~\ref{sec:sample} and \citetalias{Ruffa20}). Flux density measurements including emission from single components (i.e.\,core or jet) are instead taken only from spatially-resolved and matched-resolution data (i.e.\,the 5 and 10~GHz VLA and all ALMA continuum data from \citetalias{Ruffa19a}, \citetalias{Ruffa20} and this work).

To obtain spectral indices, the data points of each component are fitted with a single power-law function (of the form $S_{\rm \nu} = A\nu^{\rm \alpha}$, where $S_\nu$ is the flux density as a function of frequency $\nu$, $A$ is the power-law normalisation, and $\alpha$ is the spectral index) using the least-squares minimization Python routine \texttt{lmfit} \citep{Newville14}. The spectral index and power-law normalization are left free to vary in the fitting process. The resulting best-fitting spectral indices are $\alpha_{\rm tot}=-0.49\pm0.03$ for the total continuum, $\alpha_{\rm core}=-0.3\pm0.2$ for the core component, and $\alpha_{\rm jet}=-0.73\pm0.01$ for the jet component. 

The presence of a flatter and steeper component associated with the core and jet, respectively, was already reported in \citetalias{Ruffa19a}, based on a spatially-resolved spectral index analysis carried out using the ALMA 235~GHz and VLA 5~GHz data available at that time. The obtained $\alpha_{\rm jet}$ is consistent with optically thin synchrotron emission, but is slightly steeper than that typically observed in FR\,I jets ($\alpha \approx -0.6$ between 1.4 and 4.9\,GHz; \citealt{LB13}). Nevertheless, there is no clear sign of an abrupt high-frequency steepening due to synchrotron or inverse Compton losses.
The best-fitting core spectral index, $\alpha_{\rm core}$, is consistent with partially optically-thick (self-absorbed) emission from a compact component. The observed scatter in the data points, however, results in large fitting errors, making $\alpha_{\rm core}$ poorly constrained. Since the various flux measurements are taken at different epochs, we may be observing the effects of core variability. %indicating that a more complex modelling would be required in this case. Such detailed continuum analysis is beyond the purpose of this paper. We nevertheless note that similar variabilities have been often observed in the cores of nearby low-luminosity (LINER-like) AGN and attributed to the advection-dominated accretion flows (ADAF; \citealp[]{Nara95}) thought to occur in LERGs \citep[see e.g.][]{Doi05a,Doi05b}.

In general, the radio--(sub-)mm morphological and spectral properties of the radio source in NGC\,3100 (i.e.\,compact linear size and overall steep spectrum) support a scenario (already outlined in \citetalias{Ruffa20}) in which it has likely been caught in the early phases of its evolution \citep[see e.g.][for a recent review on this subject]{ODea20}. %CSS are intrinsically compact (i.e.\,linear sizes between 500~pc and 20~kpc) radio sources characterized by a steep radio spectrum (i.e.\,$\alpha_{\rm tot}\leq-0.5$) that flattens and turns over at frequencies $\leq400$~MHz. The origin of CSS sources is still debated, but there is a consensus that the majority are young radio sources. A large population of nearby ($z<0.1$), low-luminosity ($P_{\rm 1.4GHz}<10^{25}$~W~Hz$^{-1}$) CSS sources exists, most of which ($\approx$75\%) are classified as compact LERGs \citep[e.g.][]{Sadler14,Sadler16}. Given its general (see Table~\ref{tab:ngc3100}) and radio-mm continuum properties, NGC\,3100 is likely one of these examples.

\subsection{Jet-cold gas interaction}\label{sec:interaction_discuss}
\subsubsection{CO excitation conditions and kinematics}
The analysis of the multiple CO line detections presented in this paper allows us to more robustly confirm and complement the results obtained from the sole CO(2-1) transition in NGC\,3100 (\citetalias[]{Ruffa19b}). First of all, our analysis clearly shows that a simple model assuming gas in pure circular motions well reproduces most of the observed kinematic features, once accounting for the presence of both a PA and inclination warp (Figures~\ref{fig:CO_moments} and \ref{fig:Res_spectra}). This confirms that the bulk of the CO gas in NGC\,3100 is regularly rotating (at least at the spatial resolution of our ALMA observations; see Table~\ref{tab:ALMA observations summary}). Nevertheless, a separate kinematic component can be clearly identified from the mean line-of-sight velocity residuals along the CO kinematic major axis (Figure~\ref{fig:Res_spectra}) and ascribed to inflow/outflow streaming motions induced by an inner two-armed spiral perturbation (already tentatively inferred from the previous CO(2-1) data; \citetalias[]{Ruffa19b}). The maximum inflow/outflow velocities are $\approx +90$~km~s$^{-1}$ and $\approx -250$~km~s$^{-1}$, respectively (see Section~\ref{sec:kin_mod} and Figure~\ref{fig:Res_spectra}). These numbers indicate that streaming motions induced by sub-kpc scale spiral shocks can be a significant fraction of the underlying circular velocity (as suggested by - e.g.\,- \citealt[][]{deVen10}). As extensively discussed in \citetalias{Ruffa19b}, the significance of this result relies on the fact that nuclear spiral perturbations and associated non-circular motions are considered as an efficient mechanism for transporting gas from kpc scales to the nucleus and feeding the central SMBH \citep[e.g.][]{Fathi13,Combes13}. Theoretical studies, however, demonstrate that nuclear spiral instabilities can produce large accretion rates ($1-10$~M$_{\odot}$~yr$^{-1}$; \citealp[e.g.][]{Hopkins10}), as opposed to the strongly sub-Eddington rates estimated in LERGs such as NGC\,3100 (see below). This suggests that other physical processes intervene in the very inner regions of these objects, keeping the accretion rate low. Higher resolution observations probing the gas at scales relevant for the accretion process ($\ll100$~pc) are needed to examine the gas kinematics around the SMBH and shed light on this issue. %A detailed discussion on this aspect, however, has been already presented in  and thus not repeated here.

Beyond that, the results presented in Sections~\ref{sec:ratios_results} and \ref{sec:kin_mod} clearly show that an additional ongoing perturbation is affecting both the physical conditions and kinematics of the CO gas located in the inner regions. Indeed, another separate kinematic component can be clearly distinguished along the CO kinematic minor axis and associated to an outflow (see Section~\ref{sec:kin_mod} and Figure~\ref{fig:Res_spectra}). The spatial correlation between the outflow region and the location of the radio lobes strongly suggests that the former is jet-induced. The maximum observed outflow velocity is $\approx 200$~km~s$^{-1}$ and the mass outflow rate is estimated to be $\lesssim 0.12$~M$_{\odot}$~yr$^{-1}$ (see Section~\ref{sec:kin_mod}), indicating that the jet impact on global (galactic) scales is likely negligible at this stage. These values are also significantly lower than those expected from available simulations \citep[e.g.][]{Wagner12,Wagner16,Mukherjee16,Mukherjee18a,Mukherjee18b} and estimated in the few existing spatially-resolved studies of (sub-)kpc scale jet-cold gas interactions \citep[e.g.][]{Combes13,Garcia14,Morganti15,Mahony16,Oosterloo17}. On the other hand, the line ratios of the outflowing CO gas are drastically enhanced, implying optically thin conditions and high excitation temperatures ($\gtrsim50$~K; see Section~\ref{sec:ratios_results}). Similar modifications to the CO physics have been observed in the most extreme cases of sub-kpc scale jet-cold gas interactions \citep[e.g.][]{Oosterloo17}. In short, our results indicate that the jet-CO coupling ongoing in NGC\,3100 is strongly altering the physical conditions of the involved CO gas fraction, but only mildly affecting its kinematics. 

The fact that the kinematic perturbations in the interacting CO gas are significantly lower compared to those expected from simulations and observations can depend on various factors. Among these, the age of the radio source may play an important role. Indeed, both simulations and observations have so far taken snapshots of the radio jet-cold gas interplay mostly during the very early phases of the jet evolution \citep[i.e.\,within less than 1 Myr from their first ignition, corresponding to linear sizes $<1$~kpc; e.g.][]{Oosterloo17,Mukherjee18a,Mukherjee18b,Zovaro19}. Based on observations of a handful of objects with jets at slightly different stages of their {\it initial} evolution (but no more extended than 1~kpc), \citet{Morganti21} recently suggested that the impact of the radio jets may change over time. In this scenario, the kinematics of the interacting cold gas should become less perturbed as the jets break out from the core region and start to expand on larger scales (typically at ages of few Myr, corresponding to linear sizes of the radio source $\gtrsim1$~kpc). The maximum linear extent of the core-double lobe radio source in NGC\,3100 is 2~kpc (see Section~\ref{sec:cont_discuss} and \citetalias[]{Ruffa20}), indicating that - while still relatively young - it has been caught in a more evolved phase than those explored so far. We thus consider the possibility of a causal connection between the mild kinematic perturbations observed in the CO gas that is interacting with the radio jets and the evolutionary stage of these latter. 

As mentioned in Section~\ref{sec:intro}, the jet-cold gas coupling has also been shown to be very sensitive to the jet power \citep[][]{Mukherjee18b}. Using well-known scaling relations between the jet and radio power \citep[][]{Cavagnolo10}, we estimated $P_{\rm jet}\approx1\times10^{43}$~erg~s$^{-1}$ in NGC\,3100. Even though the large scatter in the adopted scaling relations makes such estimation uncertain up to one order of magnitude \citep[see e.g.][]{Godfrey16}, this is still fully within the range of jet powers typically associated to low-power FR\,I radio galaxies, such as the majority of LERGs \citep[e.g.][]{Godfrey13}. The extent that the jet-induced kinematic perturbations should have in such cases is not yet clear, as both simulations and observations have so far mostly focused on jets with powers $>10^{44}$~erg~s$^{-1}$ (typical of powerful radio galaxies with FR\,II morphologies; \citealp[e.g.][]{Godfrey16}). It has been suggested that the jet--ISM coupling should be most pronounced for jets with intermediate powers ($P_{\rm jet}\sim10^{43}-10^{44}$
~erg~s$^{-1}$) lying close to the CO plane, such as in the case of NGC\,3100 \citepalias[see][]{Ruffa20}. The observed kinematic features, however, do not seem to support this hypothesis.

In addition to all the above, the few objects in which jet-driven (sub-)kpc scale CO outflows have been spatially resolved are also radiatively efficient (Seyfert-like) AGN, with typical AGN bolometric luminosities between one and ten percent of their Eddington luminosities (i.e.\,$L_{\rm Bol}/L_{\rm Edd}\approx0.01-0.1$). Thus, as noted by \citet[][]{Morganti21}, there is an additional source of energy in the central regions of these objects and - although primarily jet-induced - the observed cold gas perturbations may result from a combination of processes. The nuclear luminosity in NGC\,3100 is strongly sub-Eddington\footnote{We estimated the AGN bolometric luminosity in NGC\,3100 using the [OIII]$\lambda5007$ line as a proxy and applying the bolometric correction $L_{\rm bol}/L_{\rm [OIII]}\approx3500$ \citep[][]{Heckman04}. Based on the optical spectroscopy presented by \citet[][]{Dopita15}, $L_{\rm [OIII]}\approx1.7\times10^{39}$~erg~s$^{-1}$ and thus $L_{\rm bol}=6\times10^{42}$~erg~s$^{-1}$. The mass of the SMBH in NGC\,3100 was estimated using the scaling relation $\log (M_{\rm SMBH}/M_{\rm \odot}) = 8.23 + 3.96 \log (\sigma_{*}/200~{\rm km~s^{-1}})$ \citep[][]{Tremaine02}, where we assumed a stellar velocity dispersion $\sigma_{*}=191$~km~s$^{-1}$ taken from the HyperLeda database (http://leda.univ-lyon1.fr). This results in $M_{\rm SMBH}=1.6\times10^{8}$~M$_{\odot}$ and $L_{\rm Edd}=2\times10^{46}$~erg~s$^{-1}$.}, with $L_{\rm Bol}/L_{\rm Edd}=3\times10^{-4}$. This is well within the values typically associated to radiatively inefficient AGN and compatible with advection-dominated accretion flows (ADAF) and its variants \citep[e.g.][]{Narayan02}, suggesting that the observed CO kinematic perturbations in NGC\,3100 are exclusively induced by the (low) jet kinetic power.

Appropriate models, along with a statistically-significant number of spatially-resolved studies of jet-cold gas interactions covering wide ranges of jet ages/powers and nuclear luminosities, would be crucial to better understand the physics underlying the observed CO kinematics and draw more solid conclusions.

\subsubsection{Dense and shock gas tracers}\label{sec:HCO_discuss}
As shown in Figure~\ref{fig:HCO_mom0}, the faint HCO$^{+}$(4-3) detection is restricted to a single unresolved blob-like structure at $\approx2''$ west of the nucleus. A qualitative comparison with the CO(3-2) distribution (overlaid as blue-to-warm contours in Figure~\ref{fig:HCO_mom0}) shows that the HCO$^{+}$(4-3) is not spatially correlated with the peak of the CO emission corresponding to the south-western spiral arm, where the gas is expected to be compressed and the density enhanced (\citealp[e.g.][]{deVen10}; see also \citetalias[see][]{Ruffa19b}). Figure~\ref{fig:HCO_ratios} instead shows that the faint HCO$^{+}$(4-3) emission arises from a region adjacent to the base of the northern radio lobe, co-spatial with some of the largest $R_{21}$ and $R_{31}$ values (Figure~\ref{fig:CO_ratios}) and the redshifted outflow region (Figure~\ref{fig:Res_spectra}). This suggests a correlation between the jet-ISM coupling and the detected  HCO$^{+}$ emission. Indeed, 3D hydrodynamical simulations show that the jet plasma expanding through the surrounding gaseous matter inflates pseudo-spherical bubbles driving shocks that can not only accelerate the gas outwards, but also heat (up to kinetic temperatures of $\approx 100$~K) and compress it, yielding to enhanced dense gas emissions up to distances a few kpc away from the main jet/lobe axis \citep[e.g.][]{Mukherjee16,Mukherjee18b,Mukherjee21}. Also the R$_{\rm HCO}$ flux density ratio seems to confirm this scenario, reaching peaks of $\approx0.7$ (Figure~\ref{fig:HCO_ratios}). These are consistent with those observed in regions of jet-ISM interactions and much larger than those typical of unperturbed gas regions \citep[$<0.045$; e.g.][]{Oosterloo17}. Another interesting aspect to consider is that HCO$^{+}$ emission is detected only at blueshifted velocities (with a peak at $\approx-200$~km~s$^{-1}$; see Figure~\ref{fig:HCO_spectum}), while it overlaps with a region located on the receding side of the CO disc (Figure~\ref{fig:HCO_mom0}). This suggests that the dense and tenuous gas components are kinematically decoupled, and may indicate that the former is tracing gas that is deviating from regular rotation \citep[e.g.][]{Young21}.

%The HCO$^{+}$(4-3) transition requires much larger gas densities ($>1.8\times 10^{6}$ cm$^{-3}$; \citealt[e.g.][]{Greve09}) to be excited, compared to CO transitions (typically $\sim 10^{3}$ cm$^{-3}$). As such, it is considered a very good tracer of dense molecular gas.
% We stress that also nearby spiral galaxies hosting AGN, typically characterised by a multi-transitions molecular ISM, HCO$^{+}$(4-3) and other high-density tracers remain undetected \citep[see e.g., NGC\,1433][]{Combes13}. 

As described in Section~\ref{sec:intro}, the SiO and HNCO molecules are commonly considered good tracers of shocks, with HNCO(6-5)/SiO(3-2) flux density ratios (i.e.\,the ratios of the transitions targeted in our ALMA observations) observed to vary from $\approx$0.3 to $\approx$3 from fast ($v>50$~km~s$^{-1}$) to slow ($v<20$~km~s$^{-1}$) shocks, respectively \citep[e.g.][]{Kelly17}. Strong SiO emission, in particular, has been detected in regions co-spatial with the radio jets and interpreted as an evidence of a jet-ISM interaction \citep[e.g.][]{GarciaBurillo10}. On the other hand, HNCO emission can be also observed without the need of a shock, as it has been clearly detected in unperturbed regions of nearby ETGs \citep[e.g.][]{Topal16}. Our analysis shows that both conditions (i.e.\,perturbed and unperturbed) co-exist in the cold gas component of NGC\,3100. Therefore, in this context, the non-detection of any of these two transitions is surprising. One possibility is that such non-detections, together with the faint  HCO$^{+}$(4-3) detection, trace an intrinsic deficiency of very dense ($n_{\rm crit} \gg 10^{6}$ cm$^{-3}$) cold gas in this object. This hypothesis, however, currently remains purely speculative, as observations of other HCO$^{+}$ transitions and/or of different dense/shocks gas tracers (such as HCN, HNC, CS, CH$_{3}$OH, etc.\,) would be needed to draw more solid conclusions.

\subsection{CO-dust comparison and gas-phase metallicity: implications for the gas origin}\label{sec:CO_dust_discuss}
\begin{figure}
%\begin{subfigure}[t]{.50\textwidth}
\centering
%\caption{\textbf{}}\label{fig:CO10_dust}
\includegraphics[width=\linewidth]{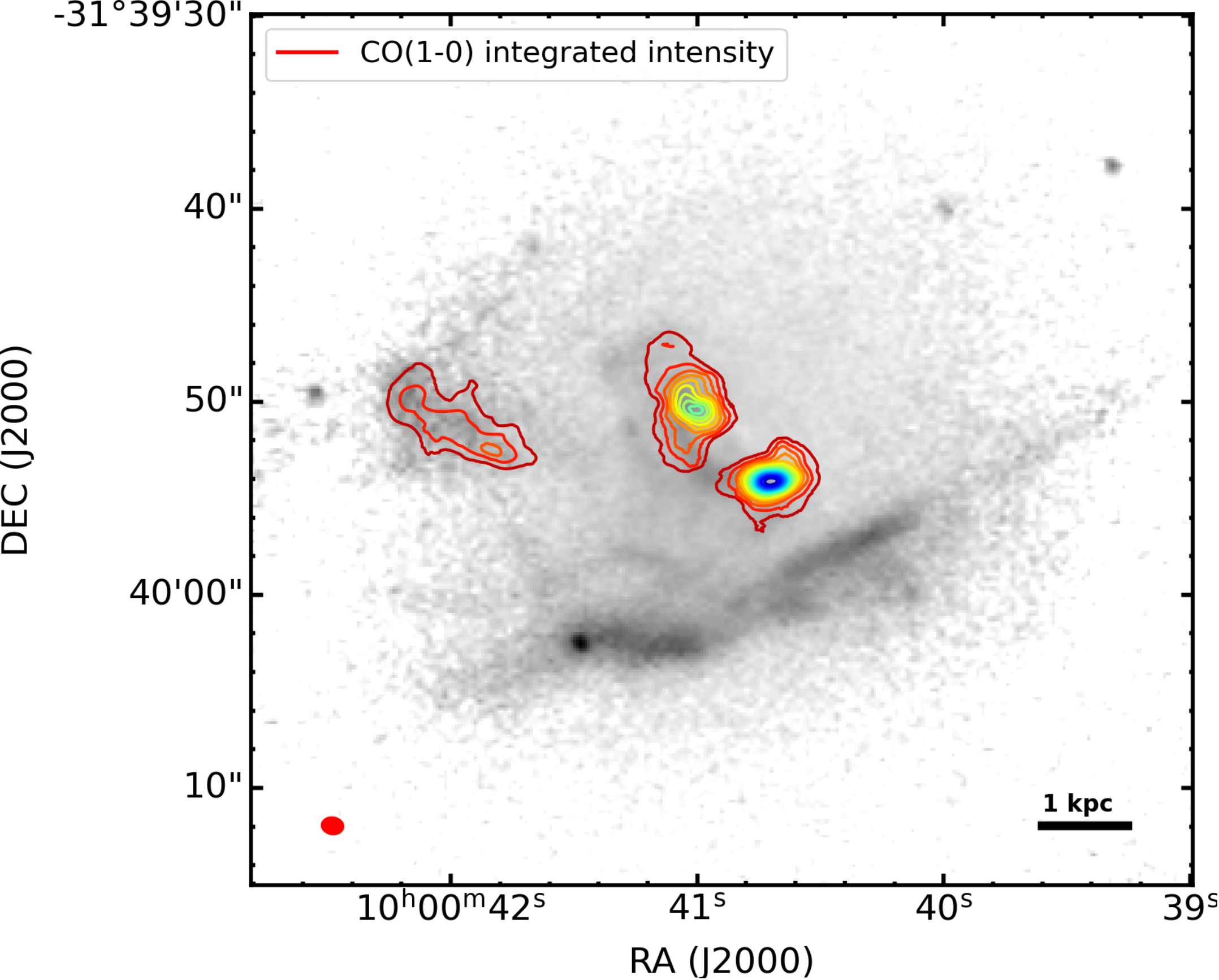}
%\end{subfigure}
%\hspace{0.05mm}
%\begin{subfigure}[t]{.48\textwidth}
%\centering
%\caption{\textbf{}}\label{fig:CO21_dust}
%\includegraphics[width=\linewidth]{NGC3100_CO21_dust.pdf}
%\end{subfigure}
\caption[]{CO(1-0) integrated intensity contours overlaid to the $B$-$I$ dust extinction map of NGC\,3100 (see \citetalias[]{Ruffa19a}) zoomed in the central $50''\times50''$ ($\sim9\times9$~kpc$^{2}$). CO contours are drawn at 1,3,9...times the 3$\sigma$ rms noise level. The CO beam size and scale bar are shown in the bottom-left and bottom-right corners, respectively; East is to the left and North to the top. \label{fig:CO_dust}}
\end{figure}

The spatial correspondence between dust and molecular gas is generally interpreted as indicating that both trace the same ISM component \citep[e.g.][]{Alatalo13}. In \citetalias{Ruffa19a} we demonstrated the tight co-spatiality between the central dust and ring-like CO(2-1) structures (see also the bottom-right panel of Figure~\ref{fig:NGC3100_optical}). Given its larger-scale sky distribution, it is interesting to repeat such comparison for the 1-0 transition. The CO(1-0) integrated intensity contours are overlaid on the $B$-$I$ colour map of NGC\,3100 in Figure~\ref{fig:CO_dust}. This image clearly shows that not only the inner gas and dust structures are spatially coincident, but also the eastern CO patch accurately traces that of the dust. Curiously, the prominent southern dust lane is not matched by any CO emission, at least at the spatial resolution and sensitivity of our CO observations. %possibly indicating that, if present, the southern CO distribution on larger scales is below the detection threshold of our observations. 

The detection of significant amounts of dust and molecular gas at the centre of nearby ETGs - the typical hosts of LERGs - points towards a recent ISM regeneration \citep{Young14}. The origin of this material, however, is still unclear \citep[e.g.][]{Davis16}: it may be either internally generated (from stellar mass loss and/or cooling from the hot halo) or externally accreted (from interactions or minor mergers). Disentangling these two scenarios is crucial for fully understanding the powering mechanism of the AGN in LERGs, which is still a matter of debate \citep[e.g.][]{Hardcastle18}. This issue was extensively discussed in \citetalias[]{Ruffa19b}. In summary, both theoretical and observational evidence suggest that the cold gas in LERGs in rich groups and clusters (where they are preferentially located) originates from their hot atmospheres, either directly and smoothly \citep[e.g.][]{Lagos15} or after chaotic cooling  (as predicted in the so-called chaotic cold accretion models, CCA; \citealp[e.g.][]{Naya12,Gaspari13,King15}). External accretion (mostly in the form of interaction with neighbouring galaxies) has been instead proposed as the dominant mechanism for the gas supply in early-type hosts in poor environments \citep[e.g.][]{Sabater15,Davis19,Storchi19}, such as the majority of the LERGs in our sample (see \citetalias[][]{Ruffa19b}). Exploring various observational and theoretical constraints (such as the (mis-)alignment between the CO and the stellar rotation axes, the presence of warps, distortions and lopsidedness in the gas morphology and kinematics, etc.\,), we found that an external gas origin is strongly favoured in sample sources with close companions, including NGC\,3100 (see Section~\ref{sec:sample}). The finding of a tight dust-molecular gas co-spatiality have also important implications in this regard, as the cold gas and dust in NGC\,3100 are clearly still in the process of settling into the host galaxy potential and ``freshly'' cooling cold gas structures are both predicted \citep[e.g.][]{Mathews03a,Hirashita17} and observed \citep[e.g.][]{Lim08,Russell16,Temi18} to be essentially dustless \citep[with typical gas-to-dust mass ratios of $\approx10^{5}$; e.g.][]{Valentini15}.

Estimating the gas-phase metallicity in the regions where dust and gas are co-spatial can provide further constraints in this regard. Indeed, dusty cold gas produced through internal processes (i.e.\,stellar mass loss and/or hot halo cooling) in ETGs is expected to have a higher abundance of heavy elements (i.e.\,super-solar metallicities) than ISM acquired from outside the galaxy \citep[e.g.][]{Martini13,Davis15}. Following \citet[][]{Draine07}, we estimated the gas-phase metallicity in the central dusty gas structure and eastern patch (i.e.\,in the regions where the CO-dust spatial correlation is observed in Figure~\ref{fig:CO_dust}) as:
\begin{eqnarray}\label{eq:metallicity}
12+\log_{10}({\rm O/H})=12+\log_{10}\left(\dfrac{4.57088\times10^{-2}}{M_{\rm gas}/M_{\rm dust}}\right).
\end{eqnarray}
The numerical factor is set as to obtain the solar metallicity ($12 + \log_{10}({\rm O/H}) = 8.66$; \citealp{Asplund04}) at a gas-to-dust mass ratio of 100. $M_{\rm gas}/M_{\rm dust}$ is the gas-to-dust mass ratio in the region(s) of interest, and $M_{\rm gas}$ refer to the total gas mass ($M_{\rm HI} + M_{\rm H_{2}}$). Spatially-resolved observations of the atomic hydrogen in NGC\,3100 are currently not available. We thus estimate a {\it molecular} gas-to-dust mass ratio, which thus yields to an upper limit of the actual gas-phase metallicity.
Based on the works of (e.g.\,) \citet[][]{Kaviraj12}, the mass of dust clumps and patches can be estimated as:
\begin{eqnarray}\label{eq:clumpy_dust_mass}
M_{\rm dust} = \Sigma \langle A_{\rm \lambda} \rangle \Gamma_{\rm \lambda}^{-1},
\end{eqnarray}
where $\Sigma$ is the area covered by the dust features, $ \langle A_{\rm \lambda} \rangle$ is the wavelength-dependent mean extinction (i.e.\,dust absorption) of that area (measured directly from the $B$-$I$ colour map and equal to 2.0 and 1.7 mag in the central structure and eastern patch, respectively), and $\Gamma_{\rm \lambda}$ is the mass absorption coefficient at the observed wavelength. Assuming that the dust properties of ETGs are similar to those of our Galaxy \citep[e.g.][]{Finkelman08}, we adopt the Milky Way $B$-band mass absorption coefficient ($\Gamma_{\rm B} = 8 \times 10^{-6}$~mag~kpc$^{2}$~M$_{\odot}^{-1}$). The estimated dust mass in the regions co-spatial with the molecular gas is then $2.5\times10^5$~M$_{\odot}$. The molecular gas mass in the same areas is $1.4\times10^8$~M$_{\odot}$. These values imply a molecular $M_{\rm gas}/M_{\rm dust}=560$ and a gas-phase metallicity $<$7.9, which about six times smaller than the solar metallicity. Such high molecular gas-to-dust mass ratio at low metallicity is consistent with that measured in galaxies within the DustPedia sample \citep[][]{Casasola20}, and when measured in ETGs is interpreted as an indication of external gas accretion \citep[e.g.][]{Kaviraj12,Davis15}. It is worth noting, however, that a major uncertainty in the estimated gas-phase metallicity is introduced by the molecular gas mass and thus by our assumption on the $X$\textsubscript{CO} conversion factor, which is in turn expected to vary with metallicity (see Section~\ref{sec:mol_masses}). A reliable independent estimate of the gas-phase metallicity is currently not available for NGC\,3100. It is nevertheless possible to qualitatively argue that even accounting for the largest $X$\textsubscript{CO} variations the above result remains unchanged. This is trivial for $X$\textsubscript{CO} larger than the Milky Way value (as typically expected in low metallicity regions; see \citealp[][]{Bolatto13}) and will be thus no further discussed. If we instead take $X$\textsubscript{CO} at the lowest end of the assumed uncertainty range (see Section~\ref{sec:mol_masses}), we obtain a molecular gas-to-dust mass ratio $M_{\rm gas}/M_{\rm dust}\approx200$ and then a gas-phase metallicity which is about 2.5 times smaller than the solar one. We thus conclude that (at least the bulk of) the cold gas and dust in NGC\,3100 originate from the interaction with the close gas-rich late-type companion, NGC\,3095. This scenario is further supported also by recently-acquired neutral hydrogen (HI) observations showing the presence of cold gas filaments in the field between the two companions (Maccagni et al.\,in preparataion).

%While such a high velocity dispersion (along with surface brightness peaks in the same regions) can be considered compatible with the two-armed spiral perturbation we found located in the corresponding areas (see Paper II), an increase in the gas flux density towards higher-J transitions implies higher gas excitation in a area that (at least in projection) appears to be located along the direction of the radio jets (see Figure~\ref{fig:CO_moments}). This supports a scenario in which the physical conditions of the gaseous medium are modified by the transit of the expanding radio jets (see Section~\ref{sec:discuss} for a detailed discussion). 

%The adopted model represents in general a good fit to both the CO(1-0) and CO(3-2) data, in agreement with what found for the CO(2-1) transition modelling. The main features observed in both the molecular gas distribution, such as the central spiral structure (i.e.\,the regions of higher surface brightness at either side of the core; see Figure~\ref{fig:CO_moments}, left panels), and rotation pattern are well reproduced.

\section{Summary}\label{sec:conclusion}
This is the fourth paper of a project aimed at undertaking an extensive multi-phase study of a complete, volume- and flux-limited ($z<0.03$, S\textsubscript{2.7 GHz}$\geq0.25$~Jy) sample of eleven LERGs in the southern sky. In this paper we presented follow-up ALMA observations of one sample source, NGC\,3100, targeting the $^{12}$CO(1-0) and $^{12}$CO(3-2) transitions, along with the dense HCO$^{+}$(4-3) and shock SiO(3-2) and HNCO(6-5) cold gas tracers. These data complemented previously-acquired $^{12}$CO(2-1) and mm-radio continuum observations of NGC\,3100 and allowed us to primarily carry out a detailed analysis of the molecular gas conditions and jet-ISM interaction in this object. 

The main results can be summarised as follows:
\begin{itemize}
\item {\bf Mm-radio continuum:} The multi-frequency (107, 136, 235 and 348~GHz) mm continuum images of NGC\,3100 show the same compact core-double lobe structure seen in the radio continuum at 5 and 10 GHz (maximum linear size $\approx$2~kpc), suggesting that the continuum emission is dominated by non-thermal synchrotron emission up to at least 348 GHz. By fitting the radio-(sub-)mm spectrum, we obtain best-fitting spectral indices $\alpha_{\rm tot}=-0.49\pm0.03$, $\alpha_{\rm core}=-0.3\pm0.2$, and $\alpha_{\rm jet}=-0.73\pm0.01$ for the total (core+jet), core and jet components, respectively. This confirms that the observed continuum arises mainly from synchrotron emission (self-absorbed in the core and optically thin in the jets). In addition, the radio to (sub-)mm morphological and spectral continuum properties (i.e.\,compact linear size and overall steep spectrum) are consistent with those of a radio galaxy in the early phases of its evolution.
\item {\bf CO morphology:} $^{12}$CO(1-0) and $^{12}$CO(3-2) lines are nicely detected and complement our previous $^{12}$CO(2-1) observations. The integrated intensity maps of these lines show that in all the cases the bulk of the emission arises from two bright peaks to the NE and SW of the core, already identified from the previous $^{12}$CO(2-1) analysis as the loci of a two-armed spiral structure. This is surrounded by fainter emission extending up to $\approx2$~kpc along the major axis. In general, substantial differences are observed in the morphology of the three lines: for instance, a gap is observed at the centre of the main CO(1-0) distribution, whereas the CO(2-1) and CO(3-2) lines shows a ring-like and full disc morphology, respectively. This suggests higher gas excitation in the central regions.
\item {\bf CO excitation conditions:} The analysis of the $R_{\rm 21} \equiv S_{\rm CO(2-1)}/S_{\rm CO(1-0)}$ and $R_{\rm 31} \equiv S_{\rm CO(3-2)}/S_{\rm CO(1-0)}$ flux density ratios confirms the presence of two distinct gas components: an outer low-excitation component with line ratios mostly between $\approx1$ and $\approx5$, and an inner high-excitation component with $R_{\rm 21}\geq9$ and $R_{\rm 31}\geq12$. The latter is consistent with gas in an optically thin regime and excitation temperatures $T_{\rm ex}\gtrsim50$~K. Such conditions are very similar to those observed in the few existing spatially-resolved, multi-line studies of jet-ISM interactions. The spatial correlation between the regions where the largest line ratios are observed and the location of the radio lobes suggests that also in NGC\,3100 the expanding radio plasma is likely responsible for the extreme gas conditions in the circumnuclear regions.
\item {\bf CO kinematics:} The mean line-of-sight velocity maps of the three CO lines show that the gas is rotating, but with evident kinematic distortions indicating the presence of unrelaxed substructures in the gas distribution. The accurate 3D kinematic modelling of the CO(3-2) transition allows us to more robustly confirm and complement the results obtained from the same analysis of the sole CO(2-1) transition: the bulk of the cold gas is regularly rotating. Nevertheless, two distinct non-rotating kinematic components can be identified in the central gas regions. The first is observed along the CO kinematic major axis and can be ascribed to inflow/outflow streaming motions (with maximum velocities of $\approx+90$km~s$^{-1}$ and $\approx-250$km~s$^{-1}$, respectively) induced by the inner spiral perturbation.
The second is observed along the kinematic minor axis and can be associated to a jet-induced outflow with $v_{\rm max}\approx 200$~km~s$^{-1}$ and $\dot{M}\lesssim 0.12$~M$_{\odot}$~yr$^{-1}$. These values suggest that the impact of the jet-ISM coupling on the host galaxy is likely negligible at this stage. The outflow kinematic properties are also significantly less extreme than those expected from jet feedback models and observed in other cases of sub-kpc scale jet-cold gas interactions. Such discrepancy may be explained in terms of the evolutionary stage of the radio source, jet power and radiative efficiency of the accretion process.
\item {\bf Dense and shock gas tracers:} Emission from the dense gas tracer HCO$^{+}$(4-3) is tentatively detected (peak S/N$=5$) and consists of a single unresolved structure at 2$''$ west of the nucleus. Such faint emission arises from a region adjacent to the base of the northern radio lobe, where also some of the largest $R_{21}$ and $R_{31}$ values and the jet-induced outflow are observed. This suggests that the detected HCO$^{+}$ emission traces a region of jet-induced gas compression. The HCO$^{+}$(4-3)/CO(2-1) flux density ratio seems consistent with this scenario, reaching peaks of $\approx0.7$ that are similar to those typically observed in regions of jet-ISM interactions. The current HCO$^{+}$ detection, however, is too marginal to carry out a detailed analysis and draw solid conclusions. The SiO(3-2) and HNCO(6-5) shock tracers are not detected. This is surprising, given the obvious perturbations affecting the inner regions of NGC\,3100, but - along with the tentative HCO$^{+}$ detection - may be consistent with an intrinsic deficiency of very dense (i.e.\,$n_{\rm crit} > 10^{6}$~cm$^{3}$) gas in the inner regions of NGC\,3100. 
\item {\bf Cold ISM origin:} The CO-dust comparison demonstrates a tight co-spatiality, indicating that they trace the same ISM component. By measuring the molecular gas-to-dust ratio in the regions where they are co-spatial, we estimate a strongly sub-solar gas-phase metallicity (i.e.\,$<7.9$). This is incompatible with gas produced through internal processes (i.e.\,stellar mass loss and/or hot halo cooling), which in ETGs is expected to have large (super-solar) metal abundances. These and other observational and theoretical constraints analysed in previous works strongly support an external origin of the cold ISM in NGC\,3100, probably via interaction with the companion gas-rich star-forming galaxy NGC\,3095. Our finding provides also support to a scenario in which galaxy-galaxy interactions have a major role in replenishing the cold ISM reservoirs of LERGs located in poor environments, such as NGC\,3100, 
\end{itemize}

\section*{Acknowledgements}
We thank the anonymous referee for useful comments that helped us improving the original manuscript. IR also thanks Q. D'Amato for useful discussions. IR and TAD acknowledge support from the UK Science and Technology Facilities Council through grants ST/S00033X/1 and ST/W000830/1. IP acknowledges support from INAF under the SKA/CTA PRIN “FORECaST” and from PRIN MIUR project “Black Hole winds and the Baryon Life Cycle of Galaxies: the stone-guest at the galaxy evolution supper”, contract \#2017PH3WAT. RP acknowledges the Italian PRIN-Miur 2017 (PI A.\,Cimatti). Part of this work has been carried out in the framework of the first author's PhD research project (IR PhD Thesis: http://amsdottorato.unibo.it/9244/1/PhDThesis\_IRuffa\_finalversion.pdf). This paper makes use of the following ALMA data: ADS/JAO.ALMA\#[2015.1.01572.S] and ADS/JAO.ALMA\#[2018.1.01095.S]. ALMA is a partnership of ESO (representing its member states), NSF (USA) and NINS (Japan), together with NRC (Canada), NSC and ASIAA (Taiwan), and KASI (Republic of Korea), in cooperation with the Republic of Chile. The Joint ALMA Observatory is operated by ESO, AUI/NRAO and NAOJ. The National Radio Astronomy Observatory is a facility of the National Science Foundation operated under cooperative agreement by Associated Universities, Inc. This paper has also made use of the NASA/IPAC Extragalactic Database (NED) which is operated by the Jet Propulsion Laboratory, California Institute of Technology under contract with NASA. This research used the facilities of the Canadian Astronomy Data Centre operated by the National Research Council of Canada with the support of the Canadian Space Agency. We acknowledge the usage of the HyperLeda database (http://leda.univ-lyon1.fr)

%%%%%%%%%%%%%%%%%%%%%%%%%%%%%%%%%%%%%%%%%%%%%%%%%%
\section*{Data Availability}
The ALMA data used in this article are available to download at the ALMA archive (https://almascience.nrao.edu/asax/; project codes: 2015.1.01572.S and 2018.1.01095.S). The calibrated data, final data products and original plots generated for the research study underlying this article will be shared upon reasonable request to the first author.



\bibliographystyle{mnras}
\bibliography{mybibliography} % if your bibtex file is called example.bib

\begin{thebibliography}{}
\makeatletter
\relax
\def\mn@urlcharsother{\let\do\@makeother \do\$\do\&\do\#\do\^\do\_\do\%\do\~}
\def\mn@doi{\begingroup\mn@urlcharsother \@ifnextchar [ {\mn@doi@}
  {\mn@doi@[]}}
\def\mn@doi@[#1]#2{\def\@tempa{#1}\ifx\@tempa\@empty \href
  {http://dx.doi.org/#2} {doi:#2}\else \href {http://dx.doi.org/#2} {#1}\fi
  \endgroup}
\def\mn@eprint#1#2{\mn@eprint@#1:#2::\@nil}
\def\mn@eprint@arXiv#1{\href {http://arxiv.org/abs/#1} {{\tt arXiv:#1}}}
\def\mn@eprint@dblp#1{\href {http://dblp.uni-trier.de/rec/bibtex/#1.xml}
  {dblp:#1}}
\def\mn@eprint@#1:#2:#3:#4\@nil{\def\@tempa {#1}\def\@tempb {#2}\def\@tempc
  {#3}\ifx \@tempc \@empty \let \@tempc \@tempb \let \@tempb \@tempa \fi \ifx
  \@tempb \@empty \def\@tempb {arXiv}\fi \@ifundefined
  {mn@eprint@\@tempb}{\@tempb:\@tempc}{\expandafter \expandafter \csname
  mn@eprint@\@tempb\endcsname \expandafter{\@tempc}}}

\bibitem[\protect\citeauthoryear{{Alatalo} et~al.,}{{Alatalo}
  et~al.}{2013}]{Alatalo13}
{Alatalo} K.,  et~al., 2013, \mn@doi [\mnras] {10.1093/mnras/sts299}, \href
  {http://adsabs.harvard.edu/abs/2013MNRAS.432.1796A} {432, 1796}

\bibitem[\protect\citeauthoryear{{Asplund}, {Grevesse}, {Sauval}, {Allende
  Prieto}  \& {Kiselman}}{{Asplund} et~al.}{2004}]{Asplund04}
{Asplund} M.,  {Grevesse} N.,  {Sauval} A.~J.,  {Allende Prieto} C.,
  {Kiselman} D.,  2004, \mn@doi [\aap] {10.1051/0004-6361:20034328}, \href
  {https://ui.adsabs.harvard.edu/abs/2004A&A...417..751A} {417, 751}

\bibitem[\protect\citeauthoryear{{Best} \& {Heckman}}{{Best} \&
  {Heckman}}{2012}]{Best12}
{Best} P.~N.,  {Heckman} T.~M.,  2012, \mn@doi [\mnras]
  {10.1111/j.1365-2966.2012.20414.x}, \href
  {http://adsabs.harvard.edu/abs/2012MNRAS.421.1569B} {421, 1569}

\bibitem[\protect\citeauthoryear{{Bieri}, {Dubois}, {Rosdahl}, {Wagner}, {Silk}
   \& {Mamon}}{{Bieri} et~al.}{2017}]{Bieri17}
{Bieri} R.,  {Dubois} Y.,  {Rosdahl} J.,  {Wagner} A.,  {Silk} J.,   {Mamon}
  G.~A.,  2017, \mn@doi [\mnras] {10.1093/mnras/stw2380}, \href
  {https://ui.adsabs.harvard.edu/abs/2017MNRAS.464.1854B} {464, 1854}

\bibitem[\protect\citeauthoryear{{Bigiel} et~al.,}{{Bigiel}
  et~al.}{2016}]{Bigiel16}
{Bigiel} F.,  et~al., 2016, \mn@doi [\apjl] {10.3847/2041-8205/822/2/L26},
  \href {https://ui.adsabs.harvard.edu/abs/2016ApJ...822L..26B} {822, L26}

\bibitem[\protect\citeauthoryear{{Bolatto}, {Wolfire}  \& {Leroy}}{{Bolatto}
  et~al.}{2013}]{Bolatto13}
{Bolatto} A.~D.,  {Wolfire} M.,   {Leroy} A.~K.,  2013, \mn@doi [\araa]
  {10.1146/annurev-astro-082812-140944}, \href
  {http://adsabs.harvard.edu/abs/2013ARA%26A..51..207B} {51, 207}

\bibitem[\protect\citeauthoryear{{Bosma}}{{Bosma}}{1981a}]{Bosma81a}
{Bosma} A.,  1981a, \mn@doi [\aj] {10.1086/113062}, \href
  {http://adsabs.harvard.edu/abs/1981AJ.....86.1791B} {86, 1791}

\bibitem[\protect\citeauthoryear{{Bosma}}{{Bosma}}{1981b}]{Bosma81b}
{Bosma} A.,  1981b, \mn@doi [\aj] {10.1086/113063}, \href
  {http://adsabs.harvard.edu/abs/1981AJ.....86.1825B} {86, 1825}

\bibitem[\protect\citeauthoryear{{Brown}, {Jannuzi}, {Floyd}  \&
  {Mould}}{{Brown} et~al.}{2011}]{Brown11}
{Brown} M. J.~I.,  {Jannuzi} B.~T.,  {Floyd} D. J.~E.,   {Mould} J.~R.,  2011,
  \mn@doi [\apjl] {10.1088/2041-8205/731/2/L41}, \href
  {https://ui.adsabs.harvard.edu/abs/2011ApJ...731L..41B} {731, L41}

\bibitem[\protect\citeauthoryear{{Carilli} \& {Walter}}{{Carilli} \&
  {Walter}}{2013}]{Carilli13}
{Carilli} C.~L.,  {Walter} F.,  2013, \mn@doi [\araa]
  {10.1146/annurev-astro-082812-140953}, \href
  {http://adsabs.harvard.edu/abs/2013ARA%26A..51..105C} {51, 105}

\bibitem[\protect\citeauthoryear{{Casasola} et~al.,}{{Casasola}
  et~al.}{2020}]{Casasola20}
{Casasola} V.,  et~al., 2020, \mn@doi [\aap] {10.1051/0004-6361/201936665},
  \href {https://ui.adsabs.harvard.edu/abs/2020A&A...633A.100C} {633, A100}

\bibitem[\protect\citeauthoryear{{Cavagnolo}, {McNamara}, {Nulsen}, {Carilli},
  {Jones}  \& {B{\^\i}rzan}}{{Cavagnolo} et~al.}{2010}]{Cavagnolo10}
{Cavagnolo} K.~W.,  {McNamara} B.~R.,  {Nulsen} P.~E.~J.,  {Carilli} C.~L.,
  {Jones} C.,   {B{\^\i}rzan} L.,  2010, \mn@doi [\apj]
  {10.1088/0004-637X/720/2/1066}, \href
  {https://ui.adsabs.harvard.edu/abs/2010ApJ...720.1066C} {720, 1066}

\bibitem[\protect\citeauthoryear{{Choi}, {Somerville}, {Ostriker}, {Naab}  \&
  {Hirschmann}}{{Choi} et~al.}{2018}]{Choi18}
{Choi} E.,  {Somerville} R.~S.,  {Ostriker} J.~P.,  {Naab} T.,   {Hirschmann}
  M.,  2018, \mn@doi [\apj] {10.3847/1538-4357/aae076}, \href
  {https://ui.adsabs.harvard.edu/abs/2018ApJ...866...91C} {866, 91}

\bibitem[\protect\citeauthoryear{{Cielo}, {Bieri}, {Volonteri}, {Wagner}  \&
  {Dubois}}{{Cielo} et~al.}{2018}]{Cielo18}
{Cielo} S.,  {Bieri} R.,  {Volonteri} M.,  {Wagner} A.~Y.,   {Dubois} Y.,
  2018, \mn@doi [\mnras] {10.1093/mnras/sty708}, \href
  {https://ui.adsabs.harvard.edu/abs/2018MNRAS.477.1336C} {477, 1336}

\bibitem[\protect\citeauthoryear{{Colless} et~al.,}{{Colless}
  et~al.}{2003}]{Colless03}
{Colless} M.,  et~al., 2003, arXiv e-prints, \href
  {https://ui.adsabs.harvard.edu/abs/2003astro.ph..6581C} {pp
  astro--ph/0306581}

\bibitem[\protect\citeauthoryear{{Collobert}, {Sarzi}, {Davies}, {Kuntschner}
  \& {Colless}}{{Collobert} et~al.}{2006}]{Coll06}
{Collobert} M.,  {Sarzi} M.,  {Davies} R.~L.,  {Kuntschner} H.,   {Colless} M.,
   2006, \mn@doi [\mnras] {10.1111/j.1365-2966.2006.10538.x}, \href
  {http://adsabs.harvard.edu/abs/2006MNRAS.370.1213C} {370, 1213}

\bibitem[\protect\citeauthoryear{{Combes}}{{Combes}}{2017}]{Combes17}
{Combes} F.,  2017, \mn@doi [Frontiers in Astronomy and Space Sciences]
  {10.3389/fspas.2017.00010}, \href
  {http://adsabs.harvard.edu/abs/2017FrASS...4...10C} {4, 10}

\bibitem[\protect\citeauthoryear{{Combes} et~al.,}{{Combes}
  et~al.}{2013}]{Combes13}
{Combes} F.,  et~al., 2013, \mn@doi [\aap] {10.1051/0004-6361/201322288}, \href
  {http://adsabs.harvard.edu/abs/2013A%26A...558A.124C} {558, A124}

\bibitem[\protect\citeauthoryear{{Cormier} et~al.,}{{Cormier}
  et~al.}{2018}]{Cormier18}
{Cormier} D.,  et~al., 2018, \mn@doi [\mnras] {10.1093/mnras/sty059}, \href
  {https://ui.adsabs.harvard.edu/abs/2018MNRAS.475.3909C} {475, 3909}

\bibitem[\protect\citeauthoryear{{Costagliola} et~al.,}{{Costagliola}
  et~al.}{2011}]{Costagliola11}
{Costagliola} F.,  et~al., 2011, \mn@doi [\aap] {10.1051/0004-6361/201015628},
  \href {https://ui.adsabs.harvard.edu/abs/2011A&A...528A..30C} {528, A30}

\bibitem[\protect\citeauthoryear{{Dame}}{{Dame}}{2011}]{Dame11}
{Dame} T.~M.,  2011, preprint, \href
  {http://adsabs.harvard.edu/abs/2011arXiv1101.1499D} {} (\mn@eprint {arXiv}
  {1101.1499})

\bibitem[\protect\citeauthoryear{{Dasyra}, {Combes}, {Oosterloo}, {Oonk},
  {Morganti}, {Salom{\'e}}  \& {Vlahakis}}{{Dasyra} et~al.}{2016}]{Dasyra16}
{Dasyra} K.~M.,  {Combes} F.,  {Oosterloo} T.,  {Oonk} J.~B.~R.,  {Morganti}
  R.,  {Salom{\'e}} P.,   {Vlahakis} N.,  2016, \mn@doi [\aap]
  {10.1051/0004-6361/201629689}, \href
  {http://adsabs.harvard.edu/abs/2016A%26A...595L...7D} {595, L7}

\bibitem[\protect\citeauthoryear{{Davis} \& {Bureau}}{{Davis} \&
  {Bureau}}{2016}]{Davis16}
{Davis} T.~A.,  {Bureau} M.,  2016, \mn@doi [\mnras] {10.1093/mnras/stv2998},
  \href {https://ui.adsabs.harvard.edu/abs/2016MNRAS.457..272D} {457, 272}

\bibitem[\protect\citeauthoryear{{Davis} et~al.,}{{Davis}
  et~al.}{2013}]{Davis13}
{Davis} T.~A.,  et~al., 2013, \mn@doi [\mnras] {10.1093/mnras/sts353}, \href
  {http://adsabs.harvard.edu/abs/2013MNRAS.429..534D} {429, 534}

\bibitem[\protect\citeauthoryear{{Davis} et~al.,}{{Davis}
  et~al.}{2015}]{Davis15}
{Davis} T.~A.,  et~al., 2015, \mn@doi [\mnras] {10.1093/mnras/stv1973}, \href
  {http://adsabs.harvard.edu/abs/2015MNRAS.454..657D} {454, 657}

\bibitem[\protect\citeauthoryear{{Davis}, {Bureau}, {Onishi}, {Cappellari},
  {Iguchi}  \& {Sarzi}}{{Davis} et~al.}{2017}]{Davis17}
{Davis} T.~A.,  {Bureau} M.,  {Onishi} K.,  {Cappellari} M.,  {Iguchi} S.,
  {Sarzi} M.,  2017, \mn@doi [\mnras] {10.1093/mnras/stw3217}, \href
  {http://adsabs.harvard.edu/abs/2017MNRAS.468.4675D} {468, 4675}

\bibitem[\protect\citeauthoryear{{Davis}, {Greene}, {Ma}, {Blakeslee},
  {Dawson}, {Pandya}, {Veale}  \& {Zabel}}{{Davis} et~al.}{2019}]{Davis19}
{Davis} T.~A.,  {Greene} J.~E.,  {Ma} C.-P.,  {Blakeslee} J.~P.,  {Dawson}
  J.~M.,  {Pandya} V.,  {Veale} M.,   {Zabel} N.,  2019, \mn@doi [\mnras]
  {10.1093/mnras/stz871}, \href
  {https://ui.adsabs.harvard.edu/abs/2019MNRAS.486.1404D} {486, 1404}

\bibitem[\protect\citeauthoryear{{De Vaucouleurs}}{{De
  Vaucouleurs}}{1976}]{Devac76}
{De Vaucouleurs} G.,  1976, in {Dickens} R.~J.,  {Perry} J.~E.,  {Smith} F.~G.,
    {King} I.~R.,  eds,  Royal Greenwich Observatory Bulletins Vol. 182, The
  Galaxy and the Local Group. p.~177

\bibitem[\protect\citeauthoryear{{Dom{\'\i}nguez-Fern{\'a}ndez}
  et~al.,}{{Dom{\'\i}nguez-Fern{\'a}ndez} et~al.}{2020}]{Dominguez20}
{Dom{\'\i}nguez-Fern{\'a}ndez} A.~J.,  et~al., 2020, \mn@doi [\aap]
  {10.1051/0004-6361/201936961}, \href
  {https://ui.adsabs.harvard.edu/abs/2020A&A...643A.127D} {643, A127}

\bibitem[\protect\citeauthoryear{{Dopita} et~al.,}{{Dopita}
  et~al.}{2015}]{Dopita15}
{Dopita} M.~A.,  et~al., 2015, \mn@doi [\apjs] {10.1088/0067-0049/217/1/12},
  \href {http://adsabs.harvard.edu/abs/2015ApJS..217...12D} {217, 12}

\bibitem[\protect\citeauthoryear{{Draine} et~al.,}{{Draine}
  et~al.}{2007}]{Draine07}
{Draine} B.~T.,  et~al., 2007, \mn@doi [\apj] {10.1086/518306}, \href
  {https://ui.adsabs.harvard.edu/abs/2007ApJ...663..866D} {663, 866}

\bibitem[\protect\citeauthoryear{{Ekers} et~al.,}{{Ekers}
  et~al.}{1989}]{Ekers89}
{Ekers} R.~D.,  et~al., 1989, \mn@doi [\mnras] {10.1093/mnras/236.4.737}, \href
  {http://adsabs.harvard.edu/abs/1989MNRAS.236..737E} {236, 737}

\bibitem[\protect\citeauthoryear{{Fanaroff} \& {Riley}}{{Fanaroff} \&
  {Riley}}{1974}]{Fanaroff74}
{Fanaroff} B.~L.,  {Riley} J.~M.,  1974, \mn@doi [\mnras]
  {10.1093/mnras/167.1.31P}, \href
  {http://adsabs.harvard.edu/abs/1974MNRAS.167P..31F} {167, 31P}

\bibitem[\protect\citeauthoryear{{Fathi} et~al.,}{{Fathi}
  et~al.}{2013}]{Fathi13}
{Fathi} K.,  et~al., 2013, \mn@doi [\apjl] {10.1088/2041-8205/770/2/L27}, \href
  {http://adsabs.harvard.edu/abs/2013ApJ...770L..27F} {770, L27}

\bibitem[\protect\citeauthoryear{{Feruglio} et~al.,}{{Feruglio}
  et~al.}{2015}]{Feruglio15}
{Feruglio} C.,  et~al., 2015, \mn@doi [\aap] {10.1051/0004-6361/201526020},
  \href {https://ui.adsabs.harvard.edu/abs/2015A&A...583A..99F} {583, A99}

\bibitem[\protect\citeauthoryear{{Finkelman} et~al.,}{{Finkelman}
  et~al.}{2008}]{Finkelman08}
{Finkelman} I.,  et~al., 2008, \mn@doi [\mnras]
  {10.1111/j.1365-2966.2008.13785.x}, \href
  {https://ui.adsabs.harvard.edu/abs/2008MNRAS.390..969F} {390, 969}

\bibitem[\protect\citeauthoryear{{Gao} \& {Solomon}}{{Gao} \&
  {Solomon}}{2004}]{Gao04}
{Gao} Y.,  {Solomon} P.~M.,  2004, \mn@doi [\apj] {10.1086/382999}, \href
  {http://adsabs.harvard.edu/abs/2004ApJ...606..271G} {606, 271}

\bibitem[\protect\citeauthoryear{{Garc{\'\i}a-Burillo}
  et~al.,}{{Garc{\'\i}a-Burillo} et~al.}{2010}]{GarciaBurillo10}
{Garc{\'\i}a-Burillo} S.,  et~al., 2010, \mn@doi [\aap]
  {10.1051/0004-6361/201014539}, \href
  {https://ui.adsabs.harvard.edu/abs/2010A&A...519A...2G} {519, A2}

\bibitem[\protect\citeauthoryear{{Garc{\'{\i}}a-Burillo}
  et~al.,}{{Garc{\'{\i}}a-Burillo} et~al.}{2014}]{Garcia14}
{Garc{\'{\i}}a-Burillo} S.,  et~al., 2014, \mn@doi [\aap]
  {10.1051/0004-6361/201423843}, \href
  {http://adsabs.harvard.edu/abs/2014A%26A...567A.125G} {567, A125}

\bibitem[\protect\citeauthoryear{{Gaspari}, {Ruszkowski}  \& {Oh}}{{Gaspari}
  et~al.}{2013}]{Gaspari13}
{Gaspari} M.,  {Ruszkowski} M.,   {Oh} S.~P.,  2013, \mn@doi [\mnras]
  {10.1093/mnras/stt692}, \href
  {http://adsabs.harvard.edu/abs/2013MNRAS.432.3401G} {432, 3401}

\bibitem[\protect\citeauthoryear{{Godfrey} \& {Shabala}}{{Godfrey} \&
  {Shabala}}{2013}]{Godfrey13}
{Godfrey} L.~E.~H.,  {Shabala} S.~S.,  2013, \mn@doi [\apj]
  {10.1088/0004-637X/767/1/12}, \href
  {https://ui.adsabs.harvard.edu/abs/2013ApJ...767...12G} {767, 12}

\bibitem[\protect\citeauthoryear{{Godfrey} \& {Shabala}}{{Godfrey} \&
  {Shabala}}{2016}]{Godfrey16}
{Godfrey} L.~E.~H.,  {Shabala} S.~S.,  2016, \mn@doi [\mnras]
  {10.1093/mnras/stv2712}, \href
  {https://ui.adsabs.harvard.edu/abs/2016MNRAS.456.1172G} {456, 1172}

\bibitem[\protect\citeauthoryear{{Greve}, {Papadopoulos}, {Gao}  \&
  {Radford}}{{Greve} et~al.}{2009}]{Greve09}
{Greve} T.~R.,  {Papadopoulos} P.~P.,  {Gao} Y.,   {Radford} S.~J.~E.,  2009,
  \mn@doi [\apj] {10.1088/0004-637X/692/2/1432}, \href
  {https://ui.adsabs.harvard.edu/abs/2009ApJ...692.1432G} {692, 1432}

\bibitem[\protect\citeauthoryear{{Hardcastle}}{{Hardcastle}}{2018}]{Hardcastle18}
{Hardcastle} M.,  2018, \mn@doi [Nature Astronomy] {10.1038/s41550-018-0424-1},
  \href {http://adsabs.harvard.edu/abs/2018NatAs...2..273H} {2, 273}

\bibitem[\protect\citeauthoryear{{Hardcastle}, {Evans}  \&
  {Croston}}{{Hardcastle} et~al.}{2007}]{Hardcastle07}
{Hardcastle} M.~J.,  {Evans} D.~A.,   {Croston} J.~H.,  2007, \mn@doi [\mnras]
  {10.1111/j.1365-2966.2007.11572.x}, \href
  {http://adsabs.harvard.edu/abs/2007MNRAS.376.1849H} {376, 1849}

\bibitem[\protect\citeauthoryear{{Harrison}, {Molyneux}, {Scholtz}  \&
  {Jarvis}}{{Harrison} et~al.}{2020}]{Harrison20}
{Harrison} C.~M.,  {Molyneux} S.~J.,  {Scholtz} J.,   {Jarvis} M.~E.,  2020,
  arXiv e-prints, \href {https://ui.adsabs.harvard.edu/abs/2020arXiv200601196H}
  {p. arXiv:2006.01196}

\bibitem[\protect\citeauthoryear{{Heckman}, {Kauffmann}, {Brinchmann},
  {Charlot}, {Tremonti}  \& {White}}{{Heckman} et~al.}{2004}]{Heckman04}
{Heckman} T.~M.,  {Kauffmann} G.,  {Brinchmann} J.,  {Charlot} S.,  {Tremonti}
  C.,   {White} S. D.~M.,  2004, \mn@doi [\apj] {10.1086/422872}, \href
  {https://ui.adsabs.harvard.edu/abs/2004ApJ...613..109H} {613, 109}

\bibitem[\protect\citeauthoryear{{Hirashita} \& {Nozawa}}{{Hirashita} \&
  {Nozawa}}{2017}]{Hirashita17}
{Hirashita} H.,  {Nozawa} T.,  2017, \mn@doi [\planss]
  {10.1016/j.pss.2017.01.009}, \href
  {https://ui.adsabs.harvard.edu/abs/2017P&SS..149...45H} {149, 45}

\bibitem[\protect\citeauthoryear{{Ho}, {Li}, {Barth}, {Seigar}  \& {Peng}}{{Ho}
  et~al.}{2011}]{Ho11}
{Ho} L.~C.,  {Li} Z.-Y.,  {Barth} A.~J.,  {Seigar} M.~S.,   {Peng} C.~Y.,
  2011, \mn@doi [\apjs] {10.1088/0067-0049/197/2/21}, \href
  {https://ui.adsabs.harvard.edu/abs/2011ApJS..197...21H} {197, 21}

\bibitem[\protect\citeauthoryear{{Hopkins} \& {Quataert}}{{Hopkins} \&
  {Quataert}}{2010}]{Hopkins10}
{Hopkins} P.~F.,  {Quataert} E.,  2010, \mn@doi [\mnras]
  {10.1111/j.1365-2966.2010.17064.x}, \href
  {http://adsabs.harvard.edu/abs/2010MNRAS.407.1529H} {407, 1529}

\bibitem[\protect\citeauthoryear{{Ishibashi}, {Fabian}  \&
  {Maiolino}}{{Ishibashi} et~al.}{2018}]{Ishibashi18}
{Ishibashi} W.,  {Fabian} A.~C.,   {Maiolino} R.,  2018, \mn@doi [\mnras]
  {10.1093/mnras/sty236}, \href
  {https://ui.adsabs.harvard.edu/abs/2018MNRAS.476..512I} {476, 512}

\bibitem[\protect\citeauthoryear{{Jim{\'e}nez-Donaire}
  et~al.,}{{Jim{\'e}nez-Donaire} et~al.}{2017}]{Jimenez17}
{Jim{\'e}nez-Donaire} M.~J.,  et~al., 2017, \mn@doi [\apjl]
  {10.3847/2041-8213/836/2/L29}, \href
  {https://ui.adsabs.harvard.edu/abs/2017ApJ...836L..29J} {836, L29}

\bibitem[\protect\citeauthoryear{{Jones} et~al.,}{{Jones}
  et~al.}{2009}]{Jones09}
{Jones} D.~H.,  et~al., 2009, \mn@doi [\mnras]
  {10.1111/j.1365-2966.2009.15338.x}, \href
  {http://adsabs.harvard.edu/abs/2009MNRAS.399..683J} {399, 683}

\bibitem[\protect\citeauthoryear{{Kamenetzky}, {Rangwala}  \&
  {Glenn}}{{Kamenetzky} et~al.}{2017}]{Kam17}
{Kamenetzky} J.,  {Rangwala} N.,   {Glenn} J.,  2017, \mn@doi [\mnras]
  {10.1093/mnras/stx1595}, \href
  {https://ui.adsabs.harvard.edu/abs/2017MNRAS.471.2917K} {471, 2917}

\bibitem[\protect\citeauthoryear{{Kamenetzky}, {Privon}  \&
  {Narayanan}}{{Kamenetzky} et~al.}{2018}]{Kam18}
{Kamenetzky} J.,  {Privon} G.~C.,   {Narayanan} D.,  2018, \mn@doi [\apj]
  {10.3847/1538-4357/aab3e2}, \href
  {https://ui.adsabs.harvard.edu/abs/2018ApJ...859....9K} {859, 9}

\bibitem[\protect\citeauthoryear{{Kaviraj} et~al.,}{{Kaviraj}
  et~al.}{2012}]{Kaviraj12}
{Kaviraj} S.,  et~al., 2012, \mn@doi [\mnras]
  {10.1111/j.1365-2966.2012.20957.x}, \href
  {https://ui.adsabs.harvard.edu/abs/2012MNRAS.423...49K} {423, 49}

\bibitem[\protect\citeauthoryear{{Kelly}, {Viti}, {Garc{\'\i}a-Burillo},
  {Fuente}, {Usero}, {Krips}  \& {Neri}}{{Kelly} et~al.}{2017}]{Kelly17}
{Kelly} G.,  {Viti} S.,  {Garc{\'\i}a-Burillo} S.,  {Fuente} A.,  {Usero} A.,
  {Krips} M.,   {Neri} R.,  2017, \mn@doi [\aap] {10.1051/0004-6361/201628946},
  \href {https://ui.adsabs.harvard.edu/abs/2017A&A...597A..11K} {597, A11}

\bibitem[\protect\citeauthoryear{{Kerr} \& {Lynden-Bell}}{{Kerr} \&
  {Lynden-Bell}}{1986}]{Kerr86}
{Kerr} F.~J.,  {Lynden-Bell} D.,  1986, \mn@doi [\mnras]
  {10.1093/mnras/221.4.1023}, \href
  {https://ui.adsabs.harvard.edu/abs/1986MNRAS.221.1023K} {221, 1023}

\bibitem[\protect\citeauthoryear{{King} \& {Nixon}}{{King} \&
  {Nixon}}{2015}]{King15}
{King} A.,  {Nixon} C.,  2015, \mn@doi [\mnras] {10.1093/mnrasl/slv098}, \href
  {http://adsabs.harvard.edu/abs/2015MNRAS.453L..46K} {453, L46}

\bibitem[\protect\citeauthoryear{{Koay}, {Vestergaard}, {Casasola}, {Lawther}
  \& {Peterson}}{{Koay} et~al.}{2016}]{Koay15}
{Koay} J.~Y.,  {Vestergaard} M.,  {Casasola} V.,  {Lawther} D.,   {Peterson}
  B.~M.,  2016, \mn@doi [\mnras] {10.1093/mnras/stv2495}, \href
  {http://adsabs.harvard.edu/abs/2016MNRAS.455.2745K} {455, 2745}

\bibitem[\protect\citeauthoryear{{K{\"o}nig} et~al.,}{{K{\"o}nig}
  et~al.}{2018}]{Konig18}
{K{\"o}nig} S.,  et~al., 2018, \mn@doi [\aap] {10.1051/0004-6361/201732436},
  \href {https://ui.adsabs.harvard.edu/abs/2018A&A...615A.122K} {615, A122}

\bibitem[\protect\citeauthoryear{{Lagos}, {Padilla}, {Davis}, {Lacey}, {Baugh},
  {Gonzalez-Perez}, {Zwaan}  \& {Contreras}}{{Lagos} et~al.}{2015}]{Lagos15}
{Lagos} C.~d.~P.,  {Padilla} N.~D.,  {Davis} T.~A.,  {Lacey} C.~G.,  {Baugh}
  C.~M.,  {Gonzalez-Perez} V.,  {Zwaan} M.~A.,   {Contreras} S.,  2015, \mn@doi
  [\mnras] {10.1093/mnras/stu2763}, \href
  {http://adsabs.harvard.edu/abs/2015MNRAS.448.1271L} {448, 1271}

\bibitem[\protect\citeauthoryear{{Laing} \& {Bridle}}{{Laing} \&
  {Bridle}}{2013}]{LB13}
{Laing} R.~A.,  {Bridle} A.~H.,  2013, \mn@doi [\mnras] {10.1093/mnras/stt531},
  \href {http://adsabs.harvard.edu/abs/2013MNRAS.432.1114L} {432, 1114}

\bibitem[\protect\citeauthoryear{{Laurikainen}, {Salo}, {Buta}, {Knapen},
  {Speltincx}  \& {Block}}{{Laurikainen} et~al.}{2006}]{Lau06}
{Laurikainen} E.,  {Salo} H.,  {Buta} R.,  {Knapen} J.,  {Speltincx} T.,
  {Block} D.,  2006, \mn@doi [\aj] {10.1086/508810}, \href
  {http://adsabs.harvard.edu/abs/2006AJ....132.2634L} {132, 2634}

\bibitem[\protect\citeauthoryear{{Lim}, {Ao}  \& {Dinh-V-Trung}}{{Lim}
  et~al.}{2008}]{Lim08}
{Lim} J.,  {Ao} Y.,   {Dinh-V-Trung} 2008, \mn@doi [\apj] {10.1086/523664},
  \href {https://ui.adsabs.harvard.edu/abs/2008ApJ...672..252L} {672, 252}

\bibitem[\protect\citeauthoryear{{Mahony}, {Oonk}, {Morganti}, {Tadhunter},
  {Bessiere}, {Short}, {Emonts}  \& {Oosterloo}}{{Mahony}
  et~al.}{2016}]{Mahony16}
{Mahony} E.~K.,  {Oonk} J.~B.~R.,  {Morganti} R.,  {Tadhunter} C.,  {Bessiere}
  P.,  {Short} P.,  {Emonts} B.~H.~C.,   {Oosterloo} T.~A.,  2016, \mn@doi
  [\mnras] {10.1093/mnras/stv2456}, \href
  {https://ui.adsabs.harvard.edu/abs/2016MNRAS.455.2453M} {455, 2453}

\bibitem[\protect\citeauthoryear{{Maiolino} et~al.,}{{Maiolino}
  et~al.}{2012}]{Maiolino12}
{Maiolino} R.,  et~al., 2012, \mn@doi [\mnras]
  {10.1111/j.1745-3933.2012.01303.x}, \href
  {https://ui.adsabs.harvard.edu/abs/2012MNRAS.425L..66M} {425, L66}

\bibitem[\protect\citeauthoryear{{Martini}, {Dicken}  \&
  {Storchi-Bergmann}}{{Martini} et~al.}{2013}]{Martini13}
{Martini} P.,  {Dicken} D.,   {Storchi-Bergmann} T.,  2013, \mn@doi [\apj]
  {10.1088/0004-637X/766/2/121}, \href
  {https://ui.adsabs.harvard.edu/abs/2013ApJ...766..121M} {766, 121}

\bibitem[\protect\citeauthoryear{{Mathews} \& {Brighenti}}{{Mathews} \&
  {Brighenti}}{2003}]{Mathews03a}
{Mathews} W.~G.,  {Brighenti} F.,  2003, \mn@doi [\araa]
  {10.1146/annurev.astro.41.090401.094542}, \href
  {https://ui.adsabs.harvard.edu/abs/2003ARA%26A..41..191M} {41, 191}

\bibitem[\protect\citeauthoryear{{McMullin}, {Waters}, {Schiebel}, {Young}  \&
  {Golap}}{{McMullin} et~al.}{2007}]{McMullin07}
{McMullin} J.~P.,  {Waters} B.,  {Schiebel} D.,  {Young} W.,   {Golap} K.,
  2007, in {Shaw} R.~A.,  {Hill} F.,   {Bell} D.~J.,  eds,  Astronomical
  Society of the Pacific Conference Series Vol. 376, Astronomical Data Analysis
  Software and Systems XVI. p.~127

\bibitem[\protect\citeauthoryear{{McNamara} \& {Nulsen}}{{McNamara} \&
  {Nulsen}}{2012}]{McNamara12}
{McNamara} B.~R.,  {Nulsen} P.~E.~J.,  2012, \mn@doi [New Journal of Physics]
  {10.1088/1367-2630/14/5/055023}, \href
  {https://ui.adsabs.harvard.edu/abs/2012NJPh...14e5023M} {14, 055023}

\bibitem[\protect\citeauthoryear{{Morganti}}{{Morganti}}{2017}]{Morganti17}
{Morganti} R.,  2017, \mn@doi [Frontiers in Astronomy and Space Sciences]
  {10.3389/fspas.2017.00042}, \href
  {https://ui.adsabs.harvard.edu/abs/2017FrASS...4...42M} {4, 42}

\bibitem[\protect\citeauthoryear{{Morganti}, {Oosterloo}, {Oonk}, {Frieswijk}
  \& {Tadhunter}}{{Morganti} et~al.}{2015}]{Morganti15}
{Morganti} R.,  {Oosterloo} T.,  {Oonk} J.~B.~R.,  {Frieswijk} W.,
  {Tadhunter} C.,  2015, \mn@doi [\aap] {10.1051/0004-6361/201525860}, \href
  {https://ui.adsabs.harvard.edu/abs/2015A&A...580A...1M} {580, A1}

\bibitem[\protect\citeauthoryear{{Morganti}, {Oosterloo}  \&
  {Tadhunter}}{{Morganti} et~al.}{2020}]{Morganti20b}
{Morganti} R.,  {Oosterloo} T.,   {Tadhunter} C.~N.,  2020, arXiv e-prints,
  \href {https://ui.adsabs.harvard.edu/abs/2020arXiv200504765M} {p.
  arXiv:2005.04765}

\bibitem[\protect\citeauthoryear{{Morganti}, {Oosterloo}  \&
  {Tadhunter}}{{Morganti} et~al.}{2021}]{Morganti21}
{Morganti} R.,  {Oosterloo} T.,   {Tadhunter} C.~N.,  2021, in {Storchi
  Bergmann} T.,  {Forman} W.,  {Overzier} R.,   {Riffel} R.,  eds,  Vol. 359,
  Galaxy Evolution and Feedback across Different Environments. pp 243--248
  (\mn@eprint {arXiv} {2005.04765}), \mn@doi{10.1017/S1743921320001775}

\bibitem[\protect\citeauthoryear{{Mukherjee}, {Bicknell}, {Sutherland }  \&
  {Wagner}}{{Mukherjee} et~al.}{2016}]{Mukherjee16}
{Mukherjee} D.,  {Bicknell} G.~V.,  {Sutherland } R.,   {Wagner} A.,  2016,
  \mn@doi [\mnras] {10.1093/mnras/stw1368}, \href
  {https://ui.adsabs.harvard.edu/abs/2016MNRAS.461..967M} {461, 967}

\bibitem[\protect\citeauthoryear{{Mukherjee}, {Wagner}, {Bicknell}, {Morganti},
  {Oosterloo}, {Nesvadba}  \& {Sutherland}}{{Mukherjee}
  et~al.}{2018a}]{Mukherjee18a}
{Mukherjee} D.,  {Wagner} A.~Y.,  {Bicknell} G.~V.,  {Morganti} R.,
  {Oosterloo} T.,  {Nesvadba} N.,   {Sutherland} R.~S.,  2018a, \mn@doi
  [\mnras] {10.1093/mnras/sty067}, \href
  {https://ui.adsabs.harvard.edu/abs/2018MNRAS.476...80M} {476, 80}

\bibitem[\protect\citeauthoryear{{Mukherjee}, {Bicknell}, {Wagner},
  {Sutherland}  \& {Silk}}{{Mukherjee} et~al.}{2018b}]{Mukherjee18b}
{Mukherjee} D.,  {Bicknell} G.~V.,  {Wagner} A.~Y.,  {Sutherland} R.~S.,
  {Silk} J.,  2018b, \mn@doi [\mnras] {10.1093/mnras/sty1776}, \href
  {https://ui.adsabs.harvard.edu/abs/2018MNRAS.479.5544M} {479, 5544}

\bibitem[\protect\citeauthoryear{{Mukherjee}, {Bicknell}  \&
  {Wagner}}{{Mukherjee} et~al.}{2021}]{Mukherjee21}
{Mukherjee} D.,  {Bicknell} G.~V.,   {Wagner} A.~Y.,  2021, arXiv e-prints,
  \href {https://ui.adsabs.harvard.edu/abs/2021arXiv211011900M} {p.
  arXiv:2110.11900}

\bibitem[\protect\citeauthoryear{{Murthy} et~al.,}{{Murthy}
  et~al.}{2019}]{Murthy19}
{Murthy} S.,  et~al., 2019, \mn@doi [\aap] {10.1051/0004-6361/201935931}, \href
  {https://ui.adsabs.harvard.edu/abs/2019A&A...629A..58M} {629, A58}

\bibitem[\protect\citeauthoryear{{Narayan}, {Quataert}, {Igumenshchev}  \&
  {Abramowicz}}{{Narayan} et~al.}{2002}]{Narayan02}
{Narayan} R.,  {Quataert} E.,  {Igumenshchev} I.~V.,   {Abramowicz} M.~A.,
  2002, \mn@doi [\apj] {10.1086/342159}, \href
  {https://ui.adsabs.harvard.edu/abs/2002ApJ...577..295N} {577, 295}

\bibitem[\protect\citeauthoryear{{Nayakshin}, {Power}  \& {King}}{{Nayakshin}
  et~al.}{2012}]{Naya12}
{Nayakshin} S.,  {Power} C.,   {King} A.~R.,  2012, \mn@doi [\apj]
  {10.1088/0004-637X/753/1/15}, \href
  {http://adsabs.harvard.edu/abs/2012ApJ...753...15N} {753, 15}

\bibitem[\protect\citeauthoryear{{Newville}, {Stensitzki}, {Allen}  \&
  {Ingargiola}}{{Newville} et~al.}{2014}]{Newville14}
{Newville} M.,  {Stensitzki} T.,  {Allen} D.~B.,   {Ingargiola} A.,  2014,
  {LMFIT: Non-Linear Least-Square Minimization and Curve-Fitting for Python},
  \mn@doi{10.5281/zenodo.11813}

\bibitem[\protect\citeauthoryear{{O'Dea} \& {Saikia}}{{O'Dea} \&
  {Saikia}}{2020}]{ODea20}
{O'Dea} C.~P.,  {Saikia} D.~J.,  2020, arXiv e-prints, \href
  {https://ui.adsabs.harvard.edu/abs/2020arXiv200902750O} {p. arXiv:2009.02750}

\bibitem[\protect\citeauthoryear{{Oosterloo}, {Raymond Oonk}, {Morganti},
  {Combes}, {Dasyra}, {Salom{\'e}}, {Vlahakis}  \& {Tadhunter}}{{Oosterloo}
  et~al.}{2017}]{Oosterloo17}
{Oosterloo} T.,  {Raymond Oonk} J.~B.,  {Morganti} R.,  {Combes} F.,  {Dasyra}
  K.,  {Salom{\'e}} P.,  {Vlahakis} N.,   {Tadhunter} C.,  2017, \mn@doi [\aap]
  {10.1051/0004-6361/201731781}, \href
  {http://adsabs.harvard.edu/abs/2017A%26A...608A..38O} {608, A38}

\bibitem[\protect\citeauthoryear{{Petrov}, {Kovalev}, {Fomalont}  \&
  {Gordon}}{{Petrov} et~al.}{2005}]{Petrov05}
{Petrov} L.,  {Kovalev} Y.~Y.,  {Fomalont} E.,   {Gordon} D.,  2005, \mn@doi
  [\aj] {10.1086/426920}, \href
  {https://ui.adsabs.harvard.edu/abs/2005AJ....129.1163P} {129, 1163}

\bibitem[\protect\citeauthoryear{{Pillepich} et~al.,}{{Pillepich}
  et~al.}{2019}]{Pillepich19}
{Pillepich} A.,  et~al., 2019, \mn@doi [\mnras] {10.1093/mnras/stz2338}, \href
  {https://ui.adsabs.harvard.edu/abs/2019MNRAS.490.3196P} {490, 3196}

\bibitem[\protect\citeauthoryear{{Rau} \& {Cornwell}}{{Rau} \&
  {Cornwell}}{2011}]{Rau11}
{Rau} U.,  {Cornwell} T.~J.,  2011, \mn@doi [\aap]
  {10.1051/0004-6361/201117104}, \href
  {http://adsabs.harvard.edu/abs/2011A%26A...532A..71R} {532, A71}

\bibitem[\protect\citeauthoryear{{Ruffa} et~al.,}{{Ruffa}
  et~al.}{2019a}]{Ruffa19a}
{Ruffa} I.,  et~al., 2019a, \mn@doi [\mnras] {10.1093/mnras/stz255}, \href
  {https://ui.adsabs.harvard.edu/abs/2019MNRAS.484.4239R} {484, 4239}

\bibitem[\protect\citeauthoryear{{Ruffa} et~al.,}{{Ruffa}
  et~al.}{2019b}]{Ruffa19b}
{Ruffa} I.,  et~al., 2019b, \mn@doi [\mnras] {10.1093/mnras/stz2368}, \href
  {https://ui.adsabs.harvard.edu/abs/2019MNRAS.489.3739R} {489, 3739}

\bibitem[\protect\citeauthoryear{{Ruffa}, {Laing}, {Prandoni}, {Paladino},
  {Parma}, {Davis}  \& {Bureau}}{{Ruffa} et~al.}{2020}]{Ruffa20}
{Ruffa} I.,  {Laing} R.~A.,  {Prandoni} I.,  {Paladino} R.,  {Parma} P.,
  {Davis} T.~A.,   {Bureau} M.,  2020, \mn@doi [\mnras]
  {10.1093/mnras/staa3166}, \href
  {https://ui.adsabs.harvard.edu/abs/2020MNRAS.499.5719R} {499, 5719}

\bibitem[\protect\citeauthoryear{{Rupen}}{{Rupen}}{1999}]{Rupen99}
{Rupen} M.~P.,  1999, in {Taylor} G.~B.,  {Carilli} C.~L.,   {Perley} R.~A.,
  eds,  Astronomical Society of the Pacific Conference Series Vol. 180,
  Synthesis Imaging in Radio Astronomy II. p.~229

\bibitem[\protect\citeauthoryear{{Russell} et~al.,}{{Russell}
  et~al.}{2016}]{Russell16}
{Russell} H.~R.,  et~al., 2016, \mn@doi [\mnras] {10.1093/mnras/stw409}, \href
  {https://ui.adsabs.harvard.edu/abs/2016MNRAS.458.3134R} {458, 3134}

\bibitem[\protect\citeauthoryear{{Sabater}, {Best}  \& {Heckman}}{{Sabater}
  et~al.}{2015}]{Sabater15}
{Sabater} J.,  {Best} P.~N.,   {Heckman} T.~M.,  2015, \mn@doi [Monthly Notices
  of the Royal Astronomical Society] {10.1093/mnras/stu2429}, \href
  {https://ui.adsabs.harvard.edu/abs/2015MNRAS.447..110S} {447, 110}

\bibitem[\protect\citeauthoryear{{Saito} et~al.,}{{Saito}
  et~al.}{2017}]{Saito17}
{Saito} T.,  et~al., 2017, \mn@doi [\apj] {10.3847/1538-4357/835/2/174}, \href
  {https://ui.adsabs.harvard.edu/abs/2017ApJ...835..174S} {835, 174}

\bibitem[\protect\citeauthoryear{{Sandage} \& {Brucato}}{{Sandage} \&
  {Brucato}}{1979}]{Sandage79}
{Sandage} A.,  {Brucato} R.,  1979, \mn@doi [\aj] {10.1086/112440}, \href
  {http://adsabs.harvard.edu/abs/1979AJ.....84..472S} {84, 472}

\bibitem[\protect\citeauthoryear{{Santoro}, {Tadhunter}, {Baron}, {Morganti}
  \& {Holt}}{{Santoro} et~al.}{2020}]{Santoro20}
{Santoro} F.,  {Tadhunter} C.,  {Baron} D.,  {Morganti} R.,   {Holt} J.,  2020,
  arXiv e-prints, \href {https://ui.adsabs.harvard.edu/abs/2020arXiv200911175S}
  {p. arXiv:2009.11175}

\bibitem[\protect\citeauthoryear{{Schilke}, {Walmsley}, {Pineau des Forets}  \&
  {Flower}}{{Schilke} et~al.}{1997}]{Schilke97}
{Schilke} P.,  {Walmsley} C.~M.,  {Pineau des Forets} G.,   {Flower} D.~R.,
  1997, \aap, \href {https://ui.adsabs.harvard.edu/abs/1997A&A...321..293S}
  {321, 293}

\bibitem[\protect\citeauthoryear{{Smith}, {Lucey}, {Hudson}, {Schlegel}  \&
  {Davies}}{{Smith} et~al.}{2000}]{Smith00}
{Smith} R.~J.,  {Lucey} J.~R.,  {Hudson} M.~J.,  {Schlegel} D.~J.,   {Davies}
  R.~L.,  2000, \mn@doi [\mnras] {10.1046/j.1365-8711.2000.03251.x}, \href
  {http://adsabs.harvard.edu/abs/2000MNRAS.313..469S} {313, 469}

\bibitem[\protect\citeauthoryear{{Storchi-Bergmann} \&
  {Schnorr-M{\"u}ller}}{{Storchi-Bergmann} \&
  {Schnorr-M{\"u}ller}}{2019}]{Storchi19}
{Storchi-Bergmann} T.,  {Schnorr-M{\"u}ller} A.,  2019, \mn@doi [Nature
  Astronomy] {10.1038/s41550-018-0611-0}, \href
  {https://ui.adsabs.harvard.edu/abs/2019NatAs...3...48S} {3, 48}

\bibitem[\protect\citeauthoryear{{Swinbank}, {Sobral}, {Smail}, {Geach},
  {Best}, {McCarthy}, {Crain}  \& {Theuns}}{{Swinbank}
  et~al.}{2012}]{Swinbank12}
{Swinbank} A.~M.,  {Sobral} D.,  {Smail} I.,  {Geach} J.~E.,  {Best} P.~N.,
  {McCarthy} I.~G.,  {Crain} R.~A.,   {Theuns} T.,  2012, \mn@doi [\mnras]
  {10.1111/j.1365-2966.2012.21774.x}, \href
  {http://adsabs.harvard.edu/abs/2012MNRAS.426..935S} {426, 935}

\bibitem[\protect\citeauthoryear{{Tadhunter}, {Morganti}, {di
  Serego-Alighieri}, {Fosbury}  \& {Danziger}}{{Tadhunter}
  et~al.}{1993}]{Tad93}
{Tadhunter} C.~N.,  {Morganti} R.,  {di Serego-Alighieri} S.,  {Fosbury}
  R.~A.~E.,   {Danziger} I.~J.,  1993, \mn@doi [\mnras]
  {10.1093/mnras/263.4.999}, \href
  {http://adsabs.harvard.edu/abs/1993MNRAS.263..999T} {263, 999}

\bibitem[\protect\citeauthoryear{{Temi}, {Amblard}, {Gitti}, {Brighenti},
  {Gaspari}, {Mathews}  \& {David}}{{Temi} et~al.}{2018}]{Temi18}
{Temi} P.,  {Amblard} A.,  {Gitti} M.,  {Brighenti} F.,  {Gaspari} M.,
  {Mathews} W.~G.,   {David} L.,  2018, \mn@doi [\apj]
  {10.3847/1538-4357/aab9b0}, \href
  {http://adsabs.harvard.edu/abs/2018ApJ...858...17T} {858, 17}

\bibitem[\protect\citeauthoryear{{Topal}, {Bureau}, {Davis}, {Krips}, {Young}
  \& {Crocker}}{{Topal} et~al.}{2016}]{Topal16}
{Topal} S.,  {Bureau} M.,  {Davis} T.~A.,  {Krips} M.,  {Young} L.~M.,
  {Crocker} A.~F.,  2016, \mn@doi [\mnras] {10.1093/mnras/stw2257}, \href
  {https://ui.adsabs.harvard.edu/abs/2016MNRAS.463.4121T} {463, 4121}

\bibitem[\protect\citeauthoryear{{Tremaine} et~al.,}{{Tremaine}
  et~al.}{2002}]{Tremaine02}
{Tremaine} S.,  et~al., 2002, \mn@doi [\apj] {10.1086/341002}, \href
  {https://ui.adsabs.harvard.edu/abs/2002ApJ...574..740T} {574, 740}

\bibitem[\protect\citeauthoryear{{Tremblay} et~al.,}{{Tremblay}
  et~al.}{2016}]{Tremblay16}
{Tremblay} G.~R.,  et~al., 2016, \mn@doi [\nat] {10.1038/nature17969}, \href
  {http://adsabs.harvard.edu/abs/2016Natur.534..218T} {534, 218}

\bibitem[\protect\citeauthoryear{{Usero}, {Garc{\'\i}a-Burillo},
  {Mart{\'\i}n-Pintado}, {Fuente}  \& {Neri}}{{Usero} et~al.}{2006}]{Usero06}
{Usero} A.,  {Garc{\'\i}a-Burillo} S.,  {Mart{\'\i}n-Pintado} J.,  {Fuente} A.,
    {Neri} R.,  2006, \mn@doi [\aap] {10.1051/0004-6361:20054033}, \href
  {https://ui.adsabs.harvard.edu/abs/2006A&A...448..457U} {448, 457}

\bibitem[\protect\citeauthoryear{{Usero} et~al.,}{{Usero}
  et~al.}{2015}]{Usero15}
{Usero} A.,  et~al., 2015, \mn@doi [\aj] {10.1088/0004-6256/150/4/115}, \href
  {https://ui.adsabs.harvard.edu/abs/2015AJ....150..115U} {150, 115}

\bibitem[\protect\citeauthoryear{{Utomo}, {Blitz}, {Davis}, {Rosolowsky},
  {Bureau}, {Cappellari}  \& {Sarzi}}{{Utomo} et~al.}{2015}]{Utomo15}
{Utomo} D.,  {Blitz} L.,  {Davis} T.,  {Rosolowsky} E.,  {Bureau} M.,
  {Cappellari} M.,   {Sarzi} M.,  2015, \mn@doi [\apj]
  {10.1088/0004-637X/803/1/16}, \href
  {https://ui.adsabs.harvard.edu/abs/2015ApJ...803...16U} {803, 16}

\bibitem[\protect\citeauthoryear{{Valentini} \& {Brighenti}}{{Valentini} \&
  {Brighenti}}{2015}]{Valentini15}
{Valentini} M.,  {Brighenti} F.,  2015, \mn@doi [\mnras]
  {10.1093/mnras/stv090}, \href
  {https://ui.adsabs.harvard.edu/abs/2015MNRAS.448.1979V} {448, 1979}

\bibitem[\protect\citeauthoryear{{Veilleux}, {Maiolino}, {Bolatto}  \&
  {Aalto}}{{Veilleux} et~al.}{2020}]{Veilleux20}
{Veilleux} S.,  {Maiolino} R.,  {Bolatto} A.~D.,   {Aalto} S.,  2020, \mn@doi
  [\aapr] {10.1007/s00159-019-0121-9}, \href
  {https://ui.adsabs.harvard.edu/abs/2020A&ARv..28....2V} {28, 2}

\bibitem[\protect\citeauthoryear{{Venturi} et~al.,}{{Venturi}
  et~al.}{2020}]{Venturi20}
{Venturi} G.,  et~al., 2020, arXiv e-prints, \href
  {https://ui.adsabs.harvard.edu/abs/2020arXiv201104677V} {p. arXiv:2011.04677}

\bibitem[\protect\citeauthoryear{{Wagner}, {Bicknell}  \& {Umemura}}{{Wagner}
  et~al.}{2012}]{Wagner12}
{Wagner} A.~Y.,  {Bicknell} G.~V.,   {Umemura} M.,  2012, \mn@doi [\apj]
  {10.1088/0004-637X/757/2/136}, \href
  {http://adsabs.harvard.edu/abs/2012ApJ...757..136W} {757, 136}

\bibitem[\protect\citeauthoryear{{Wagner}, {Bicknell}, {Umemura}, {Sutherland }
   \& {Silk}}{{Wagner} et~al.}{2016}]{Wagner16}
{Wagner} A.~Y.,  {Bicknell} G.~V.,  {Umemura} M.,  {Sutherland } R.~S.,
  {Silk} J.,  2016, \mn@doi [Astronomische Nachrichten]
  {10.1002/asna.201512287}, \href
  {https://ui.adsabs.harvard.edu/abs/2016AN....337..167W} {337, 167}

\bibitem[\protect\citeauthoryear{{Wrobel} \& {Walker}}{{Wrobel} \&
  {Walker}}{1999}]{Wrobel99}
{Wrobel} J.~M.,  {Walker} R.~C.,  1999, in {Taylor} G.~B.,  {Carilli} C.~L.,
  {Perley} R.~A.,  eds,  Astronomical Society of the Pacific Conference Series
  Vol. 180, Synthesis Imaging in Radio Astronomy II. p.~171

\bibitem[\protect\citeauthoryear{{Wylezalek} \& {Morganti}}{{Wylezalek} \&
  {Morganti}}{2018}]{Wylezalek18}
{Wylezalek} D.,  {Morganti} R.,  2018, \mn@doi [Nature Astronomy]
  {10.1038/s41550-018-0409-0}, \href
  {https://ui.adsabs.harvard.edu/abs/2018NatAs...2..181W} {2, 181}

\bibitem[\protect\citeauthoryear{{Young} et~al.,}{{Young}
  et~al.}{2014}]{Young14}
{Young} L.~M.,  et~al., 2014, \mn@doi [\mnras] {10.1093/mnras/stt2474}, \href
  {http://adsabs.harvard.edu/abs/2014MNRAS.444.3408Y} {444, 3408}

\bibitem[\protect\citeauthoryear{{Young}, {Meier}, {Bureau}, {Crocker}, {Davis}
   \& {Topal}}{{Young} et~al.}{2021}]{Young21}
{Young} L.~M.,  {Meier} D.~S.,  {Bureau} M.,  {Crocker} A.,  {Davis} T.~A.,
  {Topal} S.,  2021, \mn@doi [\apj] {10.3847/1538-4357/abe126}, \href
  {https://ui.adsabs.harvard.edu/abs/2021ApJ...909...98Y} {909, 98}

\bibitem[\protect\citeauthoryear{{Zovaro}, {Sharp}, {Nesvadba}, {Bicknell},
  {Mukherjee}, {Wagner}, {Groves}  \& {Krishna}}{{Zovaro}
  et~al.}{2019}]{Zovaro19}
{Zovaro} H. R.~M.,  {Sharp} R.,  {Nesvadba} N. P.~H.,  {Bicknell} G.~V.,
  {Mukherjee} D.,  {Wagner} A.~Y.,  {Groves} B.,   {Krishna} S.,  2019, \mn@doi
  [\mnras] {10.1093/mnras/stz233}, \href
  {https://ui.adsabs.harvard.edu/abs/2019MNRAS.484.3393Z} {484, 3393}

\bibitem[\protect\citeauthoryear{{van de Ven} \& {Fathi}}{{van de Ven} \&
  {Fathi}}{2010}]{deVen10}
{van de Ven} G.,  {Fathi} K.,  2010, \mn@doi [\apj]
  {10.1088/0004-637X/723/1/767}, \href
  {https://ui.adsabs.harvard.edu/abs/2010ApJ...723..767V} {723, 767}

\bibitem[\protect\citeauthoryear{{van de Voort} et~al.,}{{van de Voort}
  et~al.}{2018}]{Voort18}
{van de Voort} F.,  et~al., 2018, \mn@doi [\mnras] {10.1093/mnras/sty228},
  \href {http://adsabs.harvard.edu/abs/2018MNRAS.476..122V} {476, 122}

\bibitem[\protect\citeauthoryear{{van der Kruit} \& {Shostak}}{{van der Kruit}
  \& {Shostak}}{1982}]{Kruit82}
{van der Kruit} P.~C.,  {Shostak} G.~S.,  1982, \aap, \href
  {http://adsabs.harvard.edu/abs/1982A%26A...105..351V} {105, 351}

\makeatother
\end{thebibliography}

%\begin{thebibliography}{99}
%\end{thebibliography}

%%%%%%%%%%%%%%%%%%%%%%%%%%%%%%%%%%%%%%%%%%%%%%%%%%

%%%%%%%%%%%%%%%%% APPENDICES %%%%%%%%%%%%%%%%%%%%%

\appendix

%%%%%%%%%%%%%%%%%%%%%%%%%%%%%%%%%%%%%%%%%%%%%%%%%%

% Don't change these lines
\bsp	% typesetting comment
\label{lastpage}
\end{document}